\documentclass[12pt]{article}
\usepackage{amsfonts}
\usepackage{amsmath}
\usepackage{amssymb}
\usepackage{epsfig}
\usepackage{color}
\usepackage{float}

\definecolor{darkblue}{rgb}{0,0,.5}

\begin{document}

\begin{flushright}
02/2017\\
\end{flushright}
\vspace{5mm}
\begin{center}
%\large {\bf Particle Physics and Cosmology in the Microscopic Model}\\
\large {\bf Review of the Microscopic Approach to the Higgs Mechanism and to Quark and Lepton Masses and Mixings}\\
\mbox{ }\\
\normalsize
\vspace{2.0cm}
{\bf Bodo Lampe} \\              
\vspace{0.3cm}
II. Institut f\"ur theoretische Physik der Universit\"at Hamburg \\
Luruper Chaussee 149, 22761 Hamburg, Germany \\
%e-mail: Lampe.Bodo@web.de \\   

\vspace{0.8cm}

\begin{figure}[H]
%[H] in verbindung mit package float
\begin{center}
%\epsfig{file=fig1philo.eps,height=5.4cm}
%BESSER GANZ OHNE PHOTO
%\epsfig{file=Bodolampe2014a.eps,height=6.0cm}
%\epsfig{file=Bodolampe2008.eps,height=6.0cm}
\end{center}
\end{figure}

%\vspace{0.8cm}
%\vspace{3.0cm}
{\bf Abstract}\\
\end{center} 
%abstract vage halten?
This review summarizes %***This article summarizes and extends 
the results of a series of recent papers\cite{lhiggs,lmass,la4xz2,llett,lfound}, where a microscopic structure underlying the physics of elementary particles has been proposed. The 'tetron model' relies on the existence of an internal isospin space, in which an independent physical dynamics takes place. This idea is critically re-considered in the present work.
%Open problems are listed, questions are posed and, as far as possible, answers are given, some of which undecided at the present stage of knowledge.
As becomes evident in the course of discussion, the model not only describes electroweak phenomena but also modifies our understanding of other physical topics, like gravity, the big bang cosmology and the nature of the strong interactions.

%\thispagestyle{empty} %dies wuerde die erste Seite ohne Seitenzahl

%statt R(n) lieber die reellen Zahlen  \mathbb{C}
%tensor ist \otimes
%\frac{1}{2}
%statt slash bar f soll man in den nichtrelativistischen 
%Ausruecke f kreuz verwenden! \dagger

\newpage

\normalsize
%\large
%DIES GIBT GRoeSSERE SCHRIFT

%\section{nicht vergessen} 

%--------------------------------------------------------

%\newpage
%\vspace{3.0cm}

%\begin{center}
%\emph{Der Kampf der Vernunft besteht darin,}\\

%\emph{dasjenige, was der Verstand fixiert hat,}\\ 

%\emph{zu \"uberwinden.}\\

%{\sc G. W. F. Hegel}

%\end{center}

\newpage

\section{Particle Physics and Cosmology in the Tetron Model}

The Standard Model of elementary particles is very successful on the phenomenological level. The outcome of (almost) any particle physics experiment can be predicted accurately within this model, and where not, by some straightforward extension. For example, one may introduce right handed neutrinos to account for tiny neutrino masses\cite{nuemix}.

Nevertheless, it is widely believed that the SM is only an effective low-energy theory valid below a certain energy scale, which is supposed to be larger than 1 TeV. This view is based on the fact that the SM has many unknown parameters with hitherto unexplained hierarchies. Furthermore, there is one rather mysterious component, the so-called Higgs field, needed for the spontaneous symmetry breaking (SSB) to take place in the model.

In recent papers a microscopic model has been developed\cite{lhiggs,lmass,la4xz2,llett,lfound}, whose central assumption is the existence of a 3-dimensional {\it internal tetrahedral structure} attributed to each point of Minkowski space, in which an independent physical dynamics takes place. 

%die Sache mit dem Gesamtraum und der 8 koennte auch in den Frageteil, aber ich finde sie gehoert doch hierher
Under this assumption spacetime originally is 6+1 dimensional, and at the time when the tetrahedrons are formed, it fibers into internal space and Minkowski space as $R ^3\times R^{3,1}$. 

The sites $i=1,2,3,4$ of a tetrahedron in $R^{6+1}$ are populated by spinor fields $\psi$, called \textcolor{blue}{\bf{\underline{tetrons}}}. The tetron on site i will be denoted $\psi_i$. 
% or \textcolor{red}{lampeon}.
%\textcolor{cyan}{tetron} \textcolor{yellow}{tetron}{\it tetron}. 
%For reasons that will become clear later, on each site i=1,2,3,4 of the tetrahedral structure there are 2 tetron fields $\psi_i$ and $\psi'_i$). 

%Originally, the tetron $\psi$ can be interpreted as a relativistic fermion in a larger 6+1 dimensional space given by (U,D) when Minkowski and internal space separate. Namely, there is a single 8-dimensional spinor representation in SO(6,1) decomposing as
%zzz(ACHTUNG: DIE 8 IN SO61 IST NICHT REELL, IM GEGENSATZ ZU SO7 Aber es gibt nur eine, und darum vermutlich keine Antiteilchen, ODER DOCH???? zerfallen tut sie in Dirac von 3+1)
%zzz(man startet mit SO61 und einem reellen 8=(12+21,2) Gitter ohne Antiteilchen
%zzz Problem: Im Gegensatz zu 8 von SO7 ist 8 von So61 doch eine komplexe Darstellung, enthaelt also Teilchen und Antiteilchen!? SO62 entsteht wenn sich iim inneren und phys Raum die Zeiten trennen, aber dann ist es innen nicht relativistisch, was SO62 auch uninteressant macht. Man braucht es nur fue qed62)
The fundamental spinor representation in $R^{6+1}$ is of dimension 8. It decomposes as\cite{slansky}
\begin{eqnarray}
8 \rightarrow (1,2,2)+(2,1,2)=((1,2)+(2,1),2)
\label{eq8}
\end{eqnarray}
under the fibration $SO(6,1) \rightarrow SO(3,1) \times SO(3)$\footnote{
Here representations of $SO(3,1) \times SO(3)$ are denoted by a set of 3 numbers $(a,b,c)$, where $(a,b)$ are representations of the Lorentz group and $c$ is the dimension of a $SO(3)$-representation. For example, c=2 corresponds to a non-relativistic Pauli spinor in internal space, whose 2 spin orientations are identified with the SU(2) flavors U and D. It should be noted that (1,2,2) and (2,1,2) are complex conjugate with respect to each other, so one is the antiparticle representation of the other.}.

Eq. (\ref{eq8}) means that each tetron is an isospin doublet $\psi=(U,D)$ of 3+1 dimensional Dirac fermions U and D. One may write it as a 2-index object $\psi_\alpha^a$, where $\alpha=1,2,3,4$ is the Dirac index and $a=1,2$ the internal index. The internal spin will be called isospin.

Using the triplet $\vec \tau$ of internal Pauli matrices an isospin (pseudo)vector 
\begin{eqnarray}
\vec Q=\psi^\dagger \vec \tau\psi
\label{eq89p}
\end{eqnarray}
may be defined for any tetron $\psi$. It fixes a direction in the internal space and, up to an overall constant, can be interpreted as the internal angular momentum vector of the tetron $\psi$.

Since the tetrons are Dirac fermions on Minkowski space, $\vec Q$ can be written in terms of creation and annihilation operators of a tetron ($a^\dagger$ and $a$) and an antitetron ($b^\dagger$ and $b$) as 
\begin{eqnarray}
\vec Q= \psi^\dagger  \vec \tau \psi=a^\dagger \vec \tau a   - b^\dagger  \vec \tau b  
\label{m1u122}
\end{eqnarray}

For the calculation of the quark and lepton masses the chiral iso-vectors 
\begin{eqnarray}
\vec S := \vec Q_L=\frac{1}{2}\psi^\dagger (1-\gamma_5)\vec \tau\psi 
\qquad \qquad
\vec T := \vec Q_R=\frac{1}{2}\psi^\dagger (1+\gamma_5)\vec \tau\psi 
\label{eq894}
\end{eqnarray}
turn out to be of particular importance. For simplicity of notation they are called $\vec S$ and $\vec T$ in the following. Obviously, they fulfill $\vec Q=\vec S +\vec T$.
%(statt QR besser antiQR benutzen! aber QR enthaelt Antiteilchen)

%The tetrons $\psi_i$ and $\psi_{i+1}$ are assumed to be tightly bound pairs.(ich denke jetzt dass die Hälfte Antitetrons sind.) In principle, they should sit on different sites in space. However, since they are so tightly bound they are both attributed to one site i of the tetrahedron. The existence of these pairs is required in order to generate the 24 d.o.f. for the quarks and leptons, see below.

In fig. 1 the {\it local} ground state of the model is drawn, a configuration with 4 tetrons on the 4 sites of a tetrahedron, their isospin vectors $\vec Q$ pointing in radial directions away from the origin.
%their isospin vectors $Q_{Li}$ and $Q_{Ri}$ pointing in radial directions away from the origin. 
These internal vectors fulfill the commutation relations of a system of decoupled internal angular momenta. In other words, they play the role of angular momentum observables corresponding to rotations of the internal $R^3$ space.
%(in diesem Teil besser auf L und R Spins verzichten, Teilchen und Antiteilchen zusammennehmen und nur in den Fragen ausklamuesern. Vorteil waere: man kann alle Faelle abdecken, linkshaendig und rechtshaendig, Teilchen und Antiteilchen
%(with all in all 4$\times$2=8 particles $\psi_{1-8}$ on each tetrahedron together with their internal spin vectors (\ref{eq89p}). The coordinate symmetry for this system is $A_4\times Z_2$ where $A_4$ is the proper tetrahedral symmetry group and the $Z_2$ factor arises from the pairing on each tetrahedral site.)
%(aber wenn Z2 so unabhaengig ist, wie kann daraus Isospin entstehen? weil es eben doch einen kleinen Abstand im inneren Raum gibt. Damit ist aber die Theorie obsolet, dass die Verbindungslinie der Paare in den ausseren Raum geht) 

While the coordinate symmetry is $S_4$, the arrangement of isospin vectors in fig. 1 respects the Shubnikov symmetry\cite{shub,borov,white} 
\begin{eqnarray}
G_4:=A_4+S(S_4-A_4)
\label{eq8gs}
\end{eqnarray}
where $A_4 (S_4)$ is the (full) tetrahedral symmetry group and S the internal time reversal operation that changes the direction of internal spin vectors. This is equivalent to saying that S interchanges the role of the internal spinors in the following way
\begin{eqnarray}
S:(U,D)\rightarrow (-D^*,U^*)
\label{eq8it}
\end{eqnarray}
%(remembering that charge conjugation acts as $C:(U,D)\rightarrow (-\bar D,\bar U)$, S can be considered as part of C; external parity vertauscht L und R, waehrend interne Paritaet R so aehnlich wirkt; SR=laesst vec Qinvariant????)
%As shown later in (\ref{appii4}), one may actually use charge conjugation instead of the concept of an internal time to define the Shubnikov group $G_4$.

Note the arrangement fig. 1 does not respect S or internal parity $P_{in}$, but only the product S$P_{in}$. Furthermore it is chiral, the configuration with opposite internal chirality being given when the isospin vectors would point inwards instead of outwards. As will be shown in (\ref{appii5}), this internal chirality is dynamically related to the $V-A$ nature of the weak interaction.
%(frustriert nicht erwaehnen, da ja die Paare jeweils durchaus parallel stehen)
%(NICHT BRINGEN WEIL ICH JETZT EINEN KLEINEN ABSTAND INNERHALB EINES PAARES ANNEHME If $\vec Q_i$ and $\vec Q_{i+1}$ were identical observables, this configuration would possess an internal time reversal symmetry (with symmetry group the 'grey' group $S_4 \times \{ 1,S\}$), because the time reversal invariance broken by the set of vectors pointing outwards would be restored by those pointing inwards. However, since $\vec Q_i$ and $\vec Q_{i+1}$ are physically different, the ground state has still the Shubnikov group $A_4+S(S_4-A_4)$ as symmetry.)

In a relativistic environment containing antiparticles the definition (\ref{eq8gs}) of the Shubnikov group has to be modified to 
\begin{eqnarray}
%G_4^{rel}:=A_4+CP_{au}T(S_4-A_4)
G_4:=A_4+CP_{au}T(S_4-A_4)
\label{eq8rela}
\end{eqnarray}
This will be detailed later in (\ref{appiicpt}) and figs. 4 and 5. C is the charge conjugation operator of a Dirac field and $P_{au}$ the ('external') parity transformation in physical space. Since the elements of $S_4-A_4$ contain an implicit factor of internal parity $P_{in}$, the symmetry (\ref{eq8rela}) certifies CPT invariance of the local ground state in the full of $R^{6+1}$, cf. (\ref{appiicpt}). Furthermore, the concept of an internal time S is dispensible here, so instead of S ordinary time reversal T may be used in (\ref{eq8rela}).

\begin{figure}
\begin{center}
\epsfig{file=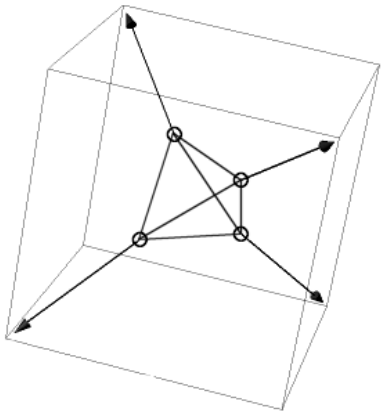,height=7.0cm}
%\bigskip
\caption{The local ground state of the model, living in a 3-dimensional isospin space called the 'fiber'. Shown are the tetron locations (open circles) and the 4 ground state isospin vectors $\langle \vec Q_i\rangle$, whose excitations will be identified with the spectrum of quarks and leptons. The origin of coordinates is taken to be the center of the tetrahedron, and is identical to the base point of the fiber in Minkowski space. The tetrahedron itself has the tetrahedral group $S_4$ as point group symmetry. However, due to the pseudovector property of the isospin vectors the whole system has the Shubnikov point symmetry (\ref{eq8gs}). The Shubnikov group is chiral, the configuration with opposite chirality being given when the 4 isospin vectors would point inwards instead of outwards. Before the formation of the chiral tetrahedron, the internal spins U and D, which according to (\ref{eq89p}) are the building blocks of the isospin vectors, can freely rotate and thus there is an internal spin SU(2) symmetry group, which however is broken to $G_4$ when the chiral tetrahedron is formed. Note there are actually 2 tetrahedrons in this figure, one with respect to the internal coordinates (tetron locations) and the other one with respect to isospin vectors, and both tetrahedrons are 'aligned', in the sense that the coordinate vectors and the isospin vectors point into the same (radial) direction. This 'alignment' of coordinate and isospin vectors within one fiber has to be distinguished from the alignment of isospin vectors with respect to the isospins of neighboring tetrahedrons, as shown in fig. 2. The latter forms the basis for the electroweak phase transition, while the coordinate alignment of neighboring tetrahedrons is relevant for crystal formation at big bang temperatures.}
%cf. the discussion in (\ref{appii206k}), (\ref{appii206m}) and before and after (\ref{eqfe}).
%(bei elastisch sind die Tetraeder evtl frei gegeneinander drehbar. ICH GLAUBE JETZT DASS DIE KOORD TETR WG IHRER kleinheit keinr rolle spielen)
%coming in pairs on each tetrahedral site.}
%On each corner point $i=1,2,3,4$ there are two internal spin vectors $Q_{Li}$ and $Q_{Ri}$ pointing outwards in the radial direction. 
%The non-zero spin vectors of the ground state corresponds to vacuum expectation values $\langle \psi^\dagger \vec \tau\psi \rangle \neq 0$. besser ohne doppelpack? weil nur fuer massenterm psibarpsi werden 8 benoetigt die auf 2 verschiedenen tetraedern sitzen und die bilden auch das higgs und w und gamma genauer erstreckt sich auch eine einzelne mignonanregung ueber 2 benachbarte tetraeder damit sie l und r und t und a abdecken kann. ICH FINDE DIE FUGUR IM MOMENT WIEDER GANZ GUT, WEIL AM BESTEN BENACHBARTE TA PAARE. DA ES PAARE AUS TEILCHEN UND ANTITEILCHEN SIND UND WENN DER ZUSTAND SOLL TINNEN BZW C VERLETZEN, MUESSEN ALLS SPINS NACH AUßEN GERICHTET SEIN (GECHECKT!!!!). Noch besser Q1-8 wo jedes Q T und A enthalten kann je nach Bedarf. Qi=QLi+QRi fuer die Massenterme wichtig!!!!!
%(ich glaube zur Zeit, dass hier 2 benachbarte Tetraeder gezeigt werden...aber dann kriegt man dof nicht hin...Antwort: q, l und VB erstrecken sich immer ueber 2 benachbarte Tetraeder, ob dies T,A,L oder R sind, kann von Fall zu Fall unterschiedlich sein. man braucht ja auch fuer die SSB mindestens 2 benachbarte Tetraeder. dies auch als Antwort auf eine Frage bringen)
\nonumber
%\label{figac1} 
\end{center}
\end{figure}

As for the {\it global} ground state the set of all tetrahedrons forms a flat 3-dimensional crystal structure within the original $R^6$, similar to what is shown in fig. 2. This structure may be called a hyper-crystal. It is our world, the space in which all physical processes take place. Actually it will turn out to resemble an elastic or even a fluid system, so that it may as well be called a hyper-plastics or, within the Lorentz covariant cosmological framework to be developed later, the discrete micro-elastic spacetime continuum, the 'DMESC'.
 
Contrary to what is drawn, the tetrahedrons extend into internal space alone, not into physical space. In other words, physical space is {\it defined} to be the 3-dimensional subspace of $R^6$ orthogonal to the 3 dimensions spanned by the aligned tetrahedrons.

\begin{figure}
\begin{center}
\epsfig{file=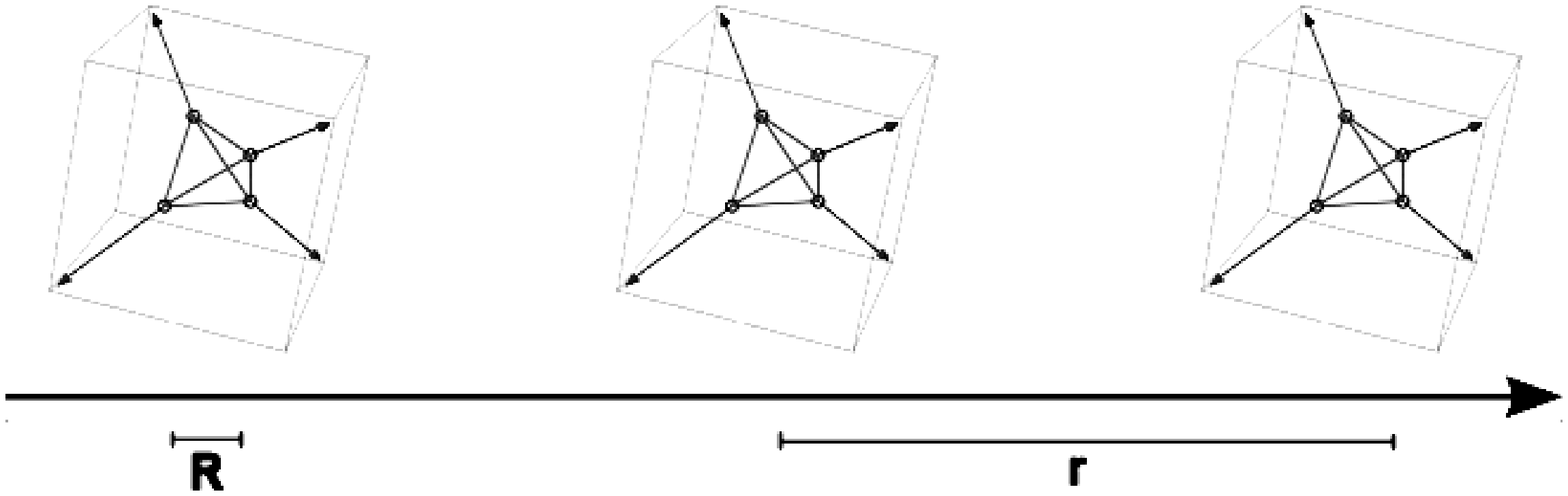,height=5.4cm}
\bigskip
\caption{The global ground state of the model after the electroweak SSB consists of an aligned system of chiral tetrahedrons over physical space (the latter represented by the long arrow). R is the internal magnitude of one tetrahedron and r the distance between two of them. The figure is a bit misleading, not only because the tetrahedrons do not have an extension into physical space, but also the relative magnitudes are not correctly drawn. While r and R are tiny (of the order of the Planck length), the tetrahedrons formed by the isospin vectors are much larger, of the order of the Fermi scale (in inverse energy units). Actually there are 2 kinds of alignment in this figure: the alignment of neighboring coordinate tetrahedrons and the alignment of isospin vectors in neighboring tetrahedrons. The isospin vector alignment is associated to the electroweak symmetry breaking, because at temperatures above the Fermi scale (before the SSB) the isospins in each tetrahedron are oriented randomly (not shown) and there is a corresponding {\it local} $SU(2)$ symmetry which gets broken when the isospin vectors align. The figure also shows how the universe looks like in the tetron model. It is a 3-dimensional 'monolayer' of internal tetrahedrons whose average distances are given by the Planck length $\langle r \rangle=L_P$. Gravity is due to the elasticity of the coordinate bonds between neighboring tetrahedrons and corresponds to tiny deviations from this average in the vertical or horizontal direction. Finally, the coordinate alignment of the tetrahedrons is related to crystal formation at the big bang. Cosmic inflation is due to the sudden release of crystallization energy. These latter issues will be discussed in detail on pages 15-28 and in section 2.5.}
%For reasons described in the main text, the covering group of this SO(3) can be identified with the group $SU(2)_L$, which gets broken spontaneously by the condensate $\langle \bar U U +\bar D D \rangle$. It is possible, that the alignment of the coordinate- as well as the spin-vectors of all tetrahedrons shown in the picture is only symbolic, not just because R is a length scale defined in internal space but also because the condensate does not define a direction in internal $R^3$ space. Though formally it can be considered as part of an SO(4) vector in the (2,2) representation of $SU(2)_L \times SU(2)_R$, physically it corresponds to a particle density, a scalar quantity giving the density of pairs of ionized tetrahedral components whose dynamical formation is responsible for the SSB at distances of order $\Lambda_F$, as explained in the main text.
\nonumber
%\label{figac1} 
\end{center}
\end{figure}

%This corresponds to the spontaneous breaking of the local internal Heisenberg $SU(2)_L$ introduced above. Before the SSB the internal spins of the chiral tetrahedrons are oriented randomly and there is a isomagnetic Heisenberg symmetry. This symmetry is local in the sense that the spins can be rotated independently within each fiber $f_x$ over each point x of Minkowski space. Furthermore, it is left-handed(linksh ist es eigentlich nur in der alten Interpretation wo die festen Sternstrukturen sich nur als ganzes drehen konnten ... aber auch wenn sie chaotisch verteilt sind dreht Heisenbergs SU2 alle Spins EINES Tetraeders um denselben Betrag. Endprodukt eine chirale Komfig ist), because the configuration fig. 1 is chiral and this internal chirality translates into a V-A structure of the weak interactions\cite{laa}.
In addition to the coordinate alignment of tetrahedrons, there is also an alignment of isospins of neighboring tetrahedrons in fig. 2. Before the appearance of this structure the internal spins U and D, which are the building blocks of the isospin vectors, can freely rotate, and thus there is an internal spin SU(2) symmetry group. In ordinary magnetism this group is usually called Heisenberg's SU(2); in sections (\ref{appii4})ff it will be shown how its isospin analogon is related to the Standard Model SU(2)$_L$ gauge group.

An important point in that consideration is that the SU(2) transformations are local symmetries in the sense that the isospin vectors can be rotated independently over each point of Minkowski space. The group gets broken to $G_4$ when the internal arrangement fig. 1 is formed. It may be given the index L, because this arrangement is chiral and because there is a dynamical relation between internal and external chirality, as explained in (\ref{appii5}).

The mixing with the electromagnetic $U(1)$ symmetry has not been introduced at this point. This omission will be clarified later in (\ref{appii216}) and (\ref{appii205}), together with the tetron model interpretation of the electroweak mixing angle.

%The iso-magnetic ground state fig. 1 is chosen so as to respect the above mentioned Shubnikov group (\ref{eq8gs}).
One wants to interpret the 3 generations of quarks and leptons as isospin wave excitations of the internal isospin structure.  These excitations will be called \textcolor{blue}{\bf{\underline{mignons}}}. They behave as quasi-particles while they travel through Minkowski space and can be classified according to representations of $G_4$, as shown below.

$G_4$ is a finite group which remains intact to the lowest energies. As shown in \cite{shub} it has only 1- and 3-dimensional representations. To generate all possible excitations describing the quarks and leptons one has to consider the vibrations of $\vec S=\vec Q_L$ and $\vec T=\vec Q_R$ for each of the 4 tetrons separately, cf. (\ref{appii51}) and (\ref{appii4}).

The isospin vibrational excitations are described by deviations $\delta$ from the ground state fig. 1, i.e.
\begin{eqnarray} 
\vec S_i=\langle\vec S_i\rangle+\delta \vec S_i \qquad \vec T_i=\langle\vec T_i\rangle+\delta \vec T_i
\label{eq723}
\end{eqnarray}
or, more precisely, by certain linear combinations of them -- the eigenmodes of the isospin Hamiltonian to be discussed later in (\ref{mm3}) and (\ref{mm3dm}).

%The excitations of the 24 d.o.f. of the 8 isospin vectors $\vec Q$ exhaust the q/l spectrum completely because 
The resulting 24 mignon states can be arranged in six singlet and six triplet representations $A_{\uparrow,\downarrow}$ and $T_{\uparrow,\downarrow}$ of $G_4$ to yield precisely the multiplet structure of the 24 fermion states of the 3 generations, not less and not more\cite{lhiggs}:
\begin{eqnarray} 
A_{\uparrow}(\nu_{e})+A_{\uparrow}(\nu_{\mu}) +A_{\uparrow}(\nu_{\tau}) &+& 
T_{\uparrow}(u)+T_{\uparrow}(c)+T_{\uparrow}(t)+ \nonumber \\
A_{\downarrow}(e)+A_{\downarrow}(\mu)+A_{\downarrow}(\tau) &+& 
T_{\downarrow}(d)+T_{\downarrow}(s)+T_{\downarrow}(b) 
\label{eq833hg}
\end{eqnarray}
Details about this arrangement will be given in (\ref{appii610}). The SM quantum numbers can be recovered from this spectrum in the following way:\\
--the $\uparrow$ representations can be obtained from the $\downarrow$ ones by the transformation $\delta \vec S \leftrightarrow \delta \vec T$ for any of the tetrons, i.e. by  interchanging left and right. As shown in (\ref{appii4}) this is precisely what is needed for a weak isospin transition on the level of mignons.\\
%It turns out that the vibrators $\vec Q_L$ and $\vec Q_R$ on each tetrahedral site i=1,2,3,4 are associated to a tetron and an antitetron, respectively. Details will be worked out in (\ref{appii51}), (\ref{appii4}), (\ref{appii610}) and (\ref{appii-ssu8}).\\
%mussen sie dann Teilchen und Antiteilchen (Dstern,-Ustern) sein um Isospin hinzukriegen?).
--singlet and triplet Shubnikov states have a different U(1) charge. The corresponding symmetry can be interpreted as gauged tetron or $B-L$ number. Details will be given in (\ref{appii216}) and (\ref{qu-charges}). The mixture with the photon and the appearance of the Weinberg angle will be discussed in (\ref{appii205}).\\
--the 3 states within each triplet $T$ in (\ref{eq833hg}) are always degenerate, because $G_4$ remains unbroken. The relation between those triplets and the QCD color triplets of quarks will be further discussed in (\ref{appiiqcd}).

Actually, to obtain the quark and lepton spectrum (\ref{eq833hg}) a discrete structure is compelling only in internal space, not in physical space. Looking at fig. 2, one could try to come along with a continuous model of Minkowski space, i.e. with $r\rightarrow 0$. However, it is tempting to assume $r\neq 0$, i.e. that there is a sort of lattice underlying spacetime, with spacings so small that Lorentz symmetry is effectively maintained for all available energies.

Details of this idea will be discussed after (\ref{eqfe}) and in section 2.5, where it will be shown that the lattice must be (i) elastic and (ii) a Planck lattice, otherwise it would contradict (i) cosmological observations and (ii) Einstein's principle of equivalence\cite{preparata,kleinertscardigli}. Due to quantum fluctuations it may be a foam\cite{foam} or a spin network\cite{loop} -- although in the tetron model there is no a priori necessity to quantize gravity, cf. (\ref{appii806}) and (\ref{appii1}).
%(hier etwas ueber raumZEITLorentz des Kristalles sagen. Dass Zeit durch Entropiewachstum entsteht, nicht diskret sein kann; dass Laengen vom Bezugssystem abhaengen und mit Zeit mischen; Lorentz allerdings sowieso nur eine Approximation. r=0 entspricht kontinuierlichem Außenraum, ohne aeußere Gitterstruktur. weiteres siehe eine der Fragen)

Can the aligned structure fig. 2 be understood heuristically?  The answer is yes, if one assumes that the arrangement of isospin vectors follows similar rules than that of spin vectors in a magnetic environment, cf.  (\ref{appiiheuri}). What matters are value and sign of (internal) exchange integrals J of tetron wave functions as a function of the distance between 2 tetrons, because these integrals will appear as couplings in the Heisenberg isospin Hamiltonian (\ref{mm3}). 

The behavior of isospins in fig. 2 can then be understood via the so-called Bethe-Slater curve shown in fig. 3. If the tetrons are part of one tetrahedron, their distance is small $\sim R$ and according to the figure J is negative. This corresponds to anti-ferromagnetic behavior and leads to the formation of the frustrated structure fig. 1 with symmetry $A_4 + S ( S_4 - A_4)$, because the spin vectors try to avoid each other as far as possible.

In contrast, if the internal spin vectors belong to different tetrahedrons, the distance of the corresponding tetrons is somewhat larger, of order r, and J is positive. This corresponds to ferromagnetic behavior
%It should be noted, that in the context of normal magnets, the Bethe-Slater curve is derived from 3-dimensional exchange integrals, while here some of the integrals are 6-dimensional. 
and leads to the isospin alignment of neighboring tetrahedrons fig. 2.

Due to the tetrahedral 'star' structure fig. 1 it is appropriate to change the notion of isospin. Usually in an (anti)ferromagnetic environment, the spin vectors align into the + or - orientation of the z(=magnetization) direction, and the corresponding Pauli spinors are given by $U=(1,0)$ and $D=(0,1)$. In the present case the (iso)magnetic structure is defined by isospin vectors either pointing outwards or inwards in the radial direction. Correspondingly, the isospinors U and D are to be understood as 'radial' spinors\cite{radsp}
\begin{eqnarray} 
U_\star= \sqrt{\frac{1}{3}}  Y_1^0 U - \sqrt{\frac{2}{3}} Y_1^1 D 
&=&\cos \frac{\vartheta}{2}\ U+ \sin \frac{\vartheta}{2} e^{i\frac{\varphi}{2}}\ D  \\
D_\star= \sqrt{\frac{2}{3}}  Y_1^{-1} U - \sqrt{\frac{1}{3}} Y_1^0 D  
&=& \sin \frac{\vartheta}{2} e^{-i\frac{\varphi}{2}}\ U -\cos \frac{\vartheta}{2}\ D
\nonumber\label{radsp1}
\end{eqnarray}
where $Y_l^m$ denote the sperical harmonics and $\vartheta$ and $\varphi$ are the angles of the radial vector w.r.t. some cartesian coordinate system. These new spinors are radial in the sense that they reproduce the unit vector in polar coordinates
\begin{eqnarray} 
\vec e_r = U^\dagger_\star \vec\tau U_\star = -D^\dagger_\star \vec\tau D_\star
\label{radsp2}
\end{eqnarray}
Furthermore they are normalized in such a way that
\begin{eqnarray} 
U^\dagger_\star U_\star + D^\dagger_\star D_\star = U^\dagger U + D^\dagger D
\label{radsp3}
\end{eqnarray}
The iso-spinor state corresponding to an isospin vector pointing outwards is denoted by $U_\star$ (and similarly $D_\star$ for inwards pointing isospins). According to figs. 1 and 2, $U_\star$ is the building block of the hyper-crystal in its ground state. As shown in sections 2.1 and (\ref{appii-qtet}) it is unpolarized and its electric charge vanishes.

Note that this presentation is equivalent to the 'universal' z-axis approach\cite{dmz1,dmz2} used in the actual mass calculations\cite{lmass}. Although according to (\ref{radsp1}) $D_\star$ has as  much to do with U as it has with D, I will often leave out the star index in the following for reasons of simplicity and understand that always $U_\star$ and $D_\star$ are meant. I will include the star only when this is needed for clarity, e.g. in (\ref{appii206}). 
%(Ur and Dr are 'radial' spinors in the sense that $\vec S_U:=U_r^\dagger \tau U_r$ points outwards as in fig. 1 (aligning with the SSB configuration), whereas $\vec S_D:=D_r^\dagger \tau D_r$ points inwards corresponding to opposite internal parity. More precisely, one has $\vec S_D=-\vec S_U$ with $\vec S_U:=\vec e_r =\bar\psi \tau_r \psi$ where $\vec \tau_r=\tau_z \cos\theta+\tau_x \sin\theta\cos\phi+\tau_y \sin\theta\sin\phi$ and $\vec e_r$ is the radial unit vector. 
%Actually, most of the formulas presented in this work are to be understood taking for U and D the radial spinors eq. Technically, this is equivalent to transforming the 4 vibrating spin vectors fig. 1 to a universal z-axis. This latter method is described in detail in \cite{festkarbeit}, and has been used to calculate the q/l mass spectrum, as described below and in \cite{laa}.

In the tetron model the SM SSB arises from the 'ferromagnetic' alignment of isospin vectors in neighboring tetrahedrons. As shown in (\ref{appii206}) and (\ref{appii-vev}), the corresponding order parameter is given by a non-vanishing vacuum expectation value 
\begin{eqnarray} 
\langle\bar U_\star U_\star\rangle \neq 0 
\label{radsp4094}
\end{eqnarray}
{\it This is the way the SM Higgs mechanism is realized on the microscopic level.} There is a pairing active comparable to the formation of Cooper pairs in a superconductor, and excitations of this tetron-antitetron pairing will appear as the physical Higgs field and the electroweak bosons.

\begin{figure}
\begin{center}
\epsfig{file=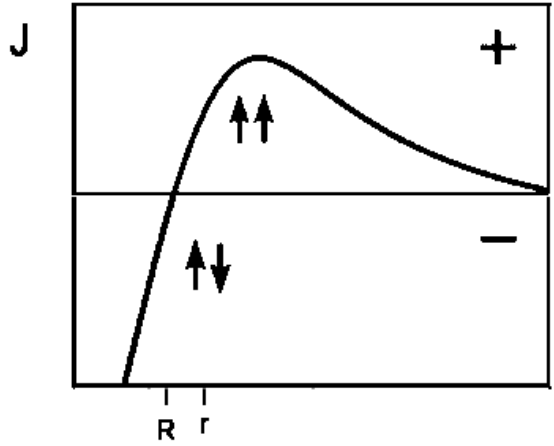,height=7cm}
%\bigskip
\caption{Qualitative behavior of the exchange integral coupling J as a function of the distance between 2 tetrons. In ordinary magnetism this is called the Bethe-Slater curve. One has anti-ferromagnetism ($J<0$) at small, ferromagnetism ($J>0$) at large distances.}
%(genauer beschreiben, dass es nur eine Analogie fuer R6 ist)}
%(ich habe jetzt bethe7 statt bethe8 weil lambdaF hier keine Rolle spielt)
\nonumber
%\label{figac1} 
\end{center}
\end{figure}

In \cite{lmass} the masses of the mignons (\ref{eq833hg}) have been calculated, and the observed hierarchy in the quark and lepton spectrum as well as the hierarchy in the CKM and non-hierarchy in the PMNS matrix elements has been reproduced. As described above, mignon masses can be identified with the eigenfrequencies of the vibrations of the isospin vectors $\vec S$ and $\vec T$. These eigenfrequencies get contributions both from inner- and from inter-tetrahedral interactions.

Firstly, the {\it inner}-tetrahedral interactions are responsible for the frustrated tetrahedral configuration fig. 1, i.e. for the structure of the local vacuum. They are small distance contributions and relatively simple to treat because they can be described by an internal Heisenberg Hamiltonian for one tetrahedron alone, with corresponding internal spin vector excitations. The most general form of this Hamiltonian is 
\begin{equation} 
H_H=- J_{SS}\sum_{i \neq j=1}^4 \vec S_{i}\vec S_{j}
- J_{TT}\sum_{i \neq j=1}^4 \vec T_{i}\vec T_{j}
- J_{ST}\sum_{i,j=1}^4 [\vec S_{i}\vec T_{j}+\vec T_{i}\vec S_{j}] 
- K_{ST}\sum_{i=1}^4 \vec S_{i}\vec T_{i} 
\label{mm3}
\end{equation}
where the couplings J are internal exchange energy densities characteristic for the internal Heisenberg interactions. By introducing $K_{ST}$, I have allowed that the coupling $\vec S_{i}\vec T_{j}$ is different within a site ($i=j$) than outside of it ($i\neq j$). 

%The isospin vectors $\vec S$ and $\vec T$ have the following meaning: to describe a quark or a lepton of a certain helicity one should in principle form eigenmodes as linear combinations of the 8 isospin vectors $\vec Q$ in fig. 1, all of them belonging to tetrons with one and the same helicity. However, q/l mass terms in the SM Lagrangian are of the form $\sim \bar\psi_R \psi_L +c.c.$ Therefore, one set $\vec S_i$ of isospin vectors are to be interpreted as left-handed $\vec Q_L=\psi_L^\dagger\vec\tau\psi_L$, the other one as right-handed. 
%($\vec T_i=\bar \psi_R^\dagger\vec\tau\bar\psi_R$ mit bar, ABER ob man bei Ti bar oder nicht verwendet sollte keinen Unterschied machen, da meine Spinvektoren immer sowohl Teilchen als auch Antiteilchen enthalten.)
%(ABER: ich dachte barpsiR sei ein linkshaendiges Teilchen. Antwort: (psiR)c=(psic)L is linksh, aber barpsiR ist rechtsh!!!!)
%(Achtung: weil man Massen Lbar*R will, soll man QL1-4 und QR1-4 verwenden mit gRR, gLL,gLR, worauf dann spaeter bei Disk der q und l Massen zureugegriffen wird. Dies ist dann zugleich auch Antiteilchen zu nehmen und selbe Struktur psibar*psi wie das Higgs - wobei sich beim Higgs selber ja die Mignonschwingunggen kompensiern sollen und ausserdem in 2 verschiedenen Tetraedern lebt ... was allerdings kein Problem waere wenn unten keine Gitterstruktur.)

Using (\ref{mm3}) and (\ref{eq723}) one is led to e.o.m. for $\delta\vec S_i$ and $\delta\vec T_i$ which can be solved in a similar way as the e.o.m. for magnons in solid state physics. On this basis the contributions from (\ref{mm3}) to the eigenfrequencies of the 24 eigenmodes were calculated in \cite{lmass}.
%and identified with the quark and lepton masses.

Using $J_{SS}=J_{TT}$ (an approximation which can be justified via the generalized NJL model discussed above\cite{lmass}) the following masses/eigenfrequencies are obtained 
\begin{eqnarray}
\pm m_\mu &=& 6 J_{ST}+ 2 K_{ST} \label{mm15f} \\
\pm m_c &=&4J_{SS} +4J_{ST} \label{mm15ff44g} \\
\pm m_s  &=&4J_{SS}+2 J_{ST} +2K_{ST} \label{mm15ffg}
\end{eqnarray} 
for the second family. Using the measured values of $m_\mu$, $m_s$ and $m_c$\cite{runningmass}, the internal exchange couplings may be determined:
\begin{eqnarray}
J_{ST}\approx -0.12 \ \text{GeV} \qquad J_{TT}=J_{SS}\approx -0.12 \ \text{GeV} \qquad K_{ST} \approx 0.32  \ \text{GeV}
\label{mm15ffff}
\end{eqnarray} 
One concludes that one has ferromagnetic coupling $K_{ST} >0$ of adjacent spin vectors $\vec S_i$ and $\vec T_i$, while the interactions with $i\neq j$ are anti-ferromagnetic. This is in accord with the heuristic expectations discussed before. Furthermore, it may be noted that all these values are rather small as compared to the SSB scale. A natural explanation for that is given in (\ref{appiimay}).

%(the {\it inter}-tetrahedral interactions  will be discussed under the question why the top quark is so heavy as compared to the other quarks and leptons. den folgeden Abschnitt da rein kopiern)  
Secondly, the {\it inter}-tetrahedral interactions: these lead to the mass of the top and bottom quark and are based on the parallel (='ferromagnetic') alignment of isospins between different tetrahedrons fig. 2. Their leading effect turns out to be a contribution of order $O(\Lambda_F)$ to the top quark mass\cite{lmass}. Physically speaking, this interaction handicaps the specific eigenmode describing the top quark, because this mode disturbs the SSB alignment in the strongest possible way.

Mathematically, the effect can be described by adding terms to the inner-tetrahedral Heisenberg interaction with a normal ferromagnetic plus a Dzyaloshinskii-Moriya (DM) component\cite{dmgut}. The sum of the 2 components will yield a quasi-democratic mass matrix\cite{democ} which in leading order only contributes a term of order $\Lambda_F$ to the top-quark mass and nothing to the masses of the other quarks and leptons.

More in detail, the Hamiltonian for the SSB interactions of neighboring tetrahedrons can be derived from the W-mass term of the SM Lagrangian. By considering a $SU(2)_L$ gauge transformation, which removes the longitudinal components of the W-bosons from the Higgs part of the SM Lagrangian, one obtains
\begin{equation} 
H_{SSB}=\;  J_{inter} \; \;  \sum_{i,j=1}^4 \; [ \vec S_{i}\vec S'_{j} \;  +\; i \; (\vec S_{i}\times \vec D_{ij})\;  \vec S'_{j}] 
\label{mm3dm}
\end{equation}
to be added to the Heisenberg Hamiltonian (\ref{mm3}). One has
\begin{equation} 
J_{inter}=\frac{\mu^2}{4\Lambda_F}
\label{mm3dm333}
\end{equation}
with $\mu$ the mass parameter of the SM Higgs potential and $\Lambda_F=\sqrt{\frac{\mu^2}{\lambda}}=246$GeV the Fermi scale (vacuum expectation value). Only terms involving the left handed isospin vectors $\vec S=\vec Q_L$ appear, as follows from (\ref{appii5}) in accordance with the $V-A$ structure of the weak interactions. In (\ref{mm3dm}) the factor $\vec S'_j$ denotes the left handed isospin vector of an adjacent tetrahedron. 
%Although it is meant to describe inter-tetrahedral interactions, this effective interaction involves only isospin vectors of one single tetrahedron.
%zzz(Es sind dann allerdings L'j weitere dof, die neu auftreten und eigentlich nicht downartigen quarks entsprechen koennen????)
The precise relationship between (\ref{mm3}) and (\ref{mm3dm}) on one side and the SM Lagrangian terms on the other are worked out in (\ref{appii207}) and \cite{lmass}.
The fact that the inter-tetrahedral coupling (\ref{mm3dm333}) is much larger than the inner-tetrahedral ones (\ref{mm15ffff}) will be naturally explained in (\ref{appiimay}).

Eq. (\ref{mm3dm}) contains a ferromagnetic interaction plus the additional DM term which is due to the non-abelian nature of the W-bosons. The overall normalization of the DM term is dictated by $SU(2)_L$ gauge invariance, while the relative values of the DM couplings $\vec D_{ij}$ are fixed by the internal $G_4$ symmetry\cite{lmass}.
%They can be interpreted as the leading anisotropic corrections to the ordinary Heisenberg equations of motion.(aber DM ist genau wie Heisenberg rot-invariant. Was ist der Unterschied zu aniso? Ausserdem kann man DM schwerlich als kleine Korrektur betrachten.)

Quite in general, a DM component stands for a tendency to form a rotational structure (instead of the ordinary ferromagnetic alignment of neighboring tetrahedrons depicted in fig. 2) simply because the DM term tends to rotate the spin vectors instead of aligning them. In the present case it appears as a consequence of the non-abelian nature of SU(2). Therefore the DM term can be interpreted quite naturally, namely by the fact that a non-vanishing $SU(2)_L$ gauge field induces a curvature of the fiber bundle formed by the system of all tetrahedrons, and the DM term simply takes care of this curvature effect to effectively maintain the aligned structure.

This argument is supported by the fact that the gauge transformation inherent in (\ref{mm3dm}) leads to the SM Lagrangian in the so-called 'unitary gauge'. This point is analyzed in detail in (\ref{appii529}).

Using (\ref{mm3}) and (\ref{mm3dm}) one can derive the e.o.m. for the isospin vectors. With the usual ansatz $\sim \exp (i\omega t)$ one obtains a 24$\times$24 eigenvalue problem.
%which has been diagonalized in\cite{lmass}. 
The eigenvalues $\omega$ correspond to the quark and lepton masses and were calculated in \cite{lmass} as a function of the exchange couplings $J_{SS}$, $J_{TT}$ etc. In that paper it was explicitly verified that the corresponding 24 eigenstates can be arranged into 6 singlets and 6 triplets as predicted by the Shubnikov symmetry analysis (\ref{eq833hg}), i.e. as 6 leptons and 6$\times$3 quarks. Each triplet (quark flavor) consists of 3 states with degenerate eigenvalues, because the Shubnikov symmetry $G_4$ is unbroken at low energies. 

The dominant contribution from (\ref{mm3dm}) gives the top quark a mass of the order of the Fermi scale while leaving the other quark and lepton masses unchanged. 
%zzz(hier KOENNTE top massenformel REIN)
As detailed in (\ref{appii614}), (\ref{appii723err}) and \cite{lmass}, b, c, s and $\tau$ get their masses mainly from (\ref{mm3}).
%zzz(hier b,c,s,tau massenformel)

In contrast, there are no contributions from (\ref{mm3}) and (\ref{mm3dm}) to u, d, e and neutrino masses. These 10 excitations remain massless on this level. To obtain their masses one has to include additional small torsional interactions\cite{lmass}.

The masses of the neutrinos are particularly suppressed because the 3 neutrino modes correspond to the vibrations of the 3 components of the total internal angular momentum vector
\begin{equation} 
\vec \Sigma :=\sum_{i=1}^4 (\vec S_{i}+\vec T_{i})=\sum_{i=1}^4 \vec Q_i 
=\sum_{i=1}^4 \psi_i^\dagger \vec \tau\psi_i
\label{tm32}
\end{equation}
%(QL+QR=Q koennte man denken, wenn man diese Formel sieht, und genau so ist es auch)
Whenever this quantity is conserved
\begin{equation} 
d\vec \Sigma/dt =0 
\label{tm31}
\end{equation}
the neutrino masses will strictly vanish ($\omega =0$). In fact, the interactions considered so far, i.e. (\ref{mm3}) and (\ref{mm3dm}), conserve total internal angular momentum. Therefore, they fulfill (\ref{tm31}) and give no contribution to the neutrino masses. Further details can be found in (\ref{appii605})-(\ref{appii-dima}) and in \cite{lmass}.

The general solution to the eigenproblem given above does not only yield the energy eigenvalues but (via the corresponding eigenvectors) can also be used to accommodate the CKM and PMNS mixing matrices\cite{lmass}. The mass eigenstates are the states corresponding to the energy eigenvalues, while the interaction eigenstates naturally correspond to the original vectors $\vec S_i$ and $\vec T_i$, cf. (\ref{appii-ck4}).

Within this framework one can understand\cite{lmass} why the CKM elements turn out to be small, whereas the PMNS matrix elements are naturally large: the lepton eigenstates (roughly given by $\vec S \pm \vec T$) are 'far away' from $\vec S$ and $\vec T$, while the up- and down-type quark eigenstates are relatively small deformations of $\vec S$ and $\vec T$, respectively. Due to the dominant contribution from (\ref{mm3dm}) the top quark triplet state has the smallest mixing matrix elements with other quarks, because it corresponds to the vibration of $\vec\Sigma_L:=\sum_i\vec S_i$ to an accuracy of less than 1\%.

In summary, the present model describes the physical world as a huge ordered crystal of internal 'molecules', each molecule of tetrahedral form and arranged in such a way that the internal Heisenberg spin symmetry is spontaneously broken. As shown below, this approach not only provides a nice microscopic understanding of particle physics phenomena but in addition substantially supplements our understanding of the inflationary big bang cosmology. In effect, it gives the phase transitions in the early universe a microscopic meaning.

%{\bf THE EVOLUTION OF THE UNIVERSE IN THE MICROSCOPIC MODEL}

To comprehend this fact, it is appropriate to redevelop the full history of the early universe within the assumptions of the tetron model: 
%Starting below the Planck scale 10$^18$GeV(wie groß ist die Planckskala mit den Extrapolationen c=a, G=a und h=a**4? Mpl=a**2 quadratisch mit a laut Adler, Lpl=a linear und Tpl bleibt konstant. Epl=Mpl*c2=a**4 ist das Objekt was heute 10**19GeV hat), 
before the 'big bang' there were the free tetrons $\psi$ floating around as a Fermi gas in $R^{6+1}$ space at extremely high pressure and temperature. 
%(At this stage the relevant symmetries were either a Galilean or a SO(6,1) Lorentz symmetry (with possibly a limiting speed $c'$ different from of c, $c < c' < \infty$, as explained below. ABER ES GIBT KEINE PHOTONEN 
%fuer endliches cstrich annihilieren sich TA-Paare? von mir aus sollen sie doch
While the universe was cooling down, 3 fundamental transitions occurred: 
%(aber warum hat es sich abgekuehlt, wenn es sich doch noch gar nicht ausgedehnt hat? Weil es irgendwo eine Wand hat, wohin es Waerme abgeben kann
%expected from the present model, which the 6+1 dimensional universe has run through one after the other since the big bang:

\par
\begingroup
\leftskip=0.3cm % Parameter anpassen
\noindent 
{\bf I. the formation of tetrahedral 'molecules'} from tetrons at very high temperature of order $\Lambda_R$, where the scale $R$ is roughly given by the extension of one molecule. Although this process is not a phase transition in the strict sense it has certainly released a large amount of energy which has amplified the initial temperature of the universe.\\
%(eigentlich it r=LPl bei uns die big bang skala, denn bei R kann noch gar nicht von 3+1 Universum gesprochen werden. Also kann es sich auch nicht amplified haben. Ich rede jetzt erst bei der Kristallisation von big bang und ist gleich Planck Skala und Inflationsskale)
Note that with 4 molecular sites each molecule 'fills' only 3 of the 6 spatial dimensions.
%(bilden sich hier schon 4 Paare in einem 8er Tetraeder? Oder haben wir nur 4er Tetrader, die mit einem Antitetraeder zu einem 8er System zusammengefasst werden)
%(inwzischen glaube ich lambda(r)=lambda(Planck). Dann ist lambda(R) noch groesser und bei beiden Skalen r und R gelten die Gesetze der Physik nicht mehr. Daher ist es einfacher, anzunehmen, dass sich der kristall aus freien tetronen bildete. Dann sind auch die inneren und aeusseren Abstaende der Tetronen in natuerlicher weise von derselben groessenordnung. Hierzu habe ich unter II. einen kleinen Abschnitt reingeschrieben)
\par
\endgroup

\par
\begingroup
\leftskip=0.3cm % Parameter anpassen
\noindent 
{\bf II. the formation of the 'hyper-crystal'} from tetrahedrons takes place at somewhat lower temperatures $T \sim \Lambda_r$, where $r$ is the 'lattice spacing', i.e. is roughly given by the distance between 2 tetrahedrons. This alignment of all tetrahedral structures is a coordinate alignment and to be distinguished from the isospin vector alignment (item III) describing the electroweak phase transition. It puts all 3-dimensional molecular structures in parallel thus separating an internal 3-dimensional space from the rest. In other words, the crystal expands into a 3+1-dimensional subspace of $R^{6+1}$, while the tetrahedrons extend into what becomes the 3 internal dimensions. 
%The natural scale for this phase transition is the distance r between tetrahedrons as introduced in fig. 2. 
\par
\endgroup

%Note, in some parts of the review I will follow a slightly different scenario in which there is no step I of tetrahedral molecule formation, i.e. the crystal spontaneously forms out of free tetrons below the critical temperature $\Lambda_r$. In that case the distances $R \lesssim r$ initially are of the same order of magnitude, both representing the length of the tetron bonds.
%zzz(nb: ich brauche R kleiner r fuer das Bethe Slater Argument)

\par
\begingroup
\leftskip=0.3cm % Parameter anpassen
\noindent 
Since II corresponds to the process, in which our 3+1 dimensional universe was born, it may rightfully be called the big bang. As a crystallization process it is a first order phase transition associated with the sudden release of a large amount of energy. As will be explained later, the coordinate interactions among the tetrahedrons are of elastic type. Under this condition the outcome of phase transition II is not a crystal in the strict sense, and one may \textit{as well call it a condensation of a hyper-plastics instead of a crystallization}, cf. (\ref{appii7a7}). In any case the release of crystallization/condensation energy naturally drives an inflationary expansion of the system and the corresponding metric. Therefore, within the framework of the tetron model, the big bang and the beginning of inflation are more or less identical. As argued in (\ref{appii7a7}) and further below in this section, the characteristic scale $\Lambda_r$ can be identified to be of the order of the Planck scale $\Lambda_P$.
\par
\endgroup

\par
\begingroup
\leftskip=0.3cm % Parameter anpassen
\noindent 
{\bf III. the arrangement of isospins} at temperatures of order $\Lambda_F$. Above those temperatures the isospin vectors fluctuate randomly with an associated internal 'Heisenberg' $SU(2)$ symmetry, but at $\Lambda_F$ they arrange into the chiral isomagnetic structure figs. 1 and 2. At that point the so far freely rotatable internal spins get ordered and the $SU(2)$ is broken to the Shubnikov group (\ref{eq8gs}). Note that the $SU(2)$ is a local symmetry group, because isospins can be rotated separately over each point of the Minkowski base space.\footnote{The relation between the internal Heisenberg $SU(2)$ and weak isospin $SU_L(2)$ is clarified in sections (\ref{appii4})ff. A more detailed description of phase transition III is given in (\ref{appii529}) and (\ref{appii-vev}). The mixing with the electromagnetic $U(1)$ symmetry will be included in (\ref{appii216}) and (\ref{appii205}).}   
\par
\endgroup

%zzzIn contrast to I and II no sudden release of energy is expected from this second order process. wenn ew pt doch 1st order ist, wie Shaposh etal meinen, um baryon asymm hinzukriegen, muss doch energy released werden. aber dies ist zu frueh um dark energy zu erklaeren 

%zzz(koennte er als dark energy heute noch nachwirken ... aber das universum hat sich angeblich zwischenzeitlich zusammengezogen ausserdem hat III fuer heute viel zu frueh stattgefunden)
%zzz(die Beschreibung ist ein bisschen anders als frueher wo eine Shubnikovkonfig gegen eine andere gedreht werden konnte, was sowieso mehr einer SO3 als einer SU2 entsprach - aber im Spinorraum ist es durchaus eine SU2, und ich verstehe jetzt auch warum das Minimum SU2 fach entartet ist weil der koordinatentetraeder so klein ist)

%As discussed below, gravity cannot be described by a rigid crystalline micro-system with constant lattice spacings r and a fixed position of its micro-components (the internal tetrahedrons). It rather resembles an elastic medium, where the distances of the tetrahedrons vary around an average lattice spacing $\langle r \rangle=L_P$.

It could happen that there is no internal coordinate order right after the crystallization, in the sense that the coordinates of tetrons in neighboring tetrahedrons are aligned.\footnote{I am not talking about a regular crystal structure in physical space, which for an elastic system is missing anyhow. Instead I am talking about the alignment of the {\it tetron} coordinates as depicted in fig. 2. See also the discussion at the end of (\ref{appii529}) where the assumption of internal coordinate order is completely given up.} In that case
%one should better interpret II as a condensation instead of a crystallization of internal tetrahedrons. In particular, 
there is no global internal tetrahedral symmetry of the hyper-crystal right after the condensation. However, since internal coordinate alignment is a prerequisite for the isomagnetic alignment (= electroweak transition) III, this has to be catched up later at the Fermi temperature, i.e. it takes place at about the same time as the isospin alignment.

This possibility will be called scenario C in later discussions, and in fact it has several benefits. For example, III automatically becomes a first order phase transition (cf. \ref{appii-ewpt}) with an associated second inflationary process that removes domain walls (cf. \ref{appii535}) from the visible parts of the universe. Furthermore, it is easier to understand that the electroweak phase transition is really spontaneous (cf. \ref{appii529}) and that the ground state fig. 1 is assumed by both the left- and the right-handed isospin vectors $\vec Q_{Li}$ and $\vec Q_{Ri}$ (cf. \ref{appii4}).

Since II happens after I, i.e. at lower temperature, one naturally expects $\Lambda_R$ larger than $\Lambda_r$ (i.e. $R < r$) in agreement with fig. 2 and the Bethe-Slater curve fig. 3. As argued in (\ref{appii530}) both scales are $\geqslant \Lambda_P$ and much larger than the scale $\Lambda_F$ where the isospins align. Note that while $\Lambda_F$ approximately corresponds to the critical point of transition III, the values of the exchange integrals J and therefore the iso-magnetic behavior are determined at distances r and R, cf. (\ref{appii312}).
%zzz(r wird ja im Lauf der Expansion noch mal viel groesser, und es scheint mir unplausibel, dass 2 so kleine und weit entfernte Objekte dann noch DM-WW machen)
%fuer die Ueberlappintegrale ist nicht F, sondern r und R interessant, da F nur den Temperatureffekt markiert

%(das folgende lieber weglassen, da ich ohnehin keinen ganz normalen Kristall habe. Ich habs jetzt umgechrieben, so dass es gut auf die Situation passt laut ...lecture6.pdf)
To describe II in the framework of the Landau approach to phase transitions one should consider density fluctuations $D \exp (i\vec p\vec x)$ within the gaseous assembly of tetrahedral 'molecules' and use $D$ as the order parameter of the phase transition.
%zzz(hier spielt eine Konnektion zwischen den Fasern eine Rolle, also koennte ein superschweres SO3 Eichboson existieren. Da wir Eichbosonen zusammengesetzt denken, muesste dies eine art phinon bound state sein. Aber Phinonen sind fuer die Translationsinvarianz im Innern, nicht fuer die SO3 Rotationsinvarianz. Fluktuationen D wirken auf mich wie PhOnonen bgzl Minkraum, also wie Gravitonen!!!!)

For an ordinary crystal these fluctuations can be identified with phonons; in the general relativistic (GR)  framework of a spacetime continuum at least some of them correspond to gravitational waves. This will become clearer below in this section, where elastic deformations of the crystal will be identified with metrical changes. More information about gravitons in the microscopic model can be found in (\ref{appii806}) and (\ref{appii80d3}).
%aber wie kommt spin2 zustande
%(wg der plastics eigenschaft gibt es vielleicht keine phononen. Doch in Fluiden schon. allerdings In Fluiden nur longitudinal Orientierung der Bewegungsamplitude von Molekülen parallel zur Ausbreitungsrichtung. In Festkörpern auch transversal Bewegungsamplitude senkrecht zur Ausbreitungsrichtung)

Since the density perturbation adds to the uniform density of the tetrahedron gas, there is no symmetry under changing sign of the density wave, and so the Landau free energy expansion allows for a cubic term
\begin{eqnarray} 
\Delta F=\alpha (T-T_c)D^2+\beta D^3+\gamma D^4
\label{eqfe}
\end{eqnarray}
where $T_c \sim \Lambda_r$ is the critical temperature for phase transition II.
%zzz(gilt so ein ansatz auch fuer condensation ie landau free energy liquid gas im internet suchen. JA, SOGAR BESSER denn bei liquid-gas ist die dichte wirklich der ordnungsparamete, bei kristall ist es eher die gittersymmetrie At the time, it was known experimentally that the liquid–gas coexistence curve and the ferromagnet magnetization curve both exhibited a scaling relation of the form , where was mysteriously the same for both systems. This is the phenomenon of universality.)

Accepting the idea of an elastic spacetime continuum, the coordinate phase transition resembles a gas$\rightarrow$liquid transition rather than a crystallization process. This even more, because the tetron particle density D is an order parameter characteristic for a condensation. While during a crystallization process the lattice symmetry plays an important role, in the hyper-crystal the bonds are elastic and there is no lattice symmetry at all in the Minkowski base space. When calling our universe a hyper-'crystal' and identifying the big bang with its 'crystallization', it should therefore be kept in mind that it shares more properties with a liquid or a deformable plastics than with a real crystal. This topic will be taken up in (\ref{appii534}). 

%(alte Version: time schedule for the very early universe:\\
%1. formation of a tetrahedral molecule from tetrons shortly after the Planck epoch.\\
%2. formation of the hypercrystal (breaking of internal rotational symmetry supercooling and afterwards release of crystallization energy to drive inflation and repolulate the universe) An diesem Punkt existieren evtl noch keine quarks und leptons, ABER VERNUTLICH PHOTON...und Higgs?\\
%3. arrangement of spins within the molecules (local Shubnikov symmetry)\\
%4. aligment of the spins over all molecules (weak SU2 breaking corresponding to internal Heisenberg spin symmetry)
%PROBLEM BEI DIESER ALTVERSION: 3 und 4 scheinen identisch!!!!)

The appearance of the cubic term in (\ref{eqfe}) is characteristic for a first order phase transition where a second minimum, which develops in the potential when the temperature is lowered, for some time remains higher than the minimum at $D=0$ of the gas phase, and furthermore the two minima are separated by a potential wall. When the temperature drops below the critical value, there is a discontinuity which is not present in second order transitions.
%As well known, such a transition is a rather discontinuous process and is able to at least qualitatively describe the formation of the hyper-crystal.
%\footnote{For a first order phase transition the free energy series expansion (\ref{eqfe}) should not be truncated at the $D^4$ term, because of the discontinuity in D at the critical point.} 
%and is in contrast to the SM Higgs potential (\ref{hipo}) and its temperature dependent extension
%zzz$-\mu^2 \rightarrow \mu^2 (T-T_F)$ (ist Tc bei ew phase transition wirklich die Fermiscale?) aber es gibt auch T*fi**3 Terme laut shaposhnikov
%describing the transition III, or to other magnetization processes where the cubic term is missing. (In such cases one usually faces a second order transition.)

%In figs. 4 the thermal behavior of the Landau free energy is drawn for the 2 cases. 
%In fig. 4b (case II, coordinate alignment) the second minimum which develops when the temperature is lowered, for some time remains higher than the minimum at $D=0$ of the gas/liquid phase, and furthermore the 2 minima are separated by a potential wall. When the temperature drops below $T_c$ there is a discontinuity which is not present in fig. 4a (case III, isospin alignment).

The latent heat associated with this discontinuity is released very suddenly and can be used to explain the extreme acceleration needed for cosmic inflation\cite{baumann,guth}. As shown in (\ref{appii702}) it provides all the necessary ingredients for the inflationary process to start and to eventually stop, once the condensation of tetrahedrons is completed and 
%zzz(hier muesste eigentlich ein quant Vergleich mit gegenwaertigen Modellen der Inflation) %zzzNote that an ordinary first order phase transition as initially considered by Guth to drive the inflation was later discarded for reasons having to do with domain/bubble formation and matter re-creation. These latter items are discussed in (\ref{appii535}). 
%(nicht bringen: As a consequence this transition is first order, because there is a discontinuity in $D$ at the melting point, denn bei mir ist superheating) At $T=T_c$ the original minimum disappears altogether.
%genauer: at high temperatures there is a minimum at $\rho=0$ corresponding to the liquid state. As the temperature decreases a second minimum develops at a free energy higher than that of the liquid. erst oberhalb Tmelting hat das 2. Minimum kleinere Energie. Dann kann es zu einem jump in rho kommen, außer wenn supercooling
%Just as the formation I of tetrahedral molecules the coordinate alignment II of all internal tetrahedrons is associated with a lot of energy released. In contrast to I, however, the crystallization II is a first order phase transition with a supercooling effect and the release of energy is very sudden leading to the extreme expansion needed for cosmic inflation.
the energy is exhausted. Most of the initial molecular energy has then been transferred to the elastic energy of the crystal. However, some of it survives in the form of tetron-antitetron excitations (gauge bosons) and is converted into mignons (quarks and leptons), as the temperature decreases further.
%zzzzzz(aber muss inflation nicht baryon neu erzeugen bei reheating?)

In place of quarks and leptons, whose existence is tied to the isospin ordering fig. 2, shortly after crystallization other excitations are more important, like the internal coordinate vibrations discussed in (\ref{appii536}) and \cite{la4xz2}, or excitations of the tetron-antitetron bonds discussed in (\ref{appii206}) and (\ref{appii206f}). Most prominent among the latter are the gauge bosons and the visible and dark scalars of the 2HDM sector as described in (\ref{appii207}) and (\ref{appii811}).

Because of the dominance of the electroweak bosons this cosmological era is often called 'radiation dominated' or 'electroweak'. At temperatures far above the Fermi scale all these excitations are effectively massless states transforming under the local $SU(2)\times U(1)$ symmetry, and they dominate the universe all the way down to the electroweak SSB (=isospin vector alignment). More details about the tetron model view on this era can be found in (\ref{appii206k}), (\ref{appii206m}), (\ref{appii526aa}) and (\ref{appii527}).

%\begin{figure}
%\begin{center}
%\epsfig{file=1art2art.eps,height=7cm}
%\caption{Thermal behavior of electroweak(=isomagnetic) and inflationary(=crystallization) phase transtion.}
%\nonumber
%\end{center}
%\end{figure}
As well known, in general relativity a non-vanishing energy momentum tensor leads to a curvature of the spacetime continuum. Many authors have interpreted this on the basis of metric elasticity\cite{sak,add1,gron,bomm,hehl1}, and some of them have speculated that gravity forces might be explainable from a microscopic structure which in some sense is analogous to the atomic structures responsible for material elasticity in low-energy physics. In such a framework, GR is equivalent to an elastic continuum and the Einstein equations are not a fundamental but merely an effective description of the microscopic dynamics, only valid at distances much larger than the Planck length.

Taking this point of view one avoids the main two problems of general relativity:\\
(i) the problem of quantization - as discussed in (\ref{appii806}) and\\
(ii) the existence of singular solutions - because in the micro-elastic interpretation a solution to the Einstein equation makes no sense at distances of the order of one lattice spacing $L_P$, where the discrete nature of the hyper-crystal become apparent.

%(besser al,be statt i,j; außerdem das folgende stark kürzen; es reicht zu sagen, wir brauchen eine Metrik, und die ist via Auslenkungen definiert; sowohl cosserat als aus dislocation etc alles weglassen: im Tetronmodell sehe ich im Moment nicht, ob diese aus i. kruemmung ii. teleparallel iii. plastics distortions oder iv. kristallfehler herkommt. fuer riemann reicht wahrscheinlich einfache elastik. in any case the metrics defines the distance between 2 tetrahedrons. It is then standard techniqe to define a curvature tensor from that and formulate the Einstein equations. to do this in a Lorentz covariant way one has to extend the formalism to include time as 4. variable. This not only keeps track of the fact that any test object as well as any observer are composed of objects fulffiling klein gordon but it also allows to include energy.) 

In the following I will describe some details of this approach and adapt it to the requirements of the tetron model. 
%Actually, I will use Poincare relativity\cite{blago}, a generalization of the Einstein theory where 3+1 dimensional translations are gauged in addition to Lorentz transformations. On the level of geometric observables this means one encounters torsion in addition to curvature. Inclusion of torsion is preferable both from the conceptual point of view ('all existing global symmetries should be gauged') and also because - as already pointed out by Einstein - the effects of gravity can be described by torsion instead of curvature\cite{einst,telepar}.
The fundamental dynamical quantity in general relativity is the metric which defines the distance between 2 spacetime points (or between 2 adjacent tetrahedrons). It can be calculated e.g. from the transition function between arbitrary local coordinates $x_\mu$ on the manifold and local Lorentz coordinates $\xi^\alpha$ of an inertial system via 
\begin{equation} 
g_{\mu\nu}=\eta_{\alpha\beta} \frac{\partial \xi^\alpha}{\partial x^\mu} \frac{\partial \xi^\alpha}{ \partial x^\nu}
\label{tmet}
\end{equation}
where $\eta_{\alpha\beta}$ is the Minkowski metric valid in the inertial frame.
%The fundamental dynamical quantity in the geometrical approach to GR is the vierbein $e_\alpha^\mu$ which defines the transition between arbitrary local coordinates $x_\mu$ on the manifold and local Lorentz coordinates $\xi^\alpha$ of an inertial system via 
%\begin{equation} 
%e^\alpha_\mu= \partial \xi^\alpha / \partial x^\mu
%\label{bein4}
%\end{equation}
%Based on the vierbein, a metrical structure can be defined via
%\begin{equation} 
%g_{\mu\nu}=\eta_{\alpha\beta} e^\alpha_\mu e^\beta_\nu
%\label{tmet}
%\end{equation}
%where $\eta_{\alpha\beta}$ is the Minkowski metric valid in the inertial frame.
One may then use the resulting line element $(ds)^2= g_{\mu\nu}dx^\mu dx^\nu$ to go further ahead and write down the curvature/field strength and the Einstein equations.

%One can go further ahead and use the metric to define the curvature/field strength in order to write down the Einstein equations. It is also possible to introduce generalizations of GR, like teleparallel or Poincare gravity where in addition to Lorentz transformations 4-dimensional translations are gauged. This leads to torsion in addition to curvature as dynamical quantity\cite{blago,einst,telepar}.
%to infinitesimal local Lorentz transformations [$ \omega^{\mu\nu} (x)=- \omega^{\nu\mu} (x)$] 4-dimensional translations $ \epsilon^\mu (x)$ are considered.
%\begin{equation} 
%\delta x^\mu = \omega^\mu_\nu (x) x^\nu + \epsilon^\mu (x) 
%\label{tmet2x}
%\end{equation}
%(hangt 4bein e mit eps zusammen? antwort: die letzte gleichung beschreibt beliebige koordtrafos x-->x'=x+deltax, waehrend beim 4bein eines der beteiligten systeme ein inertialsystem sein muss!!!! aber muss man nicht unterscheiden zwischen koordtrafos, die die physik nicht aendern und displacements, die einfluss auf den kruemmungstensor haben? Antwort: fast alle haben einfluss auf den kruemmungstensor ausser den isometrien. aber mss man dann die deltax nicht allgemeiner ansetzen?)

In the micro-elastic interpretation a gravitational field induces a deformation in the medium, i.e. a displacement of the internal tetrahedrons within physical space from $x_\mu$ to $x'_\mu$. This corresponds to a modification of the metric
\begin{equation} 
%e'^\alpha_\nu =e^\alpha_\mu \frac{\partial x^\mu}{\partial x'^\nu} 
g'_{\rho\sigma}=g_{\mu\nu} \frac{\partial x^\mu}{\partial x'^\rho}\frac{\partial x^\nu}{\partial x'^\sigma} 
\label{tmet23ee}
\end{equation}
and corresponding changes in the curvature tensor. 

%of spacetime these transformations between coordinate systems can be used to describe {\it displacements of an internal tetrahedron within physical space}, i.e. from $x_\mu$ to $x_\mu + \delta x_\mu$. (unterscheide reine koordtrafos von displacements die die physik aendern. ersteres sind wohl isometrien)
%In other words, the DMESC is obtained from the hyper-crystal of internal tetrahedrons in the limit that their average distance goes to zero (while the rigid tetrahedral structure within the internal fibers is maintained).

Since GR is locally Lorentz invariant, the reader may wonder, what the physical meaning of deformations in the time direction is. In the spatial directions it is rather clear that the distances $r \sim L_P$ between neighboring tetrahedrons get modified when a gravitational field is applied. In the time direction it is the 'hopping time', which gets modified, i.e. the time a photon or some other quasi-particle needs to travel(=be emitted, run, get absorbed) from one tetrahedron to its neighbor. This modification occurs, because the presence of massive mignons and in general of any kind of mass/energy modifies the microscopic processes behind the hopping of any 'test excitation'.

Gauge invariance, i.e. the freedom to change the local Lorentz coordinate system, mixes these concepts. It means that one can use different coordinate systems for time/positions of the tetrahedrons and implies, for example, that from inside the hyper-crystal there is no possibility to distinguish 'longitudinal' from 'transversal' curvature, cf. fig. \ref{figchain2}. This point is further discussed in (\ref{appiiharm}).
%This idea will become clearer in the examples discussed below.
%(man kann entweder LP oder TP aendern, wenn man c aendern will. in beiden faellen ist dies ein long shift der tetronen innerhalb der spatially flat crystals. das entspricht der eichfreiheit. um echte spatial curvature zu bekommen, muss man die tetraeder in die inneren dimensionen schicken. von innerhalb des kristalls lassen sich transv und long shift aber nicht unterscheiden.)

%Note we are talking here about an elasticity in the plastics sense. There are no {\it R\"uckstellkr\"afte}, and the distortions are solely driven by energy as described by Einstein's equations. For example, 
Using these ideas one can understand many features of the Einstein theory. For example, the energy released during crystallization immediately blows up the distance between the tetrahedrons thus inflating the volume of the DMESC in accordance with inflationary cosmology.  The initial crystallization energy is also responsible for the subsequent FLRW expansion of the universe. More details about these issues are given in section 2.5, in particular (\ref{appii702}) and (\ref{appiiharm}). The tetron model view on inflation can be found in (\ref{appii702})ff, while the interpretation of dark energy is given in (\ref{appii81de}).

Another example is the Newtonian limit. For a spherically symmetric configuration the metric can be given via the line element
\begin{equation} 
ds^2=(1+\frac{2\phi}{c^2})(cdt)^2 - (1-\frac{2\phi}{c^2})(d\vec x)^2
\label{newt2x}
\end{equation}
%(der zweite fi term ist ambiguous und auch mit c unterdrueckt, daher ist mir nicht ganz klar wieso bei newton laengenkontr ins spiel kommt; eigentlich ist newton nur zeitdilatation)
and the Newtonian limit is defined by $|\phi| \ll c^2$. For a point mass M the gravitational potential far away from the source is given by 
\begin{equation} 
\phi=-\frac{GM}{|\vec x|}
\label{newt2xu7}
\end{equation}
The square root of the coefficients 
\begin{equation} 
\sqrt{g_{00}}=1+\frac{\phi}{c^2} \qquad \sqrt{g_{xx}}=1-\frac{\phi}{c^2}
\label{newt588x}
\end{equation}
%(nicht mit r oder R arbeiten, da dies bei mir der tetraederabstand ist)
give the general relativistic time dilation and length contraction, respectively. In the tetron model these effects are interpreted in the following way:\\
-The gravitational potential of the point source M modifies the average distance $r\sim L_P$ between 2 neighboring tetrahedrons by a factor $1+\phi/c^2$. As a consequence, any measured length of a physical object is modified by this factor.\\
-The gravitational potential of the point source M modifies the average hopping time that is needed by a hyper-crystal excitation to move from one tetrahedron to the next by a factor $1-\phi/c^2$. This applies in particular to the hopping time $T_P$ of a photon defined in (\ref{tm33b}). As a consequence, any measured time interval between physical events is modified by this factor.
%(Strictly speaking, since the Newton limit is defined via an expansion in powers of 1/c, only the modification of the time coordinate ($g_{00}$) is relevant for Newtonian gravity. This point will be further elucidated in (\ref{appiiharm}).)
%It follows from the discussion after (\ref{tmet23ee}) that these modification factors are gauge dependent, i.e. depend on the choice of the local Lorentz frame, i.e. of how one defines space and time. On the other hand, (\ref{newt2x}) implies that the speed of light $c=L_P/T_P$ gets modified by a factor $1+2\phi/c^2$, and this is a gauge invariant effect.(bin mir nicht mehr sicher, stimmt wohl nicht, klar)

More details on the status of GR in the tetron model, as well as on FLRW, gravitational waves and the interpretation of the Newton limit will be given in section 2.5. In general one has to use the ADM formalism\cite{arnodesermisner} or the approach by Carter et al.\cite{carter,pearson} to describe a general relativistic elastic system which includes arbitrary transformations of the time coordinate. I have chosen to restrict myself to the special cases (\ref{newt2x}) and (\ref{flrw3a}), because it makes the presentation much simpler and the arguments more transparent.

Furthermore, I have been intentionally vague about which version of GR must actually be chosen. There are generalizations like teleparallel, Poincare or Einstein-Cartan gravity where in addition to Lorentz transformations 4-dimensional translations are gauged. This leads to torsion in addition to curvature as dynamical quantity\cite{blago,einst,telepar}. Due to lack of experimental information on torsion and of full knowledge of the tetron dynamics it is difficult to say whether one needs a model which describes dislocations or disclinations\cite{hehl1} of tetrahedrons in a flat hyper-crystal or whether 'true' curvature effects are involved, in the sense that the tetrahedrons in fig. 2 are not only shifted by tiny amounts in the horizontal but also in the vertical, i.e. internal direction.

Personally, I give preference to the latter interpretation, because it complements Einstein's original idea of a Riemannian curvature by a physically intuitive micro-picture. One simply has to assume that there are elastic inter-tetrahedral {\it coordinate} interactions in addition to the isomagnetism describing the phenomena of particle physics. These elastic interactions allow for a buckling and bulging of the 3-dimensional DMESC within the full $R^{6+1}$ and can therefore be described in terms of a {\it non-vanishing curvature tensor}. Curvature in the time coordinate is included as described above and then patched with the spatial curvature in a Lorentz covariant way a la \cite{carter,pearson} by considering a matter manifold which is orthogonal to the time slices. More details about this issue can be found in (\ref{appiiharm}). Note that while the inter-tetrahedral coordinate interactions are elastic, the tetrahedral 'molecules' are rigid bodies which align in their internal spaces.
%It was already noted that in order to understand the forces of gravity in the tetron model the tetron model must be extended by the assumption that in addition to the isospin interactions responsible for the phenomena of particle physics  there are inter-tetrahedral {\it coordinate} interactions responsible for the forces of gravity. While the tetrahedral 'molecules' are rigid bodies which align in their internal spaces, the nature of the inter-tetrahedral {\it coordinate} interactions is elastic. Calling the universe a hyper-crystal therefore is a little misleading. One should rather call it a 'hyper-plastics' which in general can show curvature and torsion when buckling and bulging within the full $R^{6+1}$.)

In any case, the behavior of the gravitational field is determined by the form of the gravitational action $S_G$. Since the equations of motion should be of second order in field derivatives, $S_G$ must be at most quadratic in torsion and curvature
\begin{equation} 
S_G=-\frac{c^4}{16\pi G}\int d^4x \; \sqrt{\det(g)} [ \mathfrak{R} + O(\mathfrak{R}^2,\mathfrak{T}^2)]
\label{tmet5}
\end{equation}
where G is the Newton constant, $\mathfrak{T}$ the torsional and $\mathfrak{R}$ the curvature scalar. The explicit structure of $S_G$ including $\mathfrak{R}^2$ and $\mathfrak{T}^2$ terms\cite{haya} is not given here because it is rather complicated, containing the leading term (formally identical to $\mathfrak{R}$ appearing in the Einstein-Hilbert action) plus 3 independent terms quadratic in torsion and 6 quadratic in curvature, plus possibly the cosmological constant. It can be derived from an analysis which demands consistency with the principle of equivalence and the existence of second order e.o.m. and is an example of a generalized 'f(R,T)' gravity theory\cite{fr2,frgravity}.
%\footnote{WEGLda ich oben gesagt habe dass man grav auch mit torsion allein beschreiben kann und ja auch die groesser der koppkungen entscheinden It may be noted that the effects of torsion are so small that they cannot be detected by present day experiments.}
%It should be noted however that R is not identical with the Riemann scalar 

All in all 11 independent coupling constants\cite{haya} appear in (\ref{tmet5}). This large number of free parameters is in accord with the idea that the complete description of gravity must be quite complicated, because it is not more than an effective theory for an elastic system of microscopic entities (the internal tetrahedrons) that fill Minkowski space.

The tetron model allows to extend the view beyond this effective theory, to yield a new picture of material existence. According to this model, the world falls apart into 2 rather disparate pieces:\\
-firstly, {\it the realm of quasi-particles} like quarks, leptons, Higgs bosons and gauge fields. Since all these excitations fulfill Lorentz invariant wave equations, any phenomenon and signal propagation in this sphere is necessarily limited by the speed of light.\\
-secondly, {\it the realm of tetron matter}, i.e. of aligned tetrahedrons and of the hyper-crystal with its elastic/metrical structure.

Since the relevant scales $\Lambda_P\gg\Lambda_F$ are so vastly different, these two spheres do not have much in common. We ourselves exist in the sphere of quasi-particles and can perceive anything coming from the tetron sector only if suitable devices of ordinary matter are patched in between. Gravity, for example, which originally corresponds to a shift of tetrahedron locations on the DMESC, becomes visible in our physical world only due to the reaction with suitable conglomerations of quasi-particles. This is discussed in more detail in (\ref{appii80d3}), where also the role of gravitational waves is elucidated.

Since they are an independent form of matter, tetrons and tetrahedrons can propagate with velocities larger than c. Usually, this is not relevant because they are fixed by bindings within the hyper-crystal. However, the appearance of superluminal metrical velocities shortly after the big bang can be interpreted as bound tetrahedrons moving at larger than the speed of light, cf. (\ref{appii80d5}).
%(oder ist das nur ein summationseffekt durch die inzwischen großen entfernungen? das bleibt sich gleich)

If the DMESC was an ordinary crystal, one could speculate about the existence of an absolute rest system. Since it is elastic, there will only be an approximate rest system at any given cosmic time, which according to the arguments in (\ref{appii7r2}) can be identified with the comoving coordinates used in cosmology to describe the Hubble flow of galaxies.
 
At first sight the existence of such a system seems to contradict special relativity - a well established concept which I do not want to question. Indeed, in the tetron model all normal material objects are quasi-particle waves fulfilling Lorentz covariant wave equations. As such they cannot distinguish an absolute rest system, i.e. they naturally fulfill Einstein's principle of equivalence. On the other hand, the rest system fig. 2 is made of tetrons, and since it is merely the carrier of those quasi-particles, it is impossible to experimentally perceive it in Michelson Morley type of experiments. More details about this issue are given in (\ref{appii7r1})ff.

A particular consequence of tetron cosmology is that the lattice spacing r in fig. 2 is not fixed, but corresponds to an average distance between the tetrahedrons. In addition it varies with time (temperature) during cosmic expansion, simply because the properties of an elastic continuum depend on thermodynamic variables like temperature, pressure etc. The temperature dependence of c and G will be moderate, because most of it is contained in the energy momentum dependence of the Einstein equations.

For reasons explained in (\ref{appii1}), r is to be identified with a time dependent Planck length, i.e. one has 
\begin{equation} 
L_P(t)=\langle r \rangle
\label{tmet6177}
\end{equation}
with r defined in fig. 2 and $1.6\times 10^{-35}$ m being its present average value. There is clearly a relation of this quantity to the scale factor a(t) of the FLRW universe (\ref{flrw3a}) because cosmic expansion is connected to a timely increase in $L_P$.
%keine echte proportionalitaet?
%What was called the 'hyper-crystal' before, therefore is a 'Planck plastics' in the sense of \cite{preparata,kleinertscardigli}. Due to the elasticity property, the lattice spacing / Planck length $r=L_P(t)$ varies with (cosmic) time, with $1.6\times 10^{-35}$ m being its present value.
%describes the formation of the crystal, i.e. of our world. 
%As the next step, one may try to understand the connection of the present discussion with the phenomenon of gravity. Most importantly, it has to be clarified why $\Lambda_R$ and $\Lambda_r$ of figures 2 and 3 may be roughly identified with the Planck scale $\lesssim\Lambda_P$.
%In such an environment, where the inter-tetrahedral coordinate interactions are of micro-elastic type, the metric defined by the elastic medium will react to any kind of mass/energy transfer by attaining a curvature and/or acceleration. 
%In comparison to the isospin/particle interactions discussed before these gravitational effects are known to be tiny. We do not notice them in particle physics but only in macroscopic agglomerates. Furthermore, it is a kind of elasticity where the spring constant is much smaller than the damping, i.e. normally it does not lead to vibrations but to(sondern exponentielle AbklingFunktion die der Elastizitaet entspricht). In 3 dimensions such a system can be described by the differential geometric finite strain theory originally suggested by the Cosserat brothers\cite{strain,cosserat} in 1909, in 3+1 dimensions by general relativity.

By definition, the Planck length is constructed from c, G and h as one of 3 dimensionful quantities which - in the absence of SM interactions - describe all the basic properties of space[m], time[s] and matter[kg]
%Fundamental length, time and mass scales can be given in terms of these quantities as 
\begin{equation} 
L_P=\sqrt{\frac{\hbar G}{c^3}} \qquad \quad
T_P=\sqrt{\frac{\hbar G}{ c^5}} \qquad \quad
M_P=\sqrt{\frac{\hbar c}{ G}} 
\label{tm33b}
\end{equation}
One may invert these relations to obtain
\begin{eqnarray} 
c &=&\frac{L_P}{T_P} \label{tm33bxc}\\
\hbar &=&\Lambda_P T_P \label{tm33bxh} \\
\kappa &=& \frac{L_P}{\Lambda_P} \label{tm33bx}
\end{eqnarray}
where $\Lambda_P=M_Pc^2$ is the Planck energy and $\kappa =G/c^4$ the Einstein constant.

According to (\ref{tmet5}), $\kappa$ is the coupling of choice in GR. As shown below, it has a rather intuitive meaning in the micro-elastic approach, and this statement actually is true for all 3 quantities (\ref{tm33bxc})-(\ref{tm33bx}):\\
%, because they reflect the physics
%The relations (\ref{tm33bx}) have a nice interpretation in the framework of the tetron model (and more generally in any DMESC theory). 
%of space, time and matter, as inherited in the quantities c, h and G:\\ 
%Interpreted in this way, c, h and G depend on the interior conditions of the crystal (temperature, pressure etc). They are not fundamental parameters of the original $R^{6+1}$ and become relevant only after the DMESC is formed.
(i) since $L_P$ is the average distance between 2 tetrahedrons, then $c=L_P/T_P$ makes $T_P$ the 'hopping' time it takes for a photon quasi-particle to hop from one tetrahedron to the next. The question why $T_P$ is the characteristic time for the whole physical system of quasi-particles and valid even for gravitational waves is answered in (\ref{appii80d3}).\\
(ii) since $\Lambda_P\sim \Lambda_r$ is the binding energy of a tetrahedron in the DMESC, Planck's constant $\hbar =\Lambda_P T_P$ reflects the action of the binding energy during the characteristic time, cf. (\ref{appii1}).\\
(iii) finally $\kappa=L_P/\Lambda_P$ gives the disclination of a tetrahedron in the DMESC per unit energy, i.e. applying an energy $\Lambda_P$ to the tetrahedron will displace it by an amount $L_P$. In other words, the gravitational coupling quantifies the elasticity of the ground state tetrahedron material, i.e. its reaction to any kind of mass/energy influx, as described by the Einstein equations.

Formally, one may associate a Lame constant $\zeta$ to the tetrahedral material\cite{horstemeyer}
%besonders Formel 4.44 ff
and relate it to Einstein's constant via
\begin{equation} 
\zeta=\frac{1}{L_P^2\kappa} \approx 10^{112} \frac{kg}{ms^2}
\label{tm33gg8}
\end{equation}
The weakness of gravity ($\kappa$) thus corresponds to an extremely large stiffness ($\zeta$) of the DMESC, which in turn is related to the high density and the rather strong coordinate forces among tetrahedrons, cf. (\ref{appiiharm}) and (\ref{appii312}). The point is that using the value of $\zeta$ it is possible to calculate the average present-day tetrahedral density in the universe 
\begin{equation} 
\rho_T =\frac{\zeta}{c^2} \approx 10^{95} \frac{kg}{m^3}
\label{tm33chi8}
\end{equation}
%again a very large value about 
This is 121 orders of magnitude larger than the density of ordinary (=quasi-particle mignon) matter $\rho_M\approx 10^{-26}kg/m^3$. Although it is a very large value, it should not be taken as a big surprise, because after all tetrons are the omnipresent fundamental building blocks of the hyper-crystal. This issue will be further discussed in connection with (\ref{fgot82}).

It may be noted that (\ref{tm33chi8}) corresponds to the equation for the speed of sound in an ordinary elastic medium (see e.g. \cite{{pearson}}) and that the tetronic $c^2$ according to (\ref{tm38ddem2222}) appears in the energy momentum relation for all kinds of excitations on the hyper-crystal, elastic as well as isomagnetic ones, cf. (\ref{appii80d3}).

Furthermore, (\ref{tm33gg8}) may be re-expressed as
\begin{equation} 
\zeta =\frac{\Lambda_P}{L_P^3}
\label{tm33chi9}
\end{equation}
According to this formula, the Lame parameter can be interpreted as an energy density, whose numerical value is of the order of the energy density of the vacuum arising in quantum field theories, if instead of renormalization one applies the Planck scale as a cutoff for divergent integrals\cite{vacuumenergy}. This result is no accident, because the vacuum of quantum field theories is the zero point energy of all quantum oscillators, and in the framework of the tetron model is determined by the vacuum state of aligned tetrahedrons fig. 2. For details see (\ref{appiien6}).%\footnote{of why the observed value of the cosmological constant (interpreted as an energy density) is so much smaller than $\zeta$. must be noted that it reflects the ratio rhoM/rhoT wirklich???? For a historical discussion of the quantum vacuum and the cosmological constant problem see \cite{vacuumenergy}}
%More details on the physical interpretation of (\ref{tm33gg8}) and (\ref{tm33chi8})will be worked out in (\ref{appiiharm}).
%Note that for constant c one has $L_p^2 \sim T_P^2 \sim \hbar G$ and $M_P^2 \sim \hbar/G$, i.e. $\hbar \sim L_P M_P$ and $G \sim L_P/M_P$.

According to the tetron model, h, c and G are derived and (moderately) temperature dependent\footnote{It has been claimed that specifying the time evolution of these dimensional 'constants' is meaningless\cite{duff}, because the standard rulers also change with time, and that the only thing that counts in the definition of worlds are the values of the dimensionless constants. This claim has been rightfully refuted for various reasons by many authors, see \cite{moffat}.} material properties of the DMESC, not valid outside of it, the only fundamental force (valid over full $R^{6+1}$) being the unknown interaction (\ref{appiift66}) among the tetrons. According to the above h, c and G are determined by the\\
-average 'lattice spacing' $L_P$ between tetrahedrons,\\ 
-the tetrahedral density $\rho_T$\\
-and by the elastic modulus $\zeta$.\\
Therefore they are in principle calculable from the fundamental force among tetrons.

The same is true for the constants of particle physics appearing in the SM lagrangian, i.e. for the Higgs parameters, the fine structure constant $\alpha$ (electric charge e) and the weak mixing angle. While the latter will be derived in (\ref{appii206k}) and (\ref{appii215}) from the fact that the photon is of $D_\star$-tetron content only, $\alpha$ can be interpreted on a similar level as h, c and G, simply because the photon - whose coupling defines $\alpha$ - in the tetron model is not a fundamental but a quasi-particle confined to the hyper-crystal. Finally, the Higgs parameters can be traced back to isomagnetic exchange interactions of tetrons, as described in the first half of this section.

More details of the tetron model meaning of these quantities will be given in sections (\ref{appii1}) for h, (\ref{appii80d3}) for c, (\ref{appii206m}) for the Higgs potential parameters and (\ref{appii-ck4}) and (\ref{appii614}) for the Yukawa couplings. Some rudimentary ideas about the form of the fundamental tetron interaction can be found in (\ref{appiift66}).

The known particles [quarks, leptons and gauge bosons, cf. (\ref{appii726}), (\ref{appii204}) and (\ref{appii210})] are interpreted as intrinsic excitations of the hyper-crystal and as such will extend over at least one lattice spacing $\langle r\rangle = L_P$. Therefore measurements involving physical particles can never be more accurate than $L_P$. As proven in (\ref{appii1}), 
this modifies the quantum mechanical uncertainty principle and can be used to fix the value of h as 
\begin{equation} 
\hbar=\frac{c^3}{G} \langle r \rangle ^2 
\label{tm33b8}
\end{equation}
%(also muss man das Variieren von LP mit einem h(t) auffangen, so dass Schroedingerglg invariant bleibt?)
%--As argued in (\ref{appii204}) and (\ref{appii212}) the photon is an intrinsic excitation of the hyper-crystal. 
%--as an excitation the photon has a finite velocity c which is determined by properties of the hyper-crystal, just as the speed of sound in an elastic medium is determined by the properties of the medium.
%(soll man hier sagen, dass c via kleingordon die lorentzstruktur fuer uns alle erzwingt?)
Similar for the speed of light: since the photon is an excitation of the hyper-crystal, a temperature dependence as well as a dispersion of
%its velocity must be an intrinsic property of the crystal, and not of the full $R^{6+1}$ spacetime - just as the speed of sound in an elastic medium is determined by the properties of the medium. According to this picture, 
c is to be expected and calculable from crystalline parameters, like that of the speed of sound in an everyday elastic medium. A simple argument will now be given why this is not detectable in present experiments.
%(der folgende Satz verwirrt hier nur: and thus is much smaller than needed in VSL theories\cite{vsl} which try to understand inflation on the basis of a variable speed of light.) 
The point is that photons on a lattice with spacing $L_P$ have a dispersion
\begin{equation} 
c(k)=\frac{2c(0)}{L_P k} | \sin\frac{kL_P}{2} | \approx c(0) +O(kL_P)^2
\label{tm38dd}
\end{equation}
i.e. for wavelengths $\lambda = 2\pi /k$ much larger than $L_P$ the speed of light is constant to a very good approximation. Even with the hardest and 'oldest' cosmic gamma rays observed so far deviations from $c(0)=c$ cannot be tested.

Note that (\ref{tm38dd}) relies on the existence of an equilibrium state and therefore does not control the behavior of c at the time of inflation when the hyper-crystal was formed under non-equilibrium circumstances.
%(formula not true at the boundary of the Brillouin zone lamba=LP where the photon as a tetron-antitetron excitation ceases to be an ordinary wave. wie ist es bei magnon dispersion 1-coskLP? gilt die auch an der boundary der brillzone? )
%(der folgende Satz ist nicht unbedingt richtig. Phononen haben auch sinus als Dispersionsrelation und sind nur excitations. ... gilt der sin aber bis zur boundary?)\\
Furthermore, it should be mentioned that (\ref{tm38dd}) is not only valid for isomagnetic excitations like the photon but also for density fluctuations of tetrons like phinons, gravitons etc, cf. (\ref{appii80d3}). 

Actually, it can be used as a starting point to understand the dynamical background of the special-relativistic energy-momentum relation
\begin{equation} 
E^2=m^2c(0)^4+\vec p^2c(0)^2
\label{tm38ddem}
\end{equation}
within the tetron model. As well known, (\ref{tm38ddem}) is equivalent to the Klein-Gordon equation in momentum space, and since the mignons are massive isospin waves fulfilling a d'Alembert type of wave property, they clearly must respect it. Dividing by $\hbar^2$ one obtains from (\ref{tm38ddem}) the dispersion relation
\begin{equation} 
\omega(k)^2=\omega(0)^2+\vec k^2c(0)^2
\label{tm38ddem22}
\end{equation}
%=m^2\frac{c_0^4}{\bar h^2 }+\vec k^2c_0^2
with a low-frequency cutoff $\omega(0)^2=m^2c(0)^4/\hbar^2$. 

Writing $\omega(k)=kc(k)$, the second term on the rhs of (\ref{tm38ddem22}) is obtained from (\ref{tm38dd}). In other words, the propagation part $\sim \vec k^2$ in the dispersion relation for mignons is completely fixed by the coordinate interactions of the DMESC, or more precisely by the value of $c=c(0)$, which according to (\ref{tm33chi8}) is determined by the stiffness and density of tetrons in the hyper-crystal.

In this respect mignons are distinguished from the magnons in ordinary magnetism whose dispersion relation $\hbar\omega=J[1-\cos (kL_P)]$ involves the exchange coupling J even in the propagation term. In the present case, the isomagnetic exchange coupling enters the dispersion only through the mignon rest mass m which according to the calculations in \cite{lmass} is proportional to J.

J is used here as a wildcard for the various internal exchange couplings introduced in (\ref{mm3}) and (\ref{mm3dm}) and in \cite{lmass}. In the case of the strange quark mignon, for example, one would have $J=(4J_{SS}+2 J_{ST} +2K_{ST})c^2$ according to (\ref{mm15ffg}).
%and assumed to be given in units of energy.} 
%w=J(1-coska) fuer magnons, Jsinka fuer antimagnons, g*wurz(1-coska)=gsinka/2 fuer phonons

In summary the relativistic energy momentum relation (\ref{tm38ddem}) can be rewritten in terms of tetron matter properties
\begin{equation} 
E^2=J^2+\vec k^2 \frac{\zeta}{\rho_T} 
\label{tm38ddem2222}
\end{equation}
Note that the low frequency cutoff makes sense. The tetron operators want to oscillate at their natural frequency $\omega(0)\sim J$, and cannot be compelled to oscillate any slower. Such a behavior is generic in any system that has some kind of internal oscillation. The group velocity 
\begin{equation} 
v_g= \frac{d\omega}{dk}=\frac{c(0)}{\sqrt{1+\frac{\omega(0)^2}{c(0)^2k^2}}} 
\label{tm38xxem}
\end{equation}
vanishes in the long wavelength limit while at high energies, the physics of a nondispersive medium with constant group velocity $v_g=c(0)$ is recovered.
%(- the mignon is at rest in the rest system of the hyper-crystal - nur dort???)

Finally, it may be noted that there is also a high frequency cutoff. This corresponds to the Planck scale and to the appearance of the sine in (\ref{tm38dd}). The internal isomagnetic couplings J do not play a role in that regime. As the frequency of the wave is increased, one is probing the physics of an infinite system of tetrahedrons coupled with spring like forces. As it is further increased to still higher values ($kL_P\sim 1$) towards the Brillouin zone, the effects of the sine (the high energy cutoff) is seen. This issue is further discussed at the end of (\ref{appii1}).

\section{Questions and Answers}

%(SEHR GUT hyperlink funktioniert fuer pdf aber nicht dvi oben mit usepackage hyperref)
%\hyperlink{t4890}{dies ist die link caption}

In this section a list of questions and answers is presented which arise in connection with the tetron model. Open problems will be specially marked and the  more important ones reviewed in the summary section 3.

%---------------------------------------------------------

\subsection{Questions about the Gauge Sector}
According to section 1 the universe is interpreted as a discrete fiber bundle over Minkowski space $R^{3+1}$ with fibers given by the iso-magnetic tetrahedrons fig. 1. The electroweak gauge fields are to be interpreted as connections in that fiber space, i.e. they help to define what parallel alignment in between different fibers means.
\subsubsection{How can such a model lead to a local gauge theory?}\label{appii201}
The internal 3-dimensional space which hosts the tetrahedrons is naturally endowed with a $SU(2)_L\times U(1)_F$ symmetry:\\
--the $SU(2)$ factor arises from the rotational symmetry of the internal spin vectors before their alignment. Some details have already been explained in section 1. The question how this symmetry is related to the weak isospin of quarks and leptons will be analyzed in (\ref{appii4})ff. The reason for why it receives the index L in the tetron model is explained in (\ref{appii5}).\\
--the U(1) factor corresponds to tetron number conservation, which on the level of quarks and leptons translates into the $B-L$ quantum number, cf. (\ref{appii216}).\\
These groups act as {\it local} symmetries, because their elements can be chosen different for different points of the Minkowski base space. Connections can be defined for the $SU(2)_L\times U(1)_F$ bundle, which are to be interpreted as gauge bosons. As shown in (\ref{appii205}), there is a mixing of the $U(1)_F$ field with $W_z$, with mixing angle equal to the Weinberg angle. After the mixing the corresponding local gauge fields are given by the observed $W^\pm$, Z and $\gamma$.
%zzz(gibt es eine lokale Fermion number gauge theory? Ja, B-L gibt es lokal in GUTs, aber das zugehoerige VB hat sehr grosse Masse. in leftright symm models aber nicht. dort ist es Mischung mit ZR, und die Frage bei uns ist, wie kann ich WR vermeiden ohne B-L aufzugeben ... oder brauche ich doch WR? aber ich habe es unten in der Frage verneint. Und es passt auch nicht zu einem einfachen 3dim Isospinraum. Aber wenn man 2 Tetraeder definiert, einen fuer QL und einen fuer QR, dann hat man auch entsprechende Konnektionen ... nicht unbedingt denn QL und QR sind ja Vektoren im inneren R3. Wenn die QL nach aussen zeigen, muessen auch die QR nach innen zeigen denn Q=QL+QR waere sonst 0. Aber waere das so schlimm? Auf jeden Fall gilt: die psi*tau*sig*psi WW bewirkt die Auswahl der V-A WW gegenueber der V+A WW. Letztlich spricht also dies Argument gegen WR. Ich brauche es aber fuer mein B-L Argument. Evtl auch fuer die b-Masse.)
%zzz Frage: ist B-L ein sehr schweres Boson wie in GUTs? Antwort: in LR-Modellen nein. 
%normalerweise: ZL und A aus SU2LxU1Y
%SU2RxSU2LxU1BML gibt ZR, ZL und gamma
%ich sehr hier keinen Raum fuer ein schweres BML boson
%nur wenn man U1BMLxSU2LxU1Y nimmt, hat man extra BML boson
%die Frage ist bei uns: wie kommt man von B-L zu Qel, ohne WR einzuführen)
\subsubsection{Are the electroweak bosons and the Higgs field composite?}\label{appii204}
The answer is yes, but one should specify how this works out in detail.\\
--The most straightforward possibility is that they are composites of  mignon antimignon pairs. However, as will be seen in (\ref{appiiqcd}), such a pairing is more appropriate to describe chiral symmetry breaking in QCD. Furthermore, such a construction would make the top-quark content dominate the boson sector of the SM, similar to top condensate models. Since $m_t\gg m_W$, this is usually not considered a convincing scenario.\\
--Secondly, one could be tempted to insist they are fundamental objects, because they are connections of the basic $SU(2)\times U(1)$ fiber bundle in the sense of differential geometry. As such they could have been induced by curvature dynamics of the full R$^{6+1}$ geometry.
%In this case the fundamental constants h, c and N are fundamental for the whole R$^{6+1}$.
I do not think this is a very attractive option, because the flat R$^{6+1}$ knows nothing about the dynamics within the 'curved' hyper-crystal bundle. Furthermore, the Higgs field as a necessary add-on to account for the SSB does not have a simple interpretation in the pure differential geometric framework.\\
--Thirdly, they could be tetron-antitetron bound states traveling freely through the hyper-crystal. However, according to the picture developed in (\ref{appii309}) and (\ref{appii306}), tetrons are so strongly bound within the hyper-crystal with binding energies $\sim \Lambda_P$, that they cannot be split off, not even in pairs. This requirement is also dictated by the no-dissipation concept, cf. (\ref{appii726}). Namely, one has to take care that such pairs do not leave the hyper-crystal and dissipate into R$^{6+1}$, because otherwise energy would not be conserved inside of it.\\
--I adhere to the idea that they must be {\it excitations} of tetron-antitetron bonds, i.e. arise from the tetron interactions in the crystal. While mignons are defined on one tetrahedral fiber, the gauge bosons involve a system of 2 neighboring tetrahedrons. As excitations they thus consist in a global precession of the isospin 3-bein of one fiber with respect to the 3-bein of the neighboring fiber. There are 3 types of such precessions corresponding to the 3 internal Euler angles defining the d.o.f. of the differential geometric SU(2) connection among the fibers and thus to the 3 weak gauge bosons.\\
%(Since the top-quark is the mignon with a similar global rotation property W=t*bbar=sumQLi*sumQRi. die von top nach bbar uebertragene schwingung muss gleiche frequenz haben. top, bottom und w/z haben also durchaus was gemeinsam, und man kann ja auch mt=2mW oder aehnlich aus hh+dm berechnen. aber ist nicht top eher mit fermiscale, also ohne kopplungsfaktor wie beim Wboson? Ausserdem transformiert sich das top triplet nach G4, waehrend W bosonen trivial unter G4 sind)\\
%gut: mignonen sind die goldstones bzgl der inneren so3 einer faser, wzhiggs beziehen sich auf die connection zwischen 2 fasern, beim photon ist nur die tetrondichte betroffen, welche von faser zu faser schwingend variiert.
As a consequence, the gauge bosons form and travel solely inside the hyper-crystal and cannot exist outside of it.
%aber was ist mit redshift? photon kann teil seiner energie auf metrik uebertragen
'Traveling' of such a pairing excitation is meant in the sense of a quasi-particle, i.e. the excitation hops from one tetron-antitetron pair to another, while the tetrons themselves stick to their place in the crystal. It is possible to imagine it as a density wave bilinear in $\bar\psi$ and $\psi$ that travels through the crystal. This must be distinguished, however, from the density fluctuations involving $\psi^\dagger$ and $\psi$ of a single tetrahedron which are coined phinons in (\ref{appii536}) and (\ref{appii207}). 
\subsubsection{Can a stable and massless particle like the photon be an excitation?}\label{appii210}
%\label{appii212}
First of all, note that masslessness of the photon is protected by the U(1) gauge symmetry. As long as this symmetry holds, the photon remains massless, whether composite or not.\\
Furthermore, the masslessness of the photon implies its stability.\\
%\subsubsection{Is the photon really an excitation?}\label{appii210}
%The answer to the question is 'no' in the QED7 model advocated in \cite{lhiggs}, where the ordinary photon is part of a '6+1-dimensional' photon, responsible for the iso-magnetic interactions.
The answer to the given question is yes within the 'no-dissipation' hypothesis advocated in this article, cf. (\ref{appii726}). The latter has the advantage that energy is conserved for all processes inside the crystal, so no compactification of internal spaces is needed. The only objects which are not excitations are the tetrons, the building blocks of the crystal. These, however, are bound with energies $\gg 10^{10}$ GeV.\\
The photon being an internal excitation cannot be scattered away from the hyper-crystal. Since according to (\ref{qu-charges}) one has $Q(U)=0$, the photon is a $\bar D -D$ excitation of D-tetrons, conveniently abbreviated as 
%zzz(normales D oder radiales D? normales, da es auch in der symm Phase existiert, weiter unten erklaert)
\begin{equation} 
A_\mu \sim e \; Q(D) \bar D\gamma_\mu D 
\label{pho88} 
\end{equation}
As discussed after (\ref{radsp1}), in the SSB phase actually radial isospinors should be used
\begin{equation} 
A_\mu \sim e \; Q(D_\star) \bar D_\star\gamma_\mu D_\star 
\label{poo88} 
\end{equation}
%(ich denke jetzt dass die sterne in der summe eines tetraeders sich wegheben, also auch ihre Ladung)
Furthermore, it must be noted that the no-dissipation hypothesis (\ref{appii726}) has rather challenging consequences. If the photon is not a fundamental particle, it is difficult to believe that the Lorentz symmetry, valid inside the hyper-crystal with the known value of c, is a fundamental property of the original full R$^{6+1}$ spacetime. The Lorentz structure as we know it, comes into being only when the crystal is formed and holds only inside of it. This point is further elucidated in (\ref{appii703}), (\ref{appii7r1})ff and (\ref{appii80d3}). 
%ABER Frage: c is not a special property of light but of space-time as a whole the asmptotic velocity of any body with mass approaching zero. Dies interpretieren wir auch einfach als eine lattice property.
%zzz(weiterer hinweis auf compositenes: da bei mir Gravitation ein Gittereffekt ist und Photonen durch Gravi abgelenkt werden)
\subsubsection{What happens on the microscopic level when a mignon and an antimignon annihilate into an electroweak gauge boson?}\label{appii214n}
Assume the 2 mignons are located on 2 neighboring tetrahedrons. When the gauge bosons are formed, the mignons cease to exist and are replaced by an internal excitation of a 
%zzz tight(nicht sehr tight da Higgs schnell zerfaellt?) binding of 
bound $\bar\psi$-$\psi$ pair involving 2 tetrons from the neighboring tetrahedrons. This is in contrast to \cite{lhiggs} where these pairs were assumed to be made from free tetrons floating around. The latter idea has been abandoned because the binding energy of a tetron in the crystal is too large, of order $\Lambda_P$, and it would furthermore allow energy in the form of $\bar\psi$-$\psi$ pairs to dissipate away from the hyper-crystal.\\
The excitations of bound pairs of tetrons behave trivially under Shubnikov transformations (\ref{eq8rela}), i.e. the information about the discrete tetrahedral structure is washed out, because mignon and antimignon compensate each other in that respect. What remains is the transformation property under $SU(2)_L\times U(1)_F$. Since $\psi =(U,D)$ is an isospin doublet, the product of $\psi$ and $\bar\psi$ leads to $2\otimes 2=3+1$, i.e. a triplet (the weak bosons) and a singlet (the B-L photon).\\
%zzz hier passt 4x4bar von SO6 hin, also (2,2)x(2,2)=(1,1)+(1,3)+(3,1)+(3,3) dh Higgs und vecpi ist auch noch dabei
These serve as connections in the fiber space. As such they are useful to define, what alignment of adjacent tetrahedrons means, cf. the discussion after (\ref{mm3dm}).
%Frage: ist B-L wirklich Fermion number, da man ja das B-L Teilchen in GUTs eine sehr hohe Masse gibt, evtl sogas proton Zerfall???? Antwort unten diskutiert: B-L ist einge gute verallgemeinerte Fermionzahl. das b-l photon ist nur in guts sehr schwer, nicht in lr symm theories)\\
%zzz(Frage: Hat nicht das Singlet das Minus-Zeichen? Antwort: bei den zugehoerigen Spinoren, U statt S und D statt T genannt, ist klar, dass UUbar-DDbar das Singlett ist)
%Natur von WZgammaH als Teilchen-Antiteilchen? 2x2=1+3, der Rest der Mignonvib zaehlt nicht. Oder sinds die 8 Dichteschwingungen, die die 24 Spinschwingungen ergaenzen? Nein, die haben kein tau wie das W
\subsubsection{Why can mignon couplings be understood as gauge couplings?}\label{appii214}
The mignons are dynamical sections in the $SU(2)_L\times U(1)_F$ fiber bundle described above. In order to keep up gauge invariance they are naturally endowed with gauge couplings to the connections. As for the couplings of the fundamental tetron fields $\psi =(U,D)$ one may consult (\ref{qu-charges}).
%q/l sind ja die Teilchen im Faserraum, waehrend das fundamentale psi kein Faserteilchen ist, hoechstens als U und D, so dass man eine Chance hat, dass Q(U) ungleich Q(D). Braucht man das ueberhaupt?
\subsubsection{What is the meaning of the initial $U(1)_F$ symmetry?}\label{appii215}
On the tetron level F is tetron number, on the mignon level it is $B-L$.
\subsubsection{How do the electric charges of mignons arise?}\label{appii216}
According to (\ref{appii5}) parity violation of the weak interaction follows from the internal chirality of the tetrahedral 'star' configuration fig. 1. This implies that there are no separate $W_R$ bosons and that all $V+A$ couplings to mignons necessarily vanish, cf. (\ref{appii203}). Still it is possible to formally introduce a right handed isospin quantum number via $I_3=I_{3R}+I_{3L}$ (with vanishing coupling $g_R$ due to the parity violating effect).\\ 
%oft wird in LinksRechtsmodellen gL=gR angenommen. Dies entspricht einer ZUSATZannahme der P-erhaltung, die man dann durch ein Higgs brechen muss. Bei uns ist Paritaet von vornherein durch Tetraeder gerbochen, also gR kann gleich 0 sein
Furthermore, $F=B-L=B+\bar L$ is the appropriate fermion number to choose for mignons (\ref{eq833hg}), with $F(l)=-1$ for leptons and $F(q)=1/3$ for quarks. The mixing among the neutral gauge bosons can then be described by introducing the unbroken generator Q as
%Without $\gamma$Z mixing the mignon-photon coupling can be given in terms of the $U(1)_{B-L}$ charge. Note that $B-L=B+\bar L$ is the relevant (generalized) fermion number symmetry which is gauged before the mixing. If one includes mixing weak isospin comes into play. Since according to answer one can bypass any notion of right-handedness the formula for electric charge of quarks and leptons simply becomes
\begin{eqnarray} 
Q&=&I_3+\frac{F}{2} 
\label{ew93}
\end{eqnarray}
so that
\begin{eqnarray} 
Q(u)&=&\frac{1}{2}+\frac{F(q)}{2} \qquad Q(d)=-\frac{1}{2}+\frac{F(q)}{2} \\ 
Q(\nu)&=&\frac{1}{2}+\frac{F(l)}{2} \qquad Q(e)=-\frac{1}{2}+\frac{F(l)}{2} 
\label{ew94}
\end{eqnarray}
%zzz(sehr gut bemerkung: Meine Masserechnungen unterscheiden nicht L und Lquer. Daher ist gut moeglich, dass B-L=B+Lquer die natuerliche Fermionzahl ist.)\\
%zzzSEHR GUT AUFBEWAHREN, ein Faktor 1/2 andere Normierung als im gedruckten Text
%zzz'Urspruenglich' ist der neutrale Z-Strom fbar*gammue*(±1-sw2*Qf)*f/sw/cw, wobei das Vorzeichen davon abhaengt, welchen Isospin das Fermion f hat
%zzzWir haben Q=T3+BML, also fuer 
%zzzNeutrinos : 0=T3(nue)+BML, also L=T3(nue)=1/2
%zzzElektron:    -1=T3(e)+BML, also L=1+T3(e)=1-1/2=1/2
%zzzup:               2/3=T3(u)+BML, also B=2/3-T3(u)=2/3-1/2=1/6
%zzzdown:        -1/3=T3(d)+BML, alsp B=-1/3-T3(d)=-1/3+1/2=1/6
%zzzDas BML Photon hat eine psibar*gammue*psi Beimischung, das QED8 und das beobachete Photon nicht
%zzznb: baryonzahl ist 1/3 fuer alle quarks (-1/3 fuer antiq), weil Baryonen aus 3 quarks
%zzzleptonzahl ist +1 fuer Leptonen, -1 fuer antileptonen 
%zzz(aber wieso koppelt das 'urspruengliche' Photon so an die Magnonen? Die Kopplung der inneren Magnonen ans INNERE Photon geht mit kreuz Magnetfeld des inneren Photons, aber hier fragen wir nach aeußerem Photon mal innerem Magnon.) 
%zzz(Das B-L Photon scheint an die Dimensionalitaet (Singlet/Triplet) der Darstellung zu koppeln bzw einfach an die Fermionzahl!!!!)
\subsubsection{What is the tetron content of the Higgs field and of the SM vev?}\label{appii206}
To answer this question, the same idea is used which has led to the photon content (\ref{pho88}), namely that all observed scalars and vector bosons arise from correlations between tetrons and antitetrons of neighboring tetrahedrons, cf. questions (\ref{appii204}) and (\ref{appii214n}).\\
%(Problem: Das ew Kondensat kann zwar Tetron-Antitetron sein, aber Tetron-Antitetron als Teilchen ist schwer vorstellbar, wie sich das durch den Kristall bewegt. Die Bindung muss sich vom einen Paar auf das andere uebertragen und entspricht einer Konnexion zwischen den Fasern.)
One of these correlations is directly related to the electroweak SSB and is called the Higgs particle. Since it is to support the radial alignment of isospinors in fig. 2 responsible for the SSB, it can be identified as 
\begin{equation} 
H \sim \bar U_\star U_\star 
\label{pho77}
\end{equation}
where $U_\star$ is the 'radial' iso-spinor introduced in (\ref{radsp1}) corresponding to an isospin vector $\vec Q = U_\star^\dagger \vec\tau U_\star$ pointing outward as in fig. 1. The point is that the content of the Higgs particle is in one-to-one correspondence with the vev needed to stabilize the alignment of isospins in fig. 2, and isospin vectors pointing outward correspond to radial spinors $U_\star$ while those pointing inwards correspond to $D_\star$.\\
According to these considerations the SM Higgs doublet $\Phi$ must be of the form
\begin{eqnarray}
\Phi\sim \tau_2 (\bar U_{R\star} Q_{L\star})^{\dagger T} \sim
\begin{pmatrix}
-\bar D_\star (1+\gamma_5) U_\star \\
\bar U_\star (1+\gamma_5) U_\star
\end{pmatrix}
\label{pho55} 
\end{eqnarray}
i.e. not as in ordinary $SU(2)_L\times SU(2)_R$ symmetric Nambu-Jona-Lasinio (NJL) theories\cite{njl1,njl2} but formally similar to top-color models\cite{cvetic} -- provided the use of radial isospinors is understood.\\
The implication of (\ref{pho55}) on the vev and on the NJL structure inherent in the SM will be discussed in (\ref{appii207}), (\ref{appii207ccdd}) and (\ref{appii-vev}). For use in those sections I include here the definition
\begin{eqnarray}
\widetilde\Phi:=i\tau_2 \Phi^\star \sim \bar U_{R\star} Q_{L\star}
\label{pho55xyz} 
\end{eqnarray}
\subsubsection{What is the tetron content of the weak gauge bosons?}\label{appii206f}
%The answer to this question depends on whether one is talking about the ordered or about the symmetric phase. In the symmetric phase, e.g. shortly after the hyper-crystal was formed, there is only the coordinate tetrahedron but no tetrahedral 'star'-configuration of isospin vectors as in fig. 1. Therefore radial spinors (\ref{radsp1}) should play no role. 
The photon is given by (\ref{pho88}) and the $U(1)_F$ tetron number gauge boson by
\begin{equation} 
B_\mu\sim g' \; F(\psi) \ [ \ \bar U\gamma_\mu U+\bar D\gamma_\mu D \ ]
\label{pho44} 
\end{equation}
where $g'$ is the $U(1)_F$ gauge coupling. Similar formulas hold for the $SU(2)$ gauge bosons, cf. (\ref{appii205}).
\subsubsection{$\gamma$-Z mixing and the value of the Weinberg angle in the tetron model}\label{appii205}
%\subsubsection{Can $\gamma$-Z mixing and the measured value of the Weinberg angle be understood in the tetron model?}\label{appii205} Yes. 
In (\ref{qu-charges}) it is shown that $F(\psi)=-1$ and $Q(D)=-1$. Using this input one can directly infer from (\ref{pho88}) and (\ref{pho44}) that the weak mixing angle at the unification/crystallization point $\Lambda_r$ must be 45 degrees, i.e. $\sin^2 (\theta_w) =1/2$. The form of the Z-boson is
\begin{equation} 
Z_\mu\sim -\frac{e}{\sin (\theta_w)\cos (\theta_w)} \ [ I_3(U) \bar U\gamma_\mu U+I_3(D)\bar D\gamma_\mu D +Q(D)\sin^2 (\theta_w)  \bar D\gamma_\mu D \ ]
\label{phozz} 
\end{equation}
%zzz(tritt hier FD oder QD auf? da ich e vorne stehen habe, wohl QD)
%zzz(Achtung, Z content ist hier vektorartig kein Problem, da V+A Kopplung an Mignonen ist gleich 0)
which at $\Lambda_r$ reduces to $Z\sim \bar U\gamma_\mu U$, i.e. at the unification point the Z like the Higgs particle consists only of U-tetrons.
%($U_\star$, to be more precise, i.e. of tetrons with isospin vectors pointing outwards in the radial direction - aber wenn die isospins noch gar nicht ausgerichtet sind???).
%In contrast to the photon (\ref{pho88}), eq. (\ref{phozz}) shows a non-trivial dependence on $\theta_w$.
Left- and right-handed tetrons are not distinguished in these relations, because the SU(2) gauge bosons a priori contain lefthanded as well as righthanded tetrons. It is only the internal chirality of the configuration fig. 1 that prevents the $V+A$ component to become active, cf. (\ref{appii202}), (\ref{appii5}), (\ref{ew93}) and \cite{lhiggs}.\\
In the next subsection (\ref{appii206k}) the prediction $\sin^2 (\theta_w) =1/2$ at $\Lambda_r$ will be shown to agree with the present experimental value provided one uses 3 ingredients: (i) the evolution of the SM beta function as given in \cite{steinhauser}, (ii) eq. (\ref{phogg}) and (iii) a value of the electroweak unification scale relatively close to the Planck scale.
%zzz(vielleicht sollte man immer r und R etwas kleiner als P sagen.)
%The compositions (\ref{pho66}) and (\ref{pho88}) are assumed to be scale independent, i.e. they represent the tetron content for all scales up to the unification scale $\Lambda_P$. On the other hand at the particular scale $\Lambda_P$ one clearly needs for the original U(1) fermion number gauge boson
%W^z_\mu\sim \bar U_r\gamma_\mu U_r-\bar D_r\gamma_\mu D_r 
%(braucht man das nicht auch fuer alle Scales? da es ja fermionzahl sein soll)
%with $F(U)=F(D)=F(\psi)=-1$ according to question (\ref{qu-charges}).\\
%Using (\ref{pho88}) and (\ref{pho66}) one obtains $B= (Z-A)/\sqrt{2}$. This corresponds to a value of 45 degrees for the Weinberg angle at the unification scale ($\cos(\theta_W)=\sin(\theta_W)=1/\sqrt{2}$).
\subsubsection{A 'unification scale' in the framework of the tetron model}\label{appii206k}
%\subsubsection{What is meant by 'unification scale' in the framework of the tetron model?}\label{appii206k}
In the tetron model the natural electroweak unification scale is given by the energy $\Lambda_r$ at which the hyper-crystal is formed from tetrahedral 'molecules' via the phase transition II. As argued in sections 1 and (\ref{appii1}) this scale corresponds to the average distance between 2 tetrahedrons in fig. 2 and is naturally of the order of the Planck scale. As shown in (\ref{appii205}), at $\Lambda_r$ the value of the Weinberg angle must be 45 degrees. This corresponds to a relation between the $U(1)$ and $SU(2)_L$ gauge couplings
%It is simply the scale where the mixing is such that it transforms between the natural field contents of the U1F boson and the photon, as described in the last section. Since the Weinberg angle defines the relation between the U(1) and the SU(2)$_L$ gauge couplings $g'$ and $g$, using 45 degrees at the unification scale one arrives at the condition
\begin{equation} 
g'(\Lambda_r)=g(\Lambda_r)
\label{phogg} 
\end{equation}
Note that (\ref{phogg}) goes beyond the SM because the gauge group $SU(2)_L\times U(1)_F$, even in the form of a U(2) group, is not simply connected and therefore no relation between the values of $g$ and $g'$ is predicted within the SM. In contrast, in the tetron model a prediction is possible and given by (\ref{phogg}). This is based on the observation that the original U(1) gauge symmetry is tetron number and that the photon according to (\ref{pho88}) should be of D-content only.\\
Using the SM beta funtions\cite{steinhauser} one can extrapolate $g$ and $g'$ from their measured values at $m_Z$ to ultrahigh energies in order to see for which values of $\Lambda_r$ eq. (\ref{phogg}) can be satisfied. Since there is no diminishing factor 3/5 in (\ref{phogg}) like in typical GUT models\cite{rossbook}, $\Lambda_r$ comes out to be nearly equal to the Planck scale instead of $\Lambda_{GUT}\approx 10^{15}\,$GeV. Within the present model, this is a rather convenient result, because it allows to identify the electroweak unification scale with the crystallization temperature.\\
%This is in agreement with arguments in question(Planck=r), which identifies the point at which the hyper-crystal is formed, with the scale where gravity and quantum mechanics start.
%zzz(zu weitgehend) At the end of section 1 a cosmological argument was given that the Planck scale should be time dependent. This is not relevant here, because we are talking about the energy dependence of the couplings in the present-day hyper-crystal. Of course, this does not circumvent the problem whether 
It must be stressed that this is merely an order of magnitude result, because one may rightfully ask, whether the SM beta functions are really applicable up to such high energies, or whether they get appreciable corrections from other crystal excitations like the 2HDM Higgs partners discussed in (\ref{appii207}), or from phinons and isospin density waves which appear at higher energies, cf. (\ref{appii536}). Furthermore, as discussed in sections 1 and 2.5, the Planck scale is expected to be moderately cosmic time dependent in the tetron model.
%At least for the Higgs fields the above analysis is not substantially modified by including a second doublet\cite{2hdmreview}.
%16 \pi^2 \frac{g_i}{\ln Q}=-b_i g_i^3
%where the beta function coefficients are given by
%b_1=-\frac{4}{3}n_g -\frac{1}{10}n_h \qquad \qquad b_2=\frac{22}{3}-\frac{4}{3}n_g -\frac{1}{6}n_h
%nh=2 is the number of Higgs doublets and ng=3 the number of generations
%solution leads to Q0=10**13GeV. NEIN, sondern es kommt MPl=10**19 heraus, weil GUT-Faktor 3/5 fehlt. 
%Also Kristall bildet sich bei Planckskala. ALlerdings ist die, wenn man sie als Bingungsenergie eines Tetraeders im Kristall definiert, zeitabhaengig. Auch ist die Frage, ob nicht andere Kristallanreungen bei hohen Energien die beta-Funktion stoeren.
%in contrast to q/l built from a particle and antiparticle component. locally bound tetron-antitetron states. schematicalls photon=QD*Drbar*Dr, because QU=0. Der Indes r meint den Isovektor zu einem radial nach aussen (Ur) bzw innen (Dr) gerichteten Spinvektor.  Weil der GRundzustand Fig 1 nach aussen gerichtet ist, ist Urbar*Ur das Higgs und Urbar*gam5*Ur das W3. Damit laesst sich bei der ew Unifizierungsskala 10**12GeV (kurz nach Kristallbildung) die Bedingung g1=g2 erfuellen, also sintw=costh=1/w2 fuer den Weinbergwinkel, denn dieser Wert entspricht genau gamma=B0+W3=QD*Drbar*gammue*Dr und Z=B-W3=Urbar*gammue*(1-gam5)*Ur. Die Sm beta-Funktionen lassen den Weinbergwinkel dann zu seinem exp Wert laufen, wo das Photon dann anscheinend doch eine Ur-Beitrag bekommt.
\subsubsection{Connection between the tetron model unification scale and the scales relevant for the standard cosmological model}\label{appii206m}
%\subsubsection{Is there a connection between the tetron model unification scale and the scales relevant for the standard cosmological model?}\label{appii206m}
%As has been shown in question (\ref{appii206h}), in the present model (as well as in all models with $Q_U=0$ and $\bar U U$ as a Higgs) the Weinberg angle is naturally given as $\cos(\theta_W)=\sin(\theta_W)=1/\sqrt{2}$ at the tetron unification scale $g'(\Lambda_P)=g(\Lambda_P)$. It then runs to its low energy value via the SM beta functions. 
%Is there? Yes.
In the Standard cosmological model the scale at which inflation ends is usually identified as the temperature below which the radiation dominated epoque starts. This era can be described as an equilibrium state of effectively massless electroweak gauge bosons.\\
In the tetron model, inflation is associated with the release of latent heat at crystal formation time. The end of inflation is the time when crystallization(=the inflation period) has finished, and the unification of the electromagnetic and the weak interactions is naturally interpreted as happening at this point. It is the time at which our 3+1 dimensional universe started to exist. According to the analysis in (\ref{appii206k}) and (\ref{appii1}) this roughly corresponds to Planck scale energies, and therefore in the present model the electroweak era starts already at the Planck scale.
%scale $\Lambda_P$ corresponds to the beginning of the electroweak era (radiation epoque). , where no SSB and no Higgs existed, and where the original fermion number U(1) boson (\ref{pho88}) was the physical field???? Therefore, the unification scale of the SM gauge symmetries is given by the Planck scale $\Lambda_P$, cf. question(Planck=r).
%(das folgende nicht mehr relevant: Probably not. The mixing between photon and Z is related to the Lage of the tetrahedrons in SU2LxSU2R space. Probably fixed accidentally at SSB time.)
%(Das BML Photon hat eine psibar*gammue*psi Beimischung, das QED8 und das beobachete Photon nicht)
%(nb: Dies setzt U(2)=SU2xU1 Unifizierung voraus. nb: damit ist em Kopplung alpha einzig fundamentale Groesse, wird aber wohl auch eine Gittereigenschaft sein.)
%( gebrochen oder ungebrochen, bei jeder Skala, in JEDEM Fall gilt Q3=I3+F!!!!!
% g1=g2 bedeutet B0+W3 ist das Photon. Weinbergwinkel=45Grad!!!!!
% es muss einen geometrischen Grund geben, warum nicht U1Y, sondern eine Linkomb Y+I3 ungebrochen bleibt,  also das Shubnikovbild nicht genau die SU2L betrifft!!!! JETZT ERKANNT: kein geom Grund, sondern B=UU+DD, waehrend A=DD und Z=UU, also B=Z+A
%\subsubsection{Is there a connection between the isospin interactions (\ref{mm3})+(\ref{mm3dm}) of the tetron model and the SM Lagrangian?}\label{appii207}
\subsubsection{Is there a relation between the isospin interactions (\ref{mm3})+(\ref{mm3dm}) of the tetron model and the SM Lagrangian}\label{appii207}
Yes, there is. To explain this in detail, one first has to notice that 
%the mignon vibrators appearing in (\ref{mm3}) are defined in terms of $\psi^\dagger$, while the components of the Higgs doublet appearing in the SM Lagrangian are given in terms of $\bar\psi$. Even more, 
while the mignon vibrators are supposed to 'live' within one tetrahedron, the Higgs excitations according to the philosophy discussed in (\ref{appii204}) extend over two of them.\\
In accordance with this observation, two types of internal vectors should be distinguished:\\
(i) isospin vectors $\vec Q$ of type (\ref{eq89p}) and (\ref{eq894}) which are the carriers of the isospin waves (mignons). They correspond to the internal angular momentum of tetrons {\it within one tetrahedron} and are the 'charges' of the internal Noether currents. Their excitation spectrum leads to quark and lepton color and flavor. The smallness of neutrino masses is associated to the conservation of these currents, cf. the discussion after (\ref{tm32}) and in (\ref{appii605})-(\ref{appii-dima}).\\
(ii) fields like the gauge bosons or the $\vec\pi$ component of the Higgs doublet, which involve $\bar\psi$ instead of $\psi^\dagger$. Together with the vev  $\langle \bar \psi \psi \rangle \neq 0$ and the Higgs particle (\ref{pho77}) they are important for the pairing process between tetrons and antitetrons of neighboring tetrahedrons which in the tetron model is responsible for the electroweak SSB.\\
To understand this in more detail consider the SM Higgs potential with one doublet
\begin{eqnarray}  
V_{SM}(\Phi)=-\mu^2 \Phi^+ \Phi + \lambda (\Phi^+\Phi)^2= -\frac{1}{2}\mu^2 (\sigma^2 +\vec\pi^2)+\frac{1}{4}\lambda (\sigma^2 +\vec\pi^2)^2 
\label{hipo}
\end{eqnarray}
where $\sigma=\Lambda_F+H$. This potential naturally describes the alignment of neighboring tetrahedrons and anti-tetrahedrons in fig. 2, although in that figure not the $\vec\pi_i$ are drawn but the $\vec Q_i$. The point to note is that two of the $\vec \pi_i$ are in parallel iff all the corresponding $\vec Q_i$ are. Therefore the pairing force $\sim -\mu^2 \vec\pi_i\vec\pi_j$ implied by (\ref{hipo}) exactly corresponds to a 'ferromagnetic exchange coupling' of strength $\mu^2$ in the SSB interaction (\ref{mm3dm}).\\
%In other words, the ground state fig. 2 geometrically looks the same in terms of $\vec\pi$ as it does in terms of $\vec Q$.\\
There is one drawback in this argument, and this concerns the number of d.o.f. While the SM Higgs doublet only has 4 real d.o.f., the isospin vibrators in the form of $\vec Q_L$ and $\vec Q_R$ contain 8. According to (\ref{eq894}) these can be given as
\begin{eqnarray}
\psi^\dagger \psi \qquad  \psi^\dagger i \gamma_5 \vec \tau \psi \qquad \psi^\dagger i \gamma_5 \psi \qquad  \psi^\dagger \vec \tau \psi
\label{njlr1}
\end{eqnarray}
I have included $\psi^\dagger \psi$ and $\psi^\dagger i \gamma_5 \psi$ in this list albeit their vibrations do not correspond to mignons, but to phinons, cf. (\ref{appii536}).\\
The expressions (\ref{njlr1}) are adapted to the $SU(2)_L\times SU(2)_R$ symmetric limit. In more general cases, when this symmetry does not hold, the listing reads 
\begin{eqnarray}
U^\dagger U \quad  D^\dagger D \quad  U^\dagger D \quad  D^\dagger  U \quad   
U^\dagger \gamma_5 U \quad  D^\dagger \gamma_5 D \quad  U^\dagger \gamma_5 D \quad  D^\dagger \gamma_5 U 
%\\  U^\dagger (1\pm\gamma_5) U \qquad  D^\dagger (1\pm\gamma_5) D \qquad  U^\dagger (1\pm\gamma_5) D \qquad  D^\dagger (1\pm\gamma_5) U   
\label{njlr2}
\end{eqnarray}
%or
%\begin{eqnarray}
%U_R^\dagger \psi_L \pm c.c. \qquad D_R^\dagger \psi_L \pm c.c.    
%\label{njlr3}
%\end{eqnarray}
%(ist cc wirklich richtig, da bei dagger ja ein gamma0 die gamma5 kommuattion aendert. Ja, kein Problem, da es nur auf dof, nicht auf Vorzeichen von gamma5 ankommt)
By comparing with the Higgs doublet (\ref{pho55}) one sees that half of the d.o.f. are missing in (\ref{pho55}).
%only has 4 real d.o.f. It can be parametrized by
%\begin{eqnarray}
%\Phi=\frac{1}{\sqrt{2}}
%\begin{pmatrix}
%i(\pi_x-i \pi_y) \\
%\Lambda_F+H -i\pi_z
%\end{pmatrix}
%\label{hig77}
%\end{eqnarray}
%In the tetron model its explicit form is given by (\ref{pho55}) while in a $SU(2)_L\times SU(2)_R$ symmetric framework one has $\vec \pi \sim \bar \psi \vec \tau \gamma_5 \psi$.\\
%After a gauge transformation with $U=\exp (\frac{i\vec \tau \vec \pi}{2\Lambda_F})\in SU(2)_L$, $\vec\pi$ is formally removed from the Higgs doublet and reappears as the longitudinal component of the weak boson isotriplet. Within a $SU(2)_L\times SU(2)_R$ invariant approach one may identify
%\begin{eqnarray}
%\sigma \sim \bar \psi \psi \qquad \vec\pi \sim \psi i \gamma_5 \vec \tau \psi
%\label{njl09y}
%\end{eqnarray}
%while in the tetron model I tend to identify $\Phi$ according to (\ref{pho55}). 
%Obviously 4 d.o.f. are not enough to cover the 8 vibrators $\vec Q_L$ and $\vec Q_R$ needed for the spectrum of mignons (\ref{eq833hg}). 
To account for the other half one should add a second scalar doublet to the dynamics, e.g. in the form\footnote{The radial star indices (\ref{radsp1}) in these expressions are left out for simplicity.}
\begin{eqnarray}
\Phi'\sim \bar D_R Q_L \sim \begin{pmatrix}
\bar D (1-\gamma_5) U \\
\bar D (1-\gamma_5) D
\end{pmatrix}
\label{phos1} 
\end{eqnarray}
%to supplement (\ref{pho55}).
%(durch nix ist bewiesen, dass dies hier ein pseudoskalar ist aber ist eh eine mischung
Together, $\Phi$ and $\Phi'$ form the basis for an extended SM with 2 Higgs doublets. Such models are usually abbreviated as 2HDM, and have been extensively discussed in the literature\cite{2hdmreview,inert1,inert2}.\\
In the $SU(2)_L\times SU(2)_R$ symmetric limit this corresponds to adding a pseudo-scalar iso-scalar particle $\eta$ and a scalar iso-vector triplet $\vec v$ to the theory, cf. (\ref{appii-qqq5}).\\
%Note that within a chirally i.e. $SU(2)_L\times SU(2)_R$ symmetric framework one would obtain
%\begin{eqnarray}
%\Lambda_F+H \sim \bar \psi \psi \qquad \vec\pi \sim \psi i \gamma_5 \vec \tau \psi \nonumber\\
%\eta \sim \bar \psi i \gamma_5 \psi \qquad \vec v\sim \bar \psi \vec \tau \psi
%\label{njl09w}
%\end{eqnarray}
%for the particle content of $\Phi$ and $\Phi'$.\\
%zzz(es gibt auch ein Argument, das gegen 2hdm spricht, das man naemlich fuer die octonion WW, die zur PV fuehrt nur psi*psi und psi*sig*tau*psi gebrauchen kann, und auch nur in einer bestimmten linkomb - die gerade sig**2+pi**2 des Higgspotentials entspicht. auch bei 4x4bar=22x22=11+13+31+33 kriegt man nur das sm dublet, nicht das andere.)\\
The argument about the 'ferromagnetic' alignment induced by the negative mass term $-\mu^2\vec\pi\vec\pi$ in the potential can be extended to the 2HDM model where the potential contains a term $\sim \vec v \vec v$ in addition. This term, however, must not give an appreciable contribution to the SSB interaction (\ref{mm3dm}), because otherwise the b-quark mass would come out to be of order $\Lambda_F$.
%***(aber brauche ich das nicht gerade? es ist dann eben ein unnatürlich kleines v ... auch die Frage grosse Massen kleine Kopplungen ist hier interessant)
%because there is no contribution to the SSB from $V+A$ and no contribution to the inter-tetrahedral interactions (\ref{mm3dm}).
In other words, the second Higgs doublet $\Phi'$ must not take part in the SSB; its 'mass term' has to have a positive coefficient and correspondingly
\begin{eqnarray}
\langle \Phi'\rangle=0 
\label{phiss}
\end{eqnarray}
i.e. there are no quark/lepton mass contributions from $\Phi'$.\\
Eq. (\ref{phiss}) is further supported by the fact that according to (\ref{phos1}) it implies $\langle \bar D D \rangle=0$ or, more precisely, $\langle \bar D_\star D_\star \rangle=0$, i.e. no alignment of isospins in the inward direction - as should be according to fig. 1.\\
Finally, (\ref{phiss})
%This feature 
smartly agrees with the property of the inert version\cite{inert1,inert2} of the 2HDM model. It is interesting to note that in that model the $\eta$ or the $v_z$ (depending on which mass is smaller) is a serious dark matter candidate.
%because there are no couplings between q/l mignons and $\Phi'$ NUR WENN man zusaetzlich Z2 symmetrie einfuehrt.
For further details see (\ref{appii207ccdd}) and (\ref{appii811}).
%(man koennte auch fuer ein kleinen vev fistrich argumentieren. dann haette man QR*QR moriya term und damit wohl eine b-masse, kann man zb bei mb massen frage reinbringen - aber keine dark matter) ich denke jetzt dass fhi'=0 doch möglich ist und gleichzeitig mb von QR kommen kann, weil letzteres auch durch eine Mischung der beiden higgese entstehen kann ... ODER BESSER DURCH EINE MISCHUNG AUF NIVEAU DER CKM MISCHUNG
%besser habe ich es auf dem Papier bewiesen
%zzz aber einen kleinen vev, der die b-masse liefert koennte man erlauben und dann waere es nix mehr mit dem dark matter kandidaten - man koente aberargumentieren dass  <Phi'>=0 der v+a WW entspricht und die durch fig1 verboten ist
\subsubsection{What precisely is the argument in favor of the inert version of the 2HDM model?}\label{appii207ccdd}
The most general quark Yukawa Lagrangian in a 2HDM model is given by
\begin{eqnarray}  
-L_Y=\bar q_L (\Gamma \Phi  + \Gamma' \Phi') d_R + \bar q_L ( \Delta \widetilde\Phi  + \Delta'\widetilde \Phi') u_R +c.c.
\label{yuk2h}
\end{eqnarray}
where 3x3 matrices of Yukawa couplings $\Gamma$, $\Gamma'$, $\Delta$ and $\Delta'$ in family space have been introduced. The resulting quark mass matrices are then given by
\begin{eqnarray}  
M_d=  \Gamma \langle\Phi\rangle  + \Gamma' \langle\Phi'\rangle \qquad\qquad
M_u=  \Delta \langle\Phi\rangle^\star  + \Delta' \langle\Phi'\rangle^\star
\label{yuk3h}
\end{eqnarray}
Unfortunately, the diagonalization of $M_u$ and $M_d$ does not simultaneously diagonalize the quark-Higgs Yukawa interactions implied by (\ref{yuk2h}), and this leads to the problem that unwanted FCNCs are present in the most general 2HDM model\cite{2hdmreview}. This is usually handled by the ad hoc introduction of an additional $Z_2$ symmetry. For example, one may demand the 2HDM Lagrangian to be invariant under the transformation 
\begin{eqnarray}  
\Phi'\rightarrow -\Phi'
\label{yuk4h}
\end{eqnarray}
In this case all Yukawa couplings involving $\Phi'$ drop out, and all quarks and leptons couple solely to $\Phi$. Furthermore, symmetry under (\ref{yuk4h}) forbids mixing terms $\sim \Phi^\dagger\Phi' +c.c.$ in the 2HDM Higgs potential so that the Higgs field with vanishing vev can be unambiguously taken to be $\Phi'$ in accordance with the representation (\ref{phos1}).\\
In the tetron model one has explicit representations (\ref{pho55}) and (\ref{phos1}) for $\Phi$ and $\Phi'$ and may therefore ask whether the physical origin of the $Z_2$ symmetry can be understood. It is easily seen from (\ref{pho55}) and (\ref{phos1}) that (\ref{yuk4h}) corresponds to the transformation
\begin{eqnarray}  
D_R\rightarrow -D_R
\label{yuk5h}
\end{eqnarray}
among tetrons $D_R$. Indeed, this kind of symmetry naturally arises in the tetron model, because mignon interactions with the $D_R$ field in the effective tetron Lagrangian do not appear. The point is that according to fig. 1 the system's ground state is composed of tetrons U alone, and not of D. Since quarks and leptons are excitations of the ground state, it is thus understandable that their couplings to D are strongly suppressed. This reasoning applies only to $D_R$ and not to $D_L$, because parity is broken in the tetron model and $D_L$ couplings are present because $D_L$ appears in an isodoublet with $U_L$.\\
Note in this and the previous question one is always talking about the radial isospinors $U_\star$ and $D_\star$ instead of U and D.
\subsubsection{What is the tetron content of the additional scalar particles in the 2HDM?}\label{appii-qqq5}
This question is most easily answered in the $SU(2)_L\times SU(2)_R$ symmetric limit of the model. There are 5 observable Higgs scalars in the 2HDM model which are then given by
\begin{eqnarray}
H \sim \bar\psi  \psi  \qquad
\eta \sim \bar\psi i\gamma_5 \psi  \qquad
\vec v \sim \bar\psi \vec \tau \psi  
\label{e444}
\end{eqnarray}
leading to a quadratic part of the potential\cite{njl2}
\begin{eqnarray}
-\mu^2 \Phi^\dagger \Phi - \mu'^2 \Phi'^\dagger \Phi' \sim G [(\bar\psi\psi)^2 + (\bar\psi  i\gamma_5 \vec \tau \psi)^2 ] 
+ G' [(\bar\psi i\gamma_5 \psi)^2+ (\bar\psi \vec \tau \psi)^2 ]  
\label{e444bb}
\end{eqnarray}
where the terms with coupling $G$ correspond to the $\mu^2$-term in the SM Higgs potential (\ref{hipo}) and the terms with coupling $G'$ to an analogous quadratic term for the second Higgs doublet.\\
Note that although the vibrators $\vec Q_{L,R}$ in (\ref{eq894}) are chosen in a $SU(2)_L\times SU(2)_R$ symmetric manner in accordance with the discussion in (\ref{appii620}), for many other considerations it is better to use the representations (\ref{pho55}) and (\ref{phos1}) of the doublets $\Phi$ and $\Phi '$.
%calculate (\ref{e444bb}) based on the representations (\ref{pho55}) and (\ref{phos1}) for the doublets $\Phi$ and $\Phi '$. Anyway, due to the possibility of mixing, e.g. between $\eta$, H and $v_z$, the actually observed scalar excitations will in general be linear combinations of these states.
%Frage: oben habe ich gesagt dass phi und phistrich nicht mixen wg der Z2 symmetrie die den fi*fi' massenterm verbietet. Andererseits ist in allen 2hdm reviews von mixing der Higgse die rede)
\subsubsection{How can the chiral nature of the weak bosons be ensured?}\label{appii202}
The iso-magnetic tetrahedral structure in fig. 1 violates internal parity, the state with opposite internal parity being given by a system where the 4 internal spin vectors show inwards instead of outwards. In (\ref{appii5}) it will be proven that this internal parity violation triggers the violation of external parity as required for the V-A nature of the weak interactions, provided the interaction among tetrons stems from a common interaction in the full $R^{6+1}$ space, cf. (\ref{njl0gam}) and (\ref{re76}). 
\subsubsection{Are there SU(2)$_R$ gauge fields $\vec W_R$ in addition to SU(2)$_L$?}\label{appii203}
No. Z and W are originally connections of an SU(2) bundle. According to the discussion in (\ref{appii5}) it is only the formation of the chiral structure fig. 1 together with the $R^{6+1}$ origin of the interaction which forces them to couple to left-handed mignons only. Without the internal chirality fig. 1 the weak interactions would be vectorlike.
%Wir haben gar kein ZR und WR, sondern das Z ist fuer R und L gleichzeitig zustaendig und wechselwirkt nur wegen der Shubnikovgruppe linkshaendig.
\subsubsection{If there is no $\vec W_R$, why is there $\eta$ in (\ref{e444})}\label{appii208}
The 2HDM model (\ref{appii207}) naturally accompanies the vibrations of $\vec Q_L$ and $\vec Q_R$. 2HDM models do not need a $\vec W_R$-field\cite{2hdmreview}.
%zzzes gibt auch die hypothese dass die octonion-WW Higgs=1 und pi=sig*tau erlaubt, aber eta=sig und v=tau verbietet. Bei 4x4 von SO6 kommt H und pi, nicht aber v und eta raus
%da ich H als skalaren Partner von vecWL interpretiere und da es vecWR-Bosonen ja nicht gibt, erwarte ich inzwischen auch kein eta oder veca mehr! allerdings betrachte ich durchaus Schwingungen von psidagger*(1+-gam5)*tau*psi, und dazu wuerde eta und veca durchaus passen.
%In addition, one may follow the hypothesis: the non-existence of $\vec W_R$ makes the scalar field $\eta$ and $\vec v$ inert, so that they are dark matter candidates. Ja denn phistrich hat verschwindenden vev
%Oder ist das folgende richtig: wenn es kein WR gibt, darf es auch kein eta und v geben.
%Letzteres ist falsch: denn LRsymm Modell und 2HDM sind nicht dasselbe: 
%-bei LR braucht man ausser 2 Higgsdubletts (genauer 1 Bidublett LR=2,2) noch 2 Higgstripletts
%-bei LR ist die Komponente vecv des 2.Higgsdubletts VERMUTLICH Teil von WR kann also gar nicht beobachtet werden
\subsubsection{Is there a tetron interaction which gives rise to such iso-magnetic structures? Can the parity violation of the weak interactions be explained from first principles?}\label{appii5}
Before I start to discuss this question, note it is not about the fundamental coordinate forces which are responsible for the formation of (tetrahedral molecules and) the hyper-crystal at scale $\Lambda_r$, but only about the isomagnetic forces relevant for the isospin vector alignment at the Fermi scale. Of course, there is a connection between the 2 issues - to be explained in (\ref{appiift66}) and (\ref{appiiheuri}).\\
%zzz(aber wo werden die anderen behandelt?)\\
%(die folgende Notiz sagt genau dasselbe) Note this section answers the question about the origin of the forces responsible for isospin alignment of tetrons, but not about the fundamental forces that form the crystal at Planck energies, i.e. bring tetrons and tetrahedrons together on the coordinate level. That question is discussed in qu.
The model advocated in this paper consists of 6+1 dimensional spinor fields $\psi$ ('tetrons'), which form tetrahedral structures and a hyper-crystal a la fig. 2. In this world quarks and leptons propagate through spacetime as quasi-particles made of isospin precessions. The Higgs field and the observed vector bosons are excitations of tetron-antitetron bonds, and the system as a whole gives rise to the $SU(2)_L\times U(1)_F$ gauge symmetric SM.\\
The isomagnetic tetron interactions are claimed to derive from the octonion structure which is naturally inherent in a 6+1 dimensional space. The octonions form the unique non-associative, non-commutative and normed division algebra in 8 dimensions, and their imaginary units
%(eigentlich braucht man fuer oct 7 raeumliche Dims!, nein auch 6+1 ist ok).
provide for 7 of the 21 generators of SO(6,1). They are closely related to the Dirac matrices $\Gamma_\mu$ in 6+1 dimensions\cite{rossbook}. A 6+1 dimensional vector current $\bar\psi\Gamma_\mu \psi$ arises more or less directly from the product of two octonions corresponding to the spinors $\bar\psi$ and $\psi$.\\
When the hyper-crystal is formed and $R^{6+1}$ decomposes into Minkowski and internal space, the $\Gamma_\mu$ split into SO(3,1) Dirac matrices $\gamma_\mu$ and a remainder according to\cite{rossbook,dixonbook}
\begin{eqnarray}
\Gamma_{0-6}=(\gamma_{0-3},\gamma_5 \tau_{x,y,z})
\label{njl0gam}
\end{eqnarray}
where x, y and z denote the internal coordinates. This splitting has its physical origin in the coordinate interactions of the tetrons, which lead to the formation of the hyper-crystal, and mathematically it parallels the splitting of an octonion into 2 quaternions.\\
Starting from (\ref{njl0gam}) one can try to derive the parity violation of the weak interaction. The important point to note is the appearance of the product $\gamma_5 \vec \tau$ in the internal part of (\ref{njl0gam}). In principle, the presence of such a coupling corresponds already to a parity violating behavior, both in internal and Minkowski space, because $\gamma_5$ signals axial behavior in Minkowski space and $\vec \tau$ does the same job for the non-relativistic internal fiber.\\
According to (\ref{njl0gam}), any 6+1 dimensional vector coupling $\bar\psi\Gamma_\mu \psi$ reduced to internal space will induce such a term. However, for this to actually become perceivable, an additional appropriate 'chiral situation' has to be provided, again both in internal and Minkowski space. In Minkowski space this can be achieved, for example, by using polarized beams or if there is a second vertex with a $\gamma_5$-coupling in the Feynman diagram of the process.\\
An analogous requirement must be met in the internal space. In other words, a configuration with a handedness must be present, in order to pick up a non-vanishing contribution from the axial coupling, and this in the case at hand is given by the local chiral ground state structure fig. 1.\\
As a matter of fact, the non-relativistic circumstances of the internal $R^3$ space make it a similar situation as one has in optical activity of molecules, where in addition to a circularly polarized photon there must be a handed molecule in order to produce a non-vanishing effect.
%zzz(aber was ist denn die renormierbare isomagnetic WW?)
%(What is missing in the above equation, is an internal scalar potential $C_0$ to provide for the Coulomb force.)
%This is done most easily in the non-relativistic limit where it takes the form
%\begin{equation} 
%[\bar \psi_I \vec\tau  \vec\sigma  \psi_{II}][\bar \psi_{III} \vec\tau  \vec\sigma  \psi_{IV}]
%\label{nonre}
%\end{equation}
According to (\ref{njl0gam}) a 4-tetron interaction of two vector currents will induce among others a term
\begin{equation} 
\vec \pi \vec \pi \sim [\bar \psi \vec\tau \gamma_5  \psi][\bar \psi \vec\tau  \gamma_5  \psi]
\label{re76}
\end{equation}
%zzz(aber dies ist pi*pi nur im SU2xSu2 Model; bei uns pix+i*piy=U*(1+gam5)*D usw ANTWORT: Die Schwinger sind bei mir auf jeden Fall SU2xSU2 symmetrisch)\\
which agrees with the quadratic term of the Higgs potential (\ref{hipo}) responsible for the alignment of isospin vectors and on the level of isospin vectors reproduces the Heisenberg ansatz (\ref{mm3}). As discussed in connection with fig. 3 and in (\ref{appiiheuri}), the sign of the coupling must be anti-ferromagnetic for inner- and ferromagnetic for inter-tetrahedral distances. The latter is in accord with the negative mass term of the Higgs potential.\\
%for the tetrons on the 4 sites of the tetrahedron ...NEIN, sondern mehr ein 2-->2 process
An important question is whether there is a renormalizable interaction in 6+1 dimensions which can accommodate the iso-magnetic properties described here. This will partly be answered in (\ref{appiift66}).
%---------------------------------------------------------

\subsection{Questions about the fundamental Fermion $\psi$}
According to (\ref{eq8}) the tetron matter fields $\psi$ which form the sites of the ground state fig. 2 transform as a spinor 8 under SO(6,1) and decompose into an isospin doublet when the hyper-crystal is formed. In this section some further properties of the tetrons are elucidated. 
%(wenn man mit So71 arbeitet, ist die 8 gleichzeitig spinor matter und VB)
\subsubsection{What is the use of introducing an additional level of matter?}\label{appii301}
%As a student I have always felt uneasy about theories involving symmetry groups where nobody knows where they come from, beginning in nuclear physics with Heisenberg's invention of isospin SU2, going over SU3 of color and ending with GUT groups. 
There are several good reasons to do so:\\
1. the existence of 3 families of quarks and leptons with altogether 24 states (plus the corresponding mass and mixing values) strongly suggests that they are not truly elementary objects.\\
2. a material origin for the observed internal symmetry groups is highly desirable. Traditionally, they are pasted into the theory as purely abstract groups, representing a rather static behavior of the internal spaces. This line of thinking started with Heisenberg's invention of isospin SU(2), included color SU(3) and ended with the (SUSY) GUT groups. The present model works differently, color and isospin being obtained by extending spacetime by 3 internal spatial dimensions in which an independent dynamics takes place.\\
%There is not much else than this dynamical isospin space, i.e. no space for rhe large symmetry groups of GUT theories.
3. spontaneous symmetry breaking is introduced in the SM in a more or less ad hoc way by adding a scalar field to a system which otherwise is made up solely of fermions and gauge bosons. This is similar in spirit to the Ginzburg-Landau model for superconductors, extending it to a relativistic and local non-abelian symmetry. However, as well known from many branches of physics, a material background is required for a phase transition and SSB to occur. For example, in superconductivity the scalar field is provided by electrons bound as Cooper pairs.\\
In the tetron model the breaking of SU(2)$_L$ is associated to the alignment of the (material) internal spins over Minkowski space as shown in fig. 2\cite{lhiggs}. For the interpretation of the Higgs field as a tetron-antitetron correlation see (\ref{appii206}).
%the die normale Vorstellung eines leeren Raumes ist fuer SSB nicht ausreichend. 
%All in all the internal tetrahedral structure gives a material origin not only for the Higgs mechanism but to the internal symmetries themselves. 
%materieller Ursprung fuer isospin SU2, color group and perhaps GUT groups gesucht. 
%\subsubsection{How do the tetrons arise from the original 6-dimensional space?}
\subsubsection{What is so interesting about the 8 of SO(6,1)?}\label{appii302}
This question is discussed in (\ref{appii5}) and (\ref{appii711}).
\subsubsection{What are the couplings / charges of the tetrons?}\label{qu-charges}
%Within the couples with fermion number to qed8 nothing else needed in the qed8 model
After the hyper-crystal is formed, a tetron $\psi$ decomposes into its isospin components U and D. This fact fixes the weak charges
\begin{eqnarray} 
I_3(U)=+\frac{1}{2}  \qquad I_3(D)=-\frac{1}{2}
\label{ew19711}
\end{eqnarray}
Using (\ref{ew93}) one finds 
\begin{eqnarray} 
Q(U)-Q(D)=1 \qquad F(\psi)=Q(U)+Q(D)
\label{ew97}
\end{eqnarray}
where F is a U(1) charge and given by tetron(=fermion) number. Therefore it must be the same for both types of tetrons, i.e.
\begin{eqnarray} 
F(\psi)=F(U)=F(D)
\label{ew9ab7}
\end{eqnarray}
Tetrons do not have a color charge because they are not involved in interactions of triplets of the Shubnikov group, cf. (\ref{appiiqcd}). Therefore, it is appropriate to normalize F in analogy with leptons instead of quarks, i.e. to put
\begin{eqnarray} 
F(\psi)=-1
\label{ew9cd7}
\end{eqnarray}
Eq. (\ref{ew97}) then leads to
\begin{eqnarray} 
Q(U)=0 \qquad Q(D)=-1
\label{ew197}
\end{eqnarray}
In other words, the normalization (\ref{ew9cd7}) is equivalent to defining the U direction in iso-spinor space to be the one which is electrically neutral. This is in accord with the fact that there is no U component in the representation (\ref{pho88}) of the photon and that the symmetry breaking ground state and the Higgs vev (\ref{appii-vev}) are composed of U tetrons only (or actually $U_\star$).\\
Looking at (\ref{ew9cd7}) one may suspect that the result (\ref{ew197}) is just a question of normalization and therefore cannot have much physical impact. However, there are only 2 possibilities for the tetron number of one tetrons, +1 or -1. The first choice leads to Q(D)=0, the other one to Q(U)=0. This means the only freedom one has is which tetron one wants to call U and which one is called D. In this paper the electrically neutral tetron is called U and gives the dominant contribution to the Higgs particle (\ref{pho77}).
%\subsubsection{What are the charges of the tetrons relevant for the ground state fig. 1?}\label{qu-ccc44}
%According to the discussion in section 2.1 the ground state is formed by tetrons $U_\star$ whose electric charge vanishes according to (\ref{qu-charges}).
\subsubsection{How large is the mass/binding energy of a tetrahedron within the hyper-crystal. Can it be ionized?}\label{appii309}
Not with experimental means. The binding energies are extremely large, of the order of $\Lambda_r \sim \Lambda_P$, cf. (\ref{appii1}) and the discussion after (\ref{eqfe}).
\subsubsection{How large is the mass of a single tetron?}\label{appii306}
Difficult to say. If gauge bosons and Higgs scalars would be $\bar\psi$-$\psi$ bound states, the natural guess for $m_\psi$ would be in the range of 40 to 60 GeV.  However, in truth the bosons of the SM are not bound states but correlations of the $\bar\psi$-$\psi$ bonds within the hyper-crystal. Note that tetrons are even more tightly bound within the tetrahedrons than the tetrahedrons are within the hyper-crystal, cf. (\ref{appii309}).
%so tightly bound within this environment with an extremely large binding energy $\approx \Lambda_P$ which has been released during inflation/crystallization that their mass can only be given as an 'effective mass' within the crystal, with value $m_{eff}\approx M_P$.
%(Anders gesagt W, Z and Higgs are more a kind of wandernde bound states)
%zzz($m_\psi \sim 50$GeV interpreted as an effective mass, like that of an electron in an ordinary crystal. aber kristall elektr sind meist beweglich und E=p**2/2meff ist dann angemessen)
%(of a free tetron is probably much larger (oder zero????)
%In the present model, however, W, Z, $\gamma$ and H are excitations of bound $\bar\psi\psi$ pairs within the hyper-crystal, and this gives no restriction on $m_\psi$. The hyper-crystal is so tightly bound that $\psi$ (nor $\bar\psi\psi$) cannot be removed from it at any future energy, making the question of a single tetron mass rather artificial. Guess: larger than $10^{10}$ GeV.
%(Frage: geht electron mass in Magnonenergieformel ein? Nein, nur exchange integral)
\subsubsection{What is the spin/helicity of a single tetron within the internal tetrahedral ground state fig. 1?}\label{appii-qtet}
Figures 1 and 2 contain all necessary isospin information for the hyper-crystal ground state. Since the tetrons U and D are ordinary Dirac fermion in 3+1 dimensions, one may also ask what their spin direction within the hyper-crystal is.\\
First of all, the total spin of all tetrons within one tetrahedron should add up to zero, because otherwise the vacuum state (i.e. the unexcited hyper-crystal) would be polarized. I do not know exactly how strong the limits are, but I am quite sure that a polarized vacuum is not a desirable option.\\
Now assuming the spins add up to zero, there are 2 options:\\
--they do so in a similar fashion as isospins do in fig. 1, i.e. because the sum of spin vectors over all tetrahedral sites vanishes.\\
--the spins from the left-handed and the right-handed isospin vectors $\langle \vec Q_L\rangle$ and $\langle \vec Q_R\rangle$ compensate each other on each site separately.\\
As a byproduct of the considerations in (\ref{appii4}), it can be shown that the second option is fulfilled, because the spins from the tetron and the antitetron contributions in fig. 4 compensate each other.
\subsubsection{Are there $\gamma_5$ anomalies in the tetron model, which could possibly make it inconsistent?}\label{appii-anom}
There are no anomalies in the fundamental theory of tetrons, because there are no $\gamma_5$ couplings. Such couplings arise only in the effective description (the Standard Model) due to the existence of the iso-chiral tetrahedron fig. 1, cf. (\ref{appii5}). On the level of the effective theory the familiar anomaly cancellations among the quarks and leptons apply. 
\subsubsection{*How can a crystal system out of tetrons and anti-tetrons be stable?*}\label{appii304}
In other words: why is there no annihilation between its particle and antiparticle components?\\
%I have no complete answer to this question. 
One observation is that according to (\ref{pho88}) the tetrons $U_\star$ making up the ground state of the hyper-crystal have vanishing electric charge and therefore cannot annihilate into a photon.\\
Actually, this is not really a fair argument, since we are talking here about tetrons and not about their excitations.
%they can still annihilate into a Z-boson. Since the Z is massive this maybe is prevented by the strong binding between tetrons in the crystal.\\
Being the fundamental form of matter, tetrons anyhow are not expected to annihilate into quasi-particles like the photon or the weak gauge bosons. Writing down (\ref{pho88}), (\ref{pho77}) and (\ref{phozz}) refers to {\it excitations} of tetron-antitetron pairs rather than the tetrons themselves.\\
Apart from this argument, I see two further alternatives to answer the question:\\
--the hyper-crystal in its ground state consists only of tetrons and not of antitetrons, i.e. the vacuum expectations values discussed in (\ref{appii206}) and (\ref{appii-vev}) do not have an antitetron contribution.
%This can be expressed in terms of creation and annihilation operators of tetrons ($a^\dagger$ and $a$) and antitetrons ($b^\dagger$ and $b$) as
%\begin{eqnarray} 
%\langle b \rangle = 0
%\label{ew197u6}
%\end{eqnarray}
%instead of the 8 of SO(7,1) announced in eq. (\ref{eq8}) there is only the 4 of SO(6). The latter decomposes into spin and isospin (2,2) when $R^6$ is split into physical $R^3$ and into internal $R^3$.\\
Antiparticles would then only arise within the hyper-crystal on the level of excitations.
%I consider this a rather reasonable scenario, because similar to the discussion in (\ref{appii616}) one can have anti-mignon excitations on top of a purely tetronic ground state.
Although at the moment I do not completely understand all the implications, this could modify our understanding of baryogenesis, an issue to be discussed in (\ref{appiiba8}). However, because of the way weak isospin is constructed in (\ref{appii4}), with the necessity of a tetron-antitetron pair on each tetrahedral site, I do not consider this a reasonable option.\\
%(In such a scenario it is difficult to understand why the Higgs vev which after all describes the isomagnetic alignment of the ground state (independent of any excitation)
--I adhere to the possibility that no field exists into which tetrons and antitetrons annihilate. This can happen because in the original SO(6,1) there is no notion of an antiparticle, because tetron and antitetron comprise into one single representation 8 according to (\ref{eq8}).\\
Furthermore, one may argue that within the scenario developed in sections (\ref{appii7r1})ff, where the hyper-crystal including its constitutive tetron matter is a non-relativistic object with a privileged rest system, while only the quasi-particles are of relativistic nature and the familiar SO(3,1) Lorentz structure emerges but on the level of excitations, one should start with a non-relativistic spinor representation 4 of SO(6). Although antitetrons $\bar 4$ exist in addition to this 4,
%(as particles running backward in time), 
annihilation processes usually do not occur in a non-relativistic environment , cf. (\ref{appiift66}) or \cite{piso,labe}.\\
Both tetrons and antitetrons decompose into a spin and an isospin doublet according to $4\rightarrow (2,2)$, i.e.
\begin{eqnarray} 
\psi=(U_\uparrow,U_\downarrow,D_\uparrow,D_\downarrow)
\label{ew197u8}
\end{eqnarray}
when the hyper-crystal is formed. In other words, the crystallization induces a symmetry breaking
\begin{eqnarray} 
SU(4)\rightarrow SU(2)\times SU(2) 
\label{ew193u8}
\end{eqnarray}
in 6 dimensions where SU(4) is the covering group of SO(6) and the two SU(2) factors correspond to 3-dimensional rotations in internal and physical space, respectively.
\subsubsection{An intuitive understanding of the isomagnetic interactions in the hyper-crystal}\label{appiiheuri}
In section 1/fig. 3 some heuristic arguments were given as how to understand the isospin vector configurations figs. 1 and 2 in a similar way as magnetically ordered states in solid state physics. In (\ref{appii5st}) the Pauli principle will be used to prove that the tetrahedral configuration of tetrons is extremely stable, and in (\ref{appii701u}) a simple formula is given for its energy. These arguments will be extended here in order to show that the global configuration fig. 2 is by far the lowest energy state of a many tetron system.\\
%Many of them actually resemble those for an ordinary magnetic crystal of molecules, with electron spins replacing the role of tetron isospins.\\
Much of the effect can be understood by the fermion property of tetrons alone, i.e. without considering the detailed form of the fundamental tetron interaction to be discussed in (\ref{appiift66}). One may, for example, exploit the behavior of tetrons under identical particle exchange. Since anti-tetrons are different particles, one needs to consider only half of the isospinors, let's say the ones with tetrons and without anti-tetrons. According to the discussion in (\ref{appii4}) and figs. 4 and 5 one can therefore restrict attention to the lefthanded isovectors $\vec Q_L$.\\
The total wave function should be antisymmetric under tetron exchange. Since the ordinary spins are all 'parallel' (all left-handed), the ordinary-spin part of the wave function is symmetric. Concerning the spatial and the isospin part there are then 2 possibilities: either the isospin part is antisymmetric under tetron exchange and the spatial part is symmetric or vice versa. I want to argue that inside a tetrahedral 'molecule' the first and for the 'crystal' binding between tetrahedrons the second possibility is realized:\\
(i) In contrast to the inner-molecular coordinate forces, the inter-molecular coupling between 2 tetrahedrons is relatively weak, and to some extent even elastic -- though with a large stiffness (\ref{tm33gg8}). It is an open question whether this stiffness is a many particle effect or due to an additional super-strong coordinate interaction among tetrons; see the discussion below. In any case, the isomagnetic part (\ref{mm3dm}) of the inter-molecular binding corresponds to a spatial part of the wave function which is antisymmetric under tetron exchange, because under such a condition the wave function becomes small in the middle of the tetron-tetron bond, an effect which usually runs under the name 'Fermi hole' or 'exchange correlation hole'. 
%Note that in contrast to (i) the binding axis is parallel to the base space / physical space. Furthermore,
The Pauli principle then demands that the isospin part of the wave function is symmetric, which corresponds to an aligned, i.e. 'iso-ferromagnetic' configuration. This means nothing else than aligned isospin vectors $\vec Q_L$ in neighboring tetrahedral molecules as shown in fig. 2 and needed for the electroweak symmetry breaking.\\
(ii) Inside the tetrahedrons the isospin part of the wave function is antisymmetric under tetron exchange. This corresponds to an anti-aligned, i.e. 'iso-antiferromagnetic' configuration and leads to a frustrated configuration of isospin vectors $\vec Q_L$ as in fig. 1. As a consequence, the spatial part of the wave function must be symmetric, and this implies that the forces inside a tetrahedron are rather strong. In the language of molecular orbital theory they are 'sigma bonds', and the strengthening effect is referred to as 'Fermi heap'.\footnote{In a chemical bond the Fermi heap allows both electrons to be localized in the internuclear region, thus shielding the positively charged nuclei from their respective electrostatic repulsion. It is a matter of speculation whether in the present case in addition to tetrons some kind of 'nuclei' are needed to form the tetrahedral molecules.}\\
This last point, however, seems to be in contradiction to the finding that the values (\ref{mm15ffff}) of the internal exchange couplings are much smaller than the inter-tetrahedral coupling (\ref{mm3dm333}). On the basis of the preceding discussion one would expect the opposite, i.e. the anti-ferromagnetic couplings (\ref{mm3}) should be quite large, certainly larger than the O(1 GeV) values obtained in (\ref{mm15ffff}). To solve this puzzle\footnote{Another possibility is to give up the picture discussed in (\ref{appii311}) and (\ref{appii530}) and in connection with fig. 3, that the internal extension R of a tetrahedron is larger than the spatial distance $r=L_P$ between two of them.} 
%so that the ferromagnetic overlap is actually larger than anticipated.}
one has to realize that the tetrahedral arrangement of isospins fig. 1 is not an ideal antiferromagnet, but a frustrated one. This fact severely lowers the isomagnetic energy of the state, to values of about the QCD Lambda parameter, while the ferromagnetic coupling $J_{inter}$ in (\ref{mm3dm}) which is responsible for the SM SSB remains much larger.\\
Note that the small values of the internal exchange couplings call for an additional super-strong coordinate interaction among tetrons, in order to maintain strong binding within one tetrahedron. This interaction would then also determine the forces of gravity.\\
Actually, following (\ref{tm33gg8}) it has been argued that the weakness of gravity corresponds to an extremely large stiffness of the coordinate forces among tetrahedrons in the hyper-crystal. In other words, the stronger a force is among tetrons, the weaker it is among mignons. This rule can be applied to particle physics, too: while on the level of mignons the QCD interactions are much stronger than the electroweak ones, on the level of tetrons the internal isomagnetic stiffnesses $J_{SS}$, $J_{ST}$ etc are much smaller than the inter-tetrahedral stiffness $J_{inter}$, the latter being inversely proportional to the Fermi coupling $G_F$.\\
One upshot of the present discussion is that the apparent strength of the nuclear forces can be traced back to the frustrated-ness of the isospins within a tetrahedral molecule. More details on QCD aspects of the tetron model can be found in (\ref{appiiqcd})-(\ref{appii801}).  
%Note that on the level of mignon-mignon interactions the situation is reversed, i.e. the electroweak interaction is small as compared to QCD, and is in fact suppressed $\Lambda_F$ or equivalently by $J_{inter}$.\\
%In case of the isomagnetic interactions which lead to the phenomena of particle physics the situtation is similar because the isospin stiffness $J_{inter}$ is inversely proportional to the Fermi constant. In other words one has the general rule that the stronger the interactions among tetrons, the weaker the interactions among mignons. One can ausspinnen this further by arguing that the strongness of the strong interactions is related to the small internal isospin couplings (\ref{mm15ffff}) which are small due to frustration.)
\subsubsection{*What is the form of the fundamental tetron interaction?*}\label{appiift66}
Although I do not have a final answer to this question, there is a wealth of information (the SM Lagrangian, the family spectrum, the Shubnikov symmetry etc), which may be used to obtain some preliminary insights, and actually, important qualitative features of the tetron-tetron interaction have been derived in previous sections, eg. (\ref{appiiheuri}).\\
As argued in \cite{lhiggs}, rather weak many-particle correlations are sufficient to understand the frustrated arrangement of isospins and obtain the masses of quarks and leptons. The fact that the weak bosons W$^\pm$, Z and H exhibit very small lifetimes agrees with this argument and seems to indicate, that the phenomena of particle physics do not need a super-strong tetron-tetron interaction.\\
The formation of the hyper-crystal at big bang temperatures might also be describable by many-particle correlations, in a similar way as ordinary crystal structures are mainly due to electrostatic interactions. The extreme density (\ref{tm33chi8}) of tetrahedrons further strengthens the enormous stiffness (\ref{tm33chi9}) of the system. According to (\ref{tm33gg8}) it is inversely proportional to the weakness of the gravitational force, cf. (\ref{appii312}).\\
Even the initial formation of the tetrahedral molecules might be understandable in terms of rather 'traditional' forces by recourse to the Pauli principle, cf. (\ref{appiiheuri}). The DMESC is then really similar to an ordinary crystal in that the binding occurs because there is first a separation of charges due to a tetron-antitetron pairing induced by the Pauli principle and afterwards an attraction of these charges.\\
It must be noted, however, that tetron coordinate interactions typically involve binding energies that are enormous (of the order of the Planck scale), and since binding energies often are a measure of the couplings involved, at this point one may face the existence of some additional super-strong binding interaction, which holds the tetrahedral crystal and molecules together.\\
%\footnote{The situation is in contrast to higher dimensional models with compactified internal spaces. In such models extremely strong interaction effects and furthermore a large amount of energy are needed to form the compactified spaces - while in the case at hand the hyper-crystal formation is accompanied by a {\it release} of crystallization energy.} 
To make the analogy with non-relativistic QED complete, in \cite{lhiggs} the existence of an internal 'photon' as part of a 6+1 dimensional U(1) gauge theory has been assumed.\\
It should be mentioned, however, that there are several drawbacks of the QED$_{6+1}$ model:\\
-the ordinary photon interpreted as an excitation cannot be part of the fundamental 6+1 dimensional 'photon' field , cf. (\ref{appii204}) and (\ref{appii210}), because\\
-the 6+1 dimensional U(1) field must be very heavy in order that energy is not dissipated away from the hyper-crystal, cf. (\ref{appii726}).\\
-Huygens principle is not valid in 6 spatial dimensions; wave fronts of d'Alembert kind of waves do not stay sharp.\\
%(gilt das auch fuer massive klein gordon?)\\
For these reasons I follow here a more general approach and analyze various possible effective interactions among tetrons as to whether they yield the correct low energy phenomena.\\
%(In addition to the iso-magnetic forces responsible for the electroweak SSB there is another much stronger force to be considered, namely the one that keeps the skeleton of tetrons together, i.e. the tetrahedral molecules and the hyper-lattice. ABER AUCH BEI MOLEKELBILDUNG WIRD DOCH ENERIE FREIGESETZT. Since it sets up (potentially curved) spacetime, this force must be closely related to the gravitational interactions.
One could consider, for example, 4-tetron contact interactions. The SM together with the tetron description of its bosonic sector [eqs. (\ref{pho88}), (\ref{pho77}),  (\ref{pho55}), (\ref{phos1}) and (\ref{re76})] actually set important constraints on the form of such a contact Lagrangian. These can be read off e.g. from the W-mass $m_W^2 W_\mu W^\mu$ or the $\vec\pi \vec\pi$ term in the SM Lagrangian (\ref{hipo}) and, in a second step, must be adapted for use in a 6+1 dimensional Lagrangian framework.\\
In doing so, there is a scale factor to be considered when going from a 6+1 dimensional fermion $\Psi$ to a doublet $\psi=(U,D)$ of two 3+1 dimensional spinors bound in the hyper-crystal, because $\Psi$ has energy dimension 3 whereas U and D have energy dimension 3/2. I will write  
\begin{eqnarray} 
\Psi=\Lambda_R^{3/2} \; \psi
\label{ewrr80}
\end{eqnarray}
where $\Lambda_R\approx \Lambda_P$ is the scale corresponding to arranging the internal dimensions to a mono-layer of tetrahedrons of thickness R, cf. fig. 2, i.e. to the extension of an internal tetrahedron.\\ 
Consider, as an example, a contact Lagrangian formed by two 7-dimensional vector currents
\begin{eqnarray} 
L_{6+1}=\frac{1}{\Lambda_R^5} \;[\bar\Psi \Gamma_\mu \Psi] [\bar\Psi \Gamma^\mu \Psi]
\label{ewrr81}
\end{eqnarray}
%(statt lambdaR besser r(0) nehmen? und muss man nicht ueberall faktoren c und h betrachten, weil die bei 0 andere werte hatten als now ANTWORT: eh nicht wichtig weil wir diesen approach nicht weiterverfolgen)\\
In this representation $\Lambda_R$ may also be interpreted as the mass of particle exchanged between the tetrons. This could for example be the above mentioned massive QED$_{6+1}$ photon.\\
In any case one sees that $\Lambda_R$ determines the coupling of tetrons, because the Lagrangian is to describe the formation of the internal tetrahedrons at big bang times. The appearance of the fifth power has dimensional reasons and is related to the fact that a propagator of any particle exchanged behaves differently in 6+1 dimensions than in 3+1 -- namely as $r^{-4}$ instead as $r^{-1}$ at small distances (see below).\\
$\Gamma_\mu$, $\mu=0,...,6$ are the Dirac matrices in 6+1 dimensions. They are seven $8\times 8$ matrices which act on the 8 components of the tetron spinor $\Psi\sim (U,D)$. When the hyper-crystal with its 3+1 dimensional 'surface' structure is formed, they break up as\cite{rossbook}
\begin{eqnarray} 
\Gamma_{0-7}=(\gamma_{0-3},\vec\tau\gamma_5)
\label{ewrr82}
\end{eqnarray}
where according to the arguments in \cite{lhiggs} and (\ref{appii5}) the product $\vec\tau\gamma_5$ is one of the ingredients required to establish weak parity violation.\\ 
The 7-dimensional Lagrangian is related to a 4-dimensional one via
\begin{eqnarray} 
L_{3+1}=\Lambda_R^3 L_{6+1}= \frac{1}{\Lambda_R^2} \;[\bar\psi \vec\tau  \gamma_5 \psi] [\bar\psi \vec\tau  \gamma_5 \psi]
\label{ewip81}
\end{eqnarray}
This is because identifying $\int d^3x$ with the volume filled by an internal tetrahedron one has
\begin{eqnarray} 
S=\int d^7 x L_{6+1}= \int d^4x  \frac{L_{6+1}}{\Lambda_R^3} \equiv \int d^4x L_{3+1}
\label{ewrr83}
\end{eqnarray}
The result (\ref{ewip81}) formally has the same structure as (\ref{re76}). More precisely, the $\vec\pi \vec\pi$ term in (\ref{hipo}) can be rewritten as  
\begin{eqnarray} 
V_{SM}=... + \frac{\mu^2}{\Lambda_P^4} \;[\bar\psi \vec\tau  \gamma_5 \psi] [\bar\psi \vec\tau  \gamma_5 \psi]
\label{ewrr81s}
\end{eqnarray}
where $\Lambda_P$ is the Planck scale. 
%(ist es wirklich gut, dass psi und higgs via planck scale verbunden sind?)
To obtain (\ref{ewrr81s}) I have identified 
\begin{eqnarray} 
\vec\pi= \frac{1}{\Lambda_P^2} \;\bar\psi \vec\tau  \gamma_5 \psi
\label{ewrr81t}
\end{eqnarray}
In a traditional technicolor model the TC scale (=mass of the TC gauge bosons exchanged) would appear in (\ref{ewrr81t}). Here we identify it with the scale $\Lambda_P$ set by the Planck lattice constant of the hyper-crystal , cf. the discussion in (\ref{appii530}).\\
%(nur wenn man QM erklaeren will? auch in appii530 mehr auf moeglichkeit einer intermed scale eingehen).\\ 
%Unfortunately, the couplings in (\ref{ewip81}) and (\ref{ewrr81s}) do not have much in common. While (\ref{ewip81}) is true at the big bang (=crystallization point), (\ref{ewrr81s}) holds after the electroweak (=isomagnetic) SSB.\\
A disadvantage of using an approach with contact interactions like (\ref{ewrr81s}) is that it ignores the above mentioned problem with Huygens' principle as well as the Galilean nature of the hyper-crystal to be discussed in (\ref{appii703}) and (\ref{appii80d3}). According to that point of view, tetrons and tetron matter inside the hyper-crystal should not be described by d'Alembert type of wave equations, but a non-relativistic framework should better be used.\\
%(aber urspruenglich 8 von so61 mit eigener speed c' --- aber wave fronts not sharp in 6d)\\
%A related aspect of this kind of reasoning is that a Lorentz structure in SO(6,1), if it exists at all, would presumably involve a limiting speed different from the speed of light valid within the hyper-crystal.
% - simply because according to the photon is an excitation restricted to the hyper-crystal. 
%Furthermore, as  stated above, Huygens' principle is not valid in 6 spatial dimensions. 
%The wave fronts of d'Alembert kind of waves do not stay sharp. These are the reasons why one should be hesitant to extend (\ref{ewip81}) and the other equations from the hyper-crystal to the full 6+1 dimensional spacetime and why a non-relativistic approach is perhaps better suited to describe the fundamental tetron interactions.\\
It is relatively simple to write down non-relativistic 4-tetron contact terms in 6 dimensions, e.g. 
\begin{eqnarray} 
L_{nr}=\frac{1}{\Lambda_R^5} \;  [\Psi^\dagger \lambda^a \Psi] [\Psi^\dagger \lambda^a  \Psi]
%L_{nr}=\frac{1}{\Lambda_R^2} \; { 
%d_1 [\psi^\dagger  \psi] [\psi^\dagger  \psi] + 
%d_2 [\psi^\dagger \vec\tau \psi] [\psi^\dagger \vec\tau  \psi] + 
%d_3 [\psi^\dagger\vec\sigma \psi] [\psi^\dagger  \vec\sigma \psi] + 
%d_4 [\psi^\dagger \vec\tau  \vec\sigma \psi] [\psi^\dagger \vec\tau  \vec\sigma \psi] }
\label{ewip81nr}
\end{eqnarray}
where $\lambda^a$, a=1,...,15 are the generators of SO(6). $\Psi$ in this equation is formally given by (\ref{ewrr80}), but instead of the tetron field $\psi$ (= a relativistic spinor 8 of SO(6,1)) the non-relativistic 4 of SO(6) should be taken, as defined in (\ref{ew197u8}).\\
Since tetrons remain fermions in the non relativistic framework, a quantum mechanical environment must be maintained to describe their interactions, with a Planck constant $h_6$ possibly different from the ordinary one.
%one relevant within the hyper-crystal has to be chosen, cf. (\ref{ewrr86}).
%and to generalize this to 6+1 dimensions using all 15 SU(4) generators instead of only the unbroken ones in (\ref{ew197u38}).(terme mit generatoren und mit der 1 oder wie?)\\
%\begin{eqnarray} 
%\frac{\mu}{\Lambda_P}=\frac{\Lambda_P}{\Lambda_R}
%\label{ewrr84}
%\end{eqnarray}
%The next step would normally be to find a renormalizable interaction in 6+1 dimensions which reduces to (\ref{ewrr81}) at energies below the Planck mass. The appearance of the 6+1-dimensional vector current $\bar\Psi\Gamma_\mu \Psi$ seems to point directly to a 6+1 dimensional QED type of model as proposed in \cite{lhiggs}. However as explained above, this model is plagued by several drawbacks.\\
%First of all, it may suffer from the problem of energy dissipation into the internal dimensions, cf. (\ref{appii726}). Secondly, in the present article the SM gauge bosons are interpreted as tetron-antitetron excitation states which do not exist in the $R^{6+1}$ but arise only when the (3+1)-dimensional hyper-crystal is formed, cf. (\ref{appii204}) and (\ref{appii210}). Therefore they cannot be part of the original 6+1 dimensional gauge field.
%Furthermore, there is a problem connected with the tetron model interpretation of gravity in (\ref{appii703}) and (\ref{appii80d3}).  
%there is the problem that relativistic wave equations in 6+1 dimensions do not propagate as sharp wave fronts, i.e. they obey Huygens principle. Fortunately, according to the picture developed in section 2.5 - especially (\ref{appii80d3}) – 
In such a framework bound states are most conveniently analyzed using a generalization of Schr\"odinger's equation to 6 dimensions with a potential $U(r)$ between two tetrons. Spin and isospin effects can in principle be included by extending this to a 6 dimensional equation of Pauli type.\\
For many aspects of the dynamics like\\
--the coordinate formation of tetrahedrons,\\
--the big bang crystallization process and\\
--the shifts of tetrahedrons inside the elastic hyper-crystal (giving rise to the gravitational interactions)\\
it is sufficient to average over spins and isospins, so that one can do without the Pauli terms and use the 6 dimensional version of the Schr\"odinger equation. Even the exchange couplings responsible for the iso-magnetic interactions (\ref{mm3}) relevant for particle physics can be calculated with the scalar potential of the Schr\"odinger equation alone. Only when polarization effects matter, one should extend this to a 6-dimensional analog of the Pauli equation.
In many respects the 6-dimensional Schr\"odinger equation is similar to its 3-dimensional version. First, it is invariant under 6+1 dimensional Galilean transformations, if U is. The wavefunction $\Phi$ for one internal tetrahedron transforms as
\begin{eqnarray} 
\Phi \rightarrow  \Phi \exp [\frac{i}{\hbar_6} (Et-\vec p \vec r)] 
\label{ewrss2}
\end{eqnarray}
where $\vec p=m\vec v$ is the 6-dimensional momentum and $E=mv^2/2+U$ the energy.\\
Secondly, in the case of free tetrons ($U=0$) there are plane wave solutions 
\begin{eqnarray} 
\Phi \sim \exp[i(\omega t \pm \vec k \vec r)]
\label{ewrss3}
\end{eqnarray}
with a quadratic dispersion $\omega=\hbar^2 k^2/2m$. Such a relation for tetrons, derived within a 6-dimensional non-relativistic quantum mechanics and holding for the highest energies, is completely different than the $\omega \sim k$ dispersion obtained from the d'Alembert/Klein-Gordon type of equation which controls mignon behavior, cf. (\ref{tm38ddem22}).\\
The predominant question in the non-relativistic approach is what the r-dependence of the (iso)scalar potential U(r) appearing in the 6-dimensional Schr\"odinger equation is and what kind of 'charge' it contains.
%obtainable from analysis of quark and lepton mass spectrum?
%because U appears in exchange coupling abgewandelte formel ewtri3!!!!\\
%The time dependence exp$[i\omega t ]$ is actually universal even for non-trivial U. This means except for exotic cases where U itself is time dependent, the eigenfunction corresponding to an energy E is given by $\phi(\vec r,t)= \phi(\vec r) \exp(i\omega t)$ with $E=\hbar_6 \omega$.\\
Since the Green function of the Laplacian in 6 dimensions is $r^{-4}$, an educated guess is $U(r)\sim r^{-4}$. This guarantees the validity of Gauss' law and thus of charge conservation for the new tetron interaction. Furthermore, the structure of propagators is maintained, because the Fourier transform of $r^{-4}$ in 6 dimensions is $\sim \vec p^{-2}$. Nevertheless, other choices are possible, e.g. $U(r)\sim r^{-2}$ or $\sim r^{-1}$, which has been used by some authors in their studies on the stability of hydrogen like atoms in higher dimensions\cite{hydro1,hydro2,hydro3}. Another possibility is $U(r)\sim \exp (-\Lambda_R r) /r^{4}$ in case of the above mentioned massive photon model.\\
%because the corresponding 6-dimensional Schr\"odinger equation can be solved.\\
In order to obtain a tetrahedral bound state, a Newtonian attraction instead of a Coulomb repulsion among the tetrons has to be assumed, i.e. $U(\vec r_i -\vec r_j) < 0$. This could either be handled {\it en face} or as discussed above by charge separation due to the Pauli principle.\\
In any case, inverse power potentials in higher dimensions pose the additional problem that they usually lead to rather weakly bound states (as compared e.g. to hydrogen in 3 spatial dimensions)\cite{hydro1}.\\
If one does not like any of the above alternatives, one can assume instead the existence of some kind of central potential within each tetrahedron. In other words, in addition to tetrons there are yet unknown other components stabilizing the attraction among tetrahedrons within the hyper-crystal by an additional super-strong interaction, cf. discussion and footnote at the end of (\ref{appiiheuri}).
\subsubsection{*Why do the 'molecules' formed by tetrons have a tetrahedral coordinate structure?*}\label{appii307uu}
In principle this question can be answered by analyzing the fundamental tetron interaction and showing that among the 'molecules' composed of n tetrons the ones with n=4 (or n=8) are energetically most favored. Subsections (\ref{appii725}) and (\ref{appii534}) about the growth of the hyper-crystal and (\ref{appii711}) about the octonion origin of the interactions as well as (\ref{appii5st}) and (\ref{appiiheuri}) are recommended to read in this connection.
%Pairing forces among fermions are well known to arise in connection with the Pauli principle, e.g. the stability of the $\alpha$ particle
%in (\ref{appii5}) they have been shown naturally to appear as a part of the Higgs potential. The much stronger coordinate interactions among tetrons which make up the tetrahedral skeleton of the hyper-crystal remain in the dark. (oder wird auch bereits Koordinatentetraeder von re67 festgelegt?) 
%(schlechtere Antwort: urspruenglich S7 Molekuel, wo sich 4 Paare bilden. PROBLEM: WIESO IST raumzeit DANN ELASTISCH? Trennung von innerem und aeusserem Raum also durch Pauliprinzip, falls man das S7 akzeptiert. Wobei, gleichgerichtete Isospins nicht im Sinne des Pauliprinzips. Das ist nur sinnvoll bei TA-Paaren, wo (U,0) und (Dstern,0) zu Gleichrichtung fuehren. bei TA-Paaren macht der Ansatz mit S7-Molekel wenig Sinn.)
\subsubsection{Why not use an internal molecular model instead of a crystal?}\label{appii308}
According to (\ref{appii703}) Lorentz invariance can be approximately established for small enough lattice spacings (of order $L_P$). Nevertheless, some readers may find it difficult to imagine the world as an irregular elastic crystal, with every point in physical space occupied by an internal tetrahedron. So why not use a model, where the quarks and leptons are excitations of isolated tetrahedrons in an otherweise empty space? The molecules would extend into internal dimensions and have a frustrated anti-ferromagnetic structure as in fig. 1. Even an explanation of the SSB as a re-arrangement within the molecules happening below a certain temperature is feasible.\\
However, with such a picture one would run into all the known problems of classic composite models\cite{eichten}. The strongest counter argument certainly is, how one and the same molecule can sometimes have a mass larger than 100 GeV and sometimes be as light as neutrinos.
\subsubsection{Why not use a tetrahedral lattice in ordinary space, without any internal dimensions?}\label{appii-noint}
Since higher dimensions have never been observed experimentally, it is important to critically scrutinize their introduction. In this subsection I follow the idea that the tetrahedrons extend into ordinary space and only {\it mimic} the existence of internal symmetries by forming encapsulated, neutral and ordered tetrahedral systems a la figs. 1 and 2 in which mignons can be excited just as in the model with internal dimensions. As before, the extension of the tetrahedrons would have to be tiny, of order $L_P$.\\
The spin-$\frac{1}{2}$ nature of the mignons would be ensured by the spin-$\frac{1}{2}$ nature of the tetrons just as described in (\ref{appii602}), while the 'internal' quantum numbers arise from the relative angular momentum of the tetrons inside the tetrahedron, i.e. from the Shubnikov symmetry. The whole system including gravity would look similar to a Cosserat continuum, where in addition to the Cosserat deformations describing gravity\cite{hehl1} there are interactions among the encapsulated tetron spins with corresponding spin wave excitations.\\
There are then several advantages of this approach as compared to the model with internal d.o.f.:\\
--there is no problem (\ref{appii726}) with the dissipation of energy into internal dimensions.\\
--the question (\ref{appii534}) why there is no growth of the crystal into the internal dimensions, does not arise.\\
--parity violation of the weak interaction would work analogous to optical activity in molecules, without the necessity to recur to an octonion structure, cf. (\ref{appii5}) and (\ref{appii711}).\\
However, there is also a drawback. It is related to the fact that a rotation in physical space would not only flip the ordinary spin of a mignon, but also transform the 'internal' coordinates.\\
This argument seems to kill the idea. One can only come around this conclusion, if one assumes rather strange behavior of the tetrahedrons, e.g. that they are immersed into a spacetime medium in such a way that they can be rotated independently of spatial rotations, i.e. have an extremely large relaxation time against outside rotations.\\
It may be noted that color has only been observed in singlet states which would not be sensitive to rotations anyhow. As for internal SU(2), weak isospin partners like the electron and its neutrino after the SSB are Shubnikov singlets, too. Weak isospin transitions in the tetron model are constructed rather indirectly as transition between excitations of $\vec Q_L$ and $\vec Q_R$, cf. the discussion in (\ref{appii4}), and this construction can in principle be taken over to the scenario discussed in this subsection.\\
As before, the electroweak bosons would be related to $U(1)\times SU(2)$ gauge transformations referring to tetron number and the rotations of the tetron spin vectors, respectively.
\subsubsection{How large is the internal extension R of one tetrahedron? How large is the average spacing $\langle r(t) \rangle$ between 2 tetrahedrons?}\label{appii311}
In most parts of the paper the following scenario is considered: R is smaller than r, and r is of the order of the Planck length. This is a reasonable assumption because R is the scale of tetrahedral 'molecule' formation, which took place at higher temperatures than the hyper-crystallization, i.e. before our universe was born. Furthermore, the single internal tetrahedrons are rigid and strongly bound objects, while the condensed system of tetrahedrons is elastic and its 'lattice spacing' r grows with cosmic time.\\
Since $\langle r(t) \rangle$ was identified with the Planck length $L_P$ in (\ref{tmet6177}), and is thus related to the uncertainties of quantum theory, the introduction of a length R smaller than $L_P$ may look problematic. However, as discussed in (\ref{appii1}) and in connection with (\ref{tm33b8}), the quantum nature of matter arose when the hyper-crystal was formed during the big bang (via crystallization), i.e. {\it after} the time of tetrahedral molecule formation.
%(wenn die kristallbindung LP ist, warum sind dann nicht die gravikraefte von der staerke LP? weil sie wie fermi von einer Austauschmasse MP bestimmt werden)\\
%DIESEN GANZEN FRAGENKOMPLEX KANN ICH NICHT KONSISTENT BEANTWORTEN
%R<<LP bedeutet, die Drehimpuls QM im Inneren geht so nicht!? anderes hquer im Inneren?
%-ich bin jetzt soweit R<<r anzunehmen, wobei der Gitterabstand r=LP nur ein Mittelwert sein kann und kosmisch wächst, während R die Größe des starren Tetraeders ist und dem Wert von r=LP beim big bang entspricht. 
%-wächst der Gitterabstand (LP) seit dem Urknall, oder expandieren nur die Materieanregungen? Antwort: ja, denn bei jeder Gravi-WW kommt es zu einer Parallelverschiebung eines Tetraeders (Torsion). LP ist nur ein Mittelwert der Abstände in dem Plastik.
%Frage: wächst auch Radius der Tetraeder? NEIN, tetraeder sind rigid
%FRW gehört zu einem gewissen 4bein (siehe Chinesenarbeit), daher gehört die Expansion zu einer gewissen Krümmung und Torsion. Es ist aber nur die Beschleunigung für Krümmung und Torsion relevant, die gleichmäßige Vergrößerung von LP gibt weder Krümmung noch Torsion, anschaulich weil es ein voll plastisches System ist. Oder doch Torsion, weil eine Translation?
\subsubsection{Why should the distance between 2 adjacent tetrahedrons be identified with the Planck scale?}\label{appii313}
One may ask whether there is the possibility, that r und R are much larger than the Planck length. In that case tetrons would have nothing to do with general relativity, and one could forget all reasoning about gravity and cosmology presented in this work. Only the particle physics sections would apply, and r and R would be scales like appear in technicolor models. However, the arguments in (\ref{appii1}) indicate that the most consistent picture is obtained by choosing $r\approx L_P$.
\subsubsection{Is the quantum theory of angular momentum used in (\ref{mm3})-(\ref{mm3dm}) applicable in case $R < L_P$?}\label{appii311ht}
Yes, because the isomagnetic interactions responsible for the Standard Model physics take place within the hyper-crystal and at energies much smaller than $\Lambda_P$ and $\Lambda_r$.
\subsubsection{*The baryon asymmetry in the light of the tetron model*}\label{appiiba8}
%\subsubsection{*What is the tetron model explanation of the observed baryon asymmetry in the universe?*}\label{appiiba8}
Naively one could think that baryon asymmetry is due to a statistical fluctuation shortly after the big bang which makes the observable part of the universe 'baryonic' while other parts are predominantly 'antibaryonic'. However, the experts seem to agree that such an asymmetry would have long been washed out, e.g. by sphaleron effects from the non-perturbative sector of the SM. Instead they prefer to locate baryogenesis at temperatures near $\Lambda_F$\cite{bary}. In that kind of approach, it is then noted that the SM can only partly explain baryogenesis, mainly because of lack of 'enough' CP violation in the CKM sector. Additional new physics not too far above the Fermi scale is needed to remedy the situation.\\
What can be learned from the tetron model about this issue? First of all, within the tetron approach the SM is an effective theory, and it is not clear whether its equations are valid beyond the perturbative regime, i.e. whether the sphaleron argument is really true. Secondly, the low energy limit of the tetron model is a 2HDM model rather than the SM. Since the amount of CP violation is generally larger in 2HDM models than in the SM, baryogenesis can in principle be explained more easily\cite{bary2hdm}. 
%(WEGL: Thirdly, the points put forward in (\ref{appii304}) should be taken into account, and one could try to understand baryogenesis completely from the tetron level perspective, i.e. analyze whether the predicted CP violation\cite{lmass} is enough to induce the necessary imbalance between baryons and antibaryons.)
%(aber gibt es auch eine phys erklaerung auf tetron niveau? zb dass der innere Tetraeder nicht nur das innere P, sondern auch das innere T verletzt? One may try to understand these arguments on the level of tetrons. The CP violation in the 2HDM limit of the tetron model due to the P and T violation by the internal tetrahedrons and its amount calculable by simply calculation the values of the CKM matrix as described in \cite{lmass}. Origin: tetron universe without antitetrons induces a slight imbalance aber welche prozesse genau?)
\subsubsection{A simple memo to understand the role of the permutation group in the ordering of quarks and leptons}\label{appiime11}
Since the 24-dimensional representation (\ref{eq833hg}) of $G_4$ is faithful, one can assign each of the 24 quark and lepton states to an element of $G_4$, in a similar way as suggested in table 2 of \cite{lfound}.
%\begin{table}
%\label{tab2}
%\begin{center}
%\begin{tabular}{|l|c|c|c|}
%\hline
%&...1234...&...1423...&...1243... \\
%& family 1 & family 2 & family 3 \\
%\hline
%& $\tau$, $b_{1,2,3}$ & $\mu$, $s_{1,2,3}$ & $e$, $d_{1,2,3}$ \\
%\hline
%$\nu$ & $\overline{1234} $ & $\overline{2314}$ & $\overline{3124}$ \\
%$u_1$ & $\overline{2143} $ & $\overline{3241}$ & $\overline{1342}$ \\
%$u_2$ & $\overline{3412} $ & $\overline{1423}$ & $\overline{2431}$ \\
%$u_3$ & $\overline{4321} $ & $\overline{4132}$ & $\overline{4213}$ \\
%\hline
%& $\nu_\tau$, $t_{1,2,3}$ & $\nu_{\mu}$, $c_{1,2,3}$  & $\nu_e$, $u_{1,2,3}$  \\
%\hline
%$l$ & $\overline{3214} (1\leftrightarrow 3)$ & $\overline{1324} (2\leftrightarrow 3)$ & $\overline{2134} (1\leftrightarrow 2)$ \\
%$d_1$ & $\overline{2341}$ & $\overline{3142}$ & $\overline{1243} (3\leftrightarrow 4)$ \\
%$d_2$ & $\overline{1432} (2\leftrightarrow 4)$ & $\overline{2413}$ & $\overline{3421}$ \\
%$d_3$ & $\overline{4123} $ & $\overline{4231} (1\leftrightarrow 4)$ & $\overline{4312}$ \\
%\hline
%\end{tabular}
%\bigskip
%\caption{List of elements of $A_4$ ordered in 3 families. Permutations with a 4 at the last position form a $A_3$ subgroup of $A_4$ and may be thought of giving the set of lepton states.}
%\end{center}
%\end{table}
%---------------------------------------------------------

\subsection{Questions about the local tetrahedral Structure and the Nature of the SSB}
To obtain the correct mass spectrum for quarks and leptons, not only the internal geometry but also other features of the model have to fixed.
\subsubsection{Are the tetrahedrons formed by 4 or 8 tetrons, i.e. how many vibrators are needed on each lattice site?}\label{appii51}
%The short answer: according to (\ref{appii4}) the best choice is 8, with a tetron and an antitetron on each tetrahedral site i=1,2,3,4.\\
A priori one may consider several options:\\
(i) the 4-tetron option: naively, one would think that 4 tetrons can give rise to only $4\times 3=12$ excitations of their isospin vectors $\vec Q$. However, there are 2 independent vibrators $\delta\vec Q_L$ and $\delta\vec Q_R$ for each tetron, and for 4 tetrons this gives the desired $2\times 4\times 3=24$ states of eq. (\ref{eq833hg}). For this picture to work the ground state values $\langle \vec Q_L\rangle$ and $\langle \vec Q_R\rangle$ must contain antitetron contributions. This is discussed in (\ref{appii4}) and actually brings option (i) close to option (ii).\\
(ii) the 8-tetron option: here $\vec Q_{Li}$ and $\vec Q_{Ri}$ are carried by 2 different particles approximately occupying the same tetrahedral site i, as depicted in figs. 4 and 5. As discussed in (\ref{appii4}), it is most appropriate to assume this to be a particle and an antiparticle. Since these are sitting very close together, the system keeps its Shubnikov symmetry (\ref{eq8rela}), and all arguments concerning the mignon spectrum remain unchanged.\\
The reason why the options (i) and (ii) are equivalent, can also be seen in the following way: according to (\ref{eq8}) a tetron which transforms as 8 under SO(6,1) decomposes into a particle (1,2) and an antiparticle (2,1) under the Lorentz group SO(3,1), and these are just the d.o.f. required in fig. 4.\\
When assuming a hyper-crystal with an originally Galilean structure as in (\ref{appii7r1})ff, the tetron and antitetron correspond to a $4$ and $\bar 4$ of SU(4), cf. (\ref{ew197u8}). Within the above philosophy this is merely restating the fact that one may consider the particle and antiparticle contributions to the isospin vectors separately.\\
(iii) in case of 8 tetrons there is another option which however will be abandoned for reasons discussed below: namely one could consider unpolarized isospin vectors $\vec Q_{1-8}$, which appear in pairs $\vec Q_i$ and $\vec Q_{i+1}$ on each tetrahedral site i=1-4, again  to be interpreted as tetrons i and i+1 to be very close to each other (with a tiny but non-vanishing internal distance $d_8$), i.e. more tightly bound than to the others. Mathematically it corresponds to a coordinate ground state symmetry $A_4\times Z_2$ instead of $S_4$. Assuming ground state isospins $\langle \vec Q_i \rangle =\langle \vec Q_{i+1} \rangle$ on each site to be parallel, the isomagnetic ground state will again be symmetric under the Shubnikov group $A_4 + S(S_4- A_4)$. It is then straightforward to see that a spectrum of the same form (\ref{eq833hg}) as before is obtained from the vibrations of the $\vec Q_{1-8}$.\\
This scenario is clearly distinguished from those (i and ii) with chiral isospin vectors. First of all, the origin of isospin of mignons (quarks and leptons) is different in the 2 cases. According to (\ref{appii4}) the transition $L\leftrightarrow R$ can be chosen to accompany an isospin transformation. In contrast in the case with $\vec Q_{1-8}$, the $Z_2$ exchange ($i\leftrightarrow i+1$) has to provide for an isospin transition.\\
In addition, there is another, stronger disadvantage of using the approach with $\vec Q_{1-8}$, because the intimate connection between left-handed vibrators and the top-quark gets lost, cf. (\ref{mm3dm}) and (\ref{appii614}), i.e. the understanding why $m_t$ is of the order of the SSB scale while all other quarks and leptons, in particular the b-quark, have much smaller masses, cf. (\ref{appii723err}).
\subsubsection{Why is the tetrahedral 'molecule' so stable?}\label{appii5st}
The shortcut answer is that the tetrahedron is the 'helium' of the tetron model.\\
One can use the Pauli principle to understand this point, cf. (\ref{appiiheuri}). While in the case of helium, two SU(2) spinors arrange antiparallel to form the most stable and abundant element in the universe, for tetrons the spinor representation 8 of SO(6,1) is relevant. The most stable configuration corresponds to a 'shell' filled with 8 tetrons all with {\it different} SO(6,1) quantum numbers, as depicted in fig. 5.
%\\Note however, the Pauli principle can explain the extreme stability of a 8-configuration, but not its tetrahedral form.
\subsubsection{Can there be a tetron and an antitetron on one and the same tetrahedral site?}\label{appiiee5}
If we accept the idea that there are 2 independent fermions on each tetrahedral site (a tetron and an antitetron, cf. (\ref{appii51}) and figs. 4 and 5), then strictly speaking they cannot exist on exactly the same spot. In other words, there must be a tiny nonvanishing distance d between them.
%(WEGL Since according to (\ref{appii4}) and fig. 4 the tetron-antitetron pairing is to generate the weak isospin quantum number and in order to mimic the differential geometric structure of the SU(2) gauge bosons, it is a reasonable assumption, that this distance at least partly extends into physical space, not internal space.) 
For obvious reasons one should have $d\leq r, R$ where r and R were defined in fig.2 and identified as the Planck scale in sections 1 and (\ref{appii1}).
\subsubsection{Is the binding which makes up the Higgs field due to isospin or is it due to tetron coordinate interactions?}\label{appii52001}
It is due to both. The Higgs particle (\ref{pho77}) relies on the alignment of $U_\star$ isospinors and is therefore a natural part of the iso-magnetic interactions. On the other hand, the Higgs (as well as all other scalar and vector fields) is an excitation of the tetron-antitetron bonds in the crystal, and therefore controlled by the coordinate interactions.
\subsubsection{What are the unbroken symmetries of the model?}\label{appii522}
The Shubnikov group $G_4$ and the electromagnetic $U(1)_Q$. The unbroken Shubnikov group has only singlet and triplet representations and leads exactly to the observed color and flavor spectrum of 3 families of quarks and leptons (\ref{eq833hg}).
%the former leading to the spectrum of quarks and leptons with singlets and color triplets, the latter to QED and mainly gauged fermion number???
\subsubsection{Can one calculate the Fermi scale, the Weinberg angle and W/Z and Higgs mass from first principles?}\label{appii525}
The origin of the Weinberg angle was discussed in (\ref{appii205}). The Fermi scale and the Higgs mass arise from iso-magnetic exchange and pairing interactions, as discussed in (\ref{appii206}) and (\ref{appiiheuri}). Therefore if one would know the exact form of the fundamental tetron interaction (\ref{appiift66}), these quantities would be calculable from 6-dimensional exchange integrals.
\subsubsection{Did gauge bosons exist at temperatures above the Fermi scale?}\label{appii526}
Yes, they did. In the tetron model gauge bosons are particle-antiparticle correlations of crystally bound tetrons. They came into being shortly after the crystal was formed at temperature $\Lambda_r\approx \Lambda_P$ and made up for the bulk of particles in the 'radiation dominated epoque', cf. (\ref{appii206m}).
\subsubsection{Did quarks and leptons exist at temperatures above the Fermi scale?}\label{appii526aa}
%\subsubsection{Did quarks and leptons exist in the radiation dominated era shortly after the big bang?}\label{appii527aa}
%zzz(aber sollte es nicht SU2 Dubletts (e,nue) usw in der symm Phase gegeben haben? also Isospinschwingungen die sich im Chaos fortpflanzen. Vielleicht koennen sich mignons auch bei chaos der isospins fortbewegen)\\
No, because from their very nature they require the existence of the iso-magnetically ordered state fig. 2. When the tetron gas cooled down and the hyper-crystal began to form, there was only the coordinate alignment of tetrahedrons but no alignment of isospin vectors. Therefore, at that stage, at temperatures above the Fermi scale, quarks and leptons did not exist, because they could not travel as quasi-particles through the isomagnetically disordered hyper-crystal. Only gravitons, phinons (\ref{appii536}) and scalar and vector bosons (as excitations of the tetron-antitetron bonds) were present. This era is usually called radiation dominated, cf. (\ref{appii206m}).
%in den Frageteil: wieso q/l bei Energien  groesser als critical temperature 1TeV ueberhaupt existieren koennen. Im SM existieren sie schon vorher, kriegen durch die SSB nur ihre Massen. 
%Antwort: das Isospin alignment ist wohl doch stabiler wenn im Kristallverbund. Hysterese!!!!, aber auch weil nicht nur psibar*psi sondern auch fuer jedes i die pii*pii einen vev haben.)
%habe ich weiter unten: Note it is only at this point that quark and leptons exist (dann waeren q/l aber oberhalb 1 TeV nicht existent, andere Anregungen.)
%No. From their very nature they are excitations of the ordered state fig. 2, but not of the symmetric state.
\subsubsection{Then why can quarks and leptons be produced at energies above the Fermi scale?}\label{appii527}
In collider experiments they can exist at energies much larger than 1 TeV because the alignment of isospins is stable much beyond $\Lambda_F$ in the fully ordered hyper-crystal, i.e. in our universe. This is due to a collective hysteresis effect in which the crystal stabilizes itself by the concerted action of all aligned tetrahedrons to maintain the isomagnetic ordering. As a result, quarks and leptons can be produced and propagate normally, even in cases where energies locally exceed the critical temperature $\Lambda_F$.
\subsubsection{Why is $\bar\psi$ involved in the order parameter and not $\psi^\dagger$, whereas the total 'iso-magnetization' $\vec \Sigma$ eq. (\ref{tm32}) is defined just like in ordinary magnetic models?}\label{appii528}
Short answer to a long question: the ground state value of the total internal angular momentum vector $\vec \Sigma =\sum_{i=1}^4 \psi_i^\dagger \vec \tau\psi_i$ vanishes due to the tetrahedral arrangement of isospin vectors within any internal tetrahedron fig. 1, and is therefore not useful as an order parameter in the present case.\\
Furthermore, the definition of $\vec \Sigma$ involves only tetrons of one single tetrahedron, whereas the particle physics SSB consists in the isospin alignment of two {\it neighboring} tetrahedrons. 
%Twhich only indirectly influences the SSB. In contrast, the SSB is described by the alignment of two neighboring tetrahedrons over physical space in a relativistically covariant way. This means, the order parameter should be built from $\psi$ and $\bar\psi$.\\
%Furthermore, one has $\langle \vec \Sigma \rangle=0$ for the ground state fig. 1, so it is no candidate for the order parameter anyhow. 
%fuer vecpi verschwindet der vev, in einer bestimmten Eichung entsprechend Parallelitaet von psibar und psi Tetraeder. Genau in dem Moment ensteht psibar*psi als vev, weil nur wenn vecQ und vecQbar parallel sind ist es Isospin singlet UbarU+DbarD ... genauer: Urbar*Ur\\
%wenn psidagger*tau*psi parallel, so auch psibar*tau*psi, also sumvecpi=0 gilt\\
%First of all $\bar\psi$ is needed instead of $\psi^\dagger$, because it is relativistic theory. However, this is not really a good argument because one might use $\vec Q$ to include antiparticles (according to eq ). 
\subsubsection{Is the vev $\langle\bar U U +\bar D D\rangle$ or $\langle\bar U U\rangle$ or what?}\label{appii-vev}
The vev is given by $\langle\bar U_\star U_\star\rangle$ in accordance with (\ref{pho77}) where $U_\star$ is the 'radial' iso-spinor introduced in (\ref{radsp1}) corresponding to an isospin vector pointing outward as in fig. 1. Such a vev is precisely what is needed to stabilize the alignment of isospins in fig. 2.
\subsubsection{Why not use some other order parameter which is closer related to the isomagnetic alignment than the Higgs vev?}\label{appii528abc}
First of all it must be noted that according to (\ref{appii-vev}) the Higgs vev in the tetron model has a lot to do with the isomagnetic alignment.\\
Secondly, it is true that in general for a given phase transition different order parameters are possible. In the present case one may for example consider 2 neighboring tetrahedrons A and B with tetrons $\psi_{Ai}$ and $\bar\psi_{Bi}$, and $i=1,2,3,4$ counting the tetrahedral sites. Unfortunately, the aligned tetrahedral 'star' configuration of these 2 tetrahedrons not only implies $\langle \vec \Sigma_{A,B} \rangle=0$ for each tetrahedron separately, but also $\langle \vec \Xi \rangle=0$, where $\vec\Xi$ is defined as $\vec\Xi=\sum_i \bar\psi_{Bi}\vec\tau\psi_{Ai}$. This is because one can show that all vectors $\bar\psi_{Bi}\vec\tau\psi_{Ai}$ of 2 adjacent tetrahedrons are parallel, iff the corresponding isospin vectors $\vec Q_{Ai,Bi}$ are. Thus $\vec\Xi$ is not a useful order parameter either.
\subsubsection{So what is the microscopic interpretation of the Higgs particle?}\label{appii533}
%ist anders als das Photon. zunaechst psibar-psi aehnlich wie photon, aber waehrend des Phasenueberganges aendert es seine Natur.
As discussed in (\ref{appii204}) the Higgs is neither fundamental nor a bound state of mignons, but an excitation of tetron-antitetron pairs which are themselves bound within the hyper-crystal.
%and located in neighboring tetrahedrons
%zzzThe situation is similar to that of Cooper pairs in superconductors where the potential of the crystal supports the binding of two electrons. ABER Cooperpaare koennen sich frei bewegen
%(wenn also das Higgs kein Cooperpaar ist, was ist es dann? Aus den Paaren an einem Tetraederpunkt gebildete Schwingung? Es muss irgendwie die SSB beschreiben, dh Higgs-vev entspricht den aligned spin vectors. Fuer ein Paar auf einem Tetraederpunkt ist der vev das vecS*vecT mit S=(0,0,1)und T=(0,0,+1) ist psiTbar*psiS=Ubar*U+Dbar*D, weil (0,0,1) entspricht (U,D).)
%(wo habe ich 3+1 diskutiert? bei Frage: are the electroweak bosons...composite? dort habe ich geschrieben: it is assumed that when the gauge bosons are formed, mignon vibrations essentially disappear and are replaced by tight binding of $\bar\psi$-$\psi$ pairs of 2 neighboring tetrahedrons. Shubnikov ordering does not longer apply. Since $\psi$ is an isospin doublet, according to $2\times 2=3+1$ one is lead to a triplet (the weak bosons) and a singlet (the B-L photon).)
\subsubsection{Should one consider separate tetrahedrons for anti-tetrons, with isospin vectors pointing inward?}\label{appii528ta}
This question may seem justified, because anti-fermions usually react to magnetic forces with an opposite sign. However, using isospin vectors (\ref{eq89p}) one is treating the problem in a covariant way. As can be seen in (\ref{m1u122}), the isospin vectors contain particle as well as antiparticle contributions, and the antiparticle contributions have a negative sign. More details are given in (\ref{appii4}), where a $\langle \vec Q_{R}\rangle$ pointing inwards will be defined in terms of a charge conjugate tetron field. Sections (\ref{appiiee5}) and (\ref{appii304}) may also be consulted in connection with this question. 
%$\langle \vec Q_{Li}\rangle$ or $\langle \vec Q_{Ri}\rangle$ pointing outward therefore means that their antiparticle components are pointing inward. Das ist aber nicht alles: es ist bei antiteilchen (-Dstern,Ustern) zu nehmen, und das entspricht ja auch einem neg Vorzeichen bei den IsoSpinvektoren. - je nachdem ob man charge conj mit innerer Zeitumkehr zusammen definiert. Laut der pdf Datei tut man das. heben sich dann die beiden Vorzeichen weg)
\subsubsection{Is there a difference between the SM SSB and a ferromagnet, apart from the fact that the SM SSB takes place in internal space? Is the symmetry breaking in the tetron model really spontaneous?}\label{appii529}
Both cases (ferromagnet and tetron structure) are similar in that at high energies / temperatures the directions of (iso)spins are oriented randomly with an associated SU(2) Heisenberg symmetry, and this defines the symmetric state.\\
In an uni-axial ferromagnet an accidental magnetization axis usually appears spontaneously, based on a thermodynamic potential
\begin{eqnarray}  
V_{FM}(\vec M)=-a\vec M^2 +b\vec M^4 
\label{pofh}
\end{eqnarray}
where $\vec M$ is the total magnetization and the minimum of the potential is at $\langle \vec M^2 \rangle=a/2b$.\\
In the case at hand the crystallization process at scale $\Lambda_r$ is accompanied by a coordinate alignment of all tetrahedrons, i.e. there is a spontaneous selection of one global internal coordinate system for all tetrahedrons. This coordinate alignment, however,  happens prior to the alignment of isospins and has not much to do with it.\\
When the temperature decreases towards $\Lambda_F$, the anti-ferromagnetic tetrahedral 'star' configurations fig. 1 appear where the isospin vectors within one tetrahedron avoid each other as far as possible.
%zzz(haben diese alle Shubnikov symmetrie? wenn man die Vektoren im Ursprung starten laesst, dann sind sie radial - 4 Vektoren die einen Tetraeder bilden. Aber wenn man sie auf den sites des Koordinatententetraeders starten laesst, sind sie nicht radial, also kein Shubnikov. Irgendwie scheinen die Koordinatensites nicht wirklich relevant zu sein. Vielleicht weil die Koordinatentetraeder sehr klein Planckscale sind. Aber bei den shubnikov trafos transformiert man eigentlich auch die Punkte.) 
Note there is an infinite SU(2) symmetric set of such 'star' configurations just as in a ferromagnet there is an infinite set of possible magnetized states corresponding to all possible magnetization directions in $R^3$. The difference as compared to a ferromagnet is that not only the stars over one tetrahedron have to be included but also those over all the other tetrahedrons over Minkowski space, with their independent SU(2) degeneracies, and this makes the problem a local gauge symmetric one.\\
%zzz(aber ist dies nicht alles SO3 statt SU2? Nein da es letzlich auf Spinoren beruht)
%zzz(hier ist aber ausserdem noch im Nachbartetraeder noch mal ein infinite set.) 
The SSB consists in the simultaneous selection of one among all the possible star configurations over all Minkowski base points -- namely the one with $\langle \Phi \rangle \sim (0,1)$. According to (\ref{pho55}) this corresponds to a vev for the $U_\star$ isospinor component, i.e. the one with an isospin vector pointing outwards in the radial direction.
%and the corresponding isospin vector in the neighboring 'anti'-tetrahedron pointing in the same direction.
The choice of $(0,1)$ - and of U - is notational convention and, in the framework of the gauge theory, corresponds to choosing a certain gauge (the so called unitary gauge). There is again a similarity to the situation in a ferromagnet where some axis is selected by the spontaneous magnetization, and the coordinate system is then 'gauged' in such a way that this axis is called the z-axis $\sim (0,0,1)$.\\
One may ask what role the coordinate alignment of tetrahedrons at the crystallization point $\Lambda_r \sim\Lambda_P$ plays in this game, because it seems plausible that the state, where the tetrahedrons of coordinate and isospin both point in the same radial directions, is energetically preferred. (This geometry is in fact depicted in figures 1 and 2, and the conditions, under which this happens, will be called scenario Q.)\\
A similar situation is sometimes encountered in ordinary uni-axial ferromagnets in cases when the coordinate backbone of the crystal prefers one specific magnetization direction, so that the ferromagnetic phase transition is not really spontaneous. This effect can be modeled by adding a tiny explicit symmetry breaking contribution to the potential (\ref{pofh}) by hand. At high temperatures due to thermal fluctuations this structural / coordinate effect is not important. But it becomes relevant near the Curie temperature where it fixes the magnetization direction.\\
%zzz(aber wenn es in der Teilchenphysik solche Terme gaebe, muessten sie im SM Higgspot auftreten) daher habe ich den Fall hier wegdiskutiert
In the present case, however, this possibility needs no consideration. The reason is that $\Lambda_P$ is so large as compared to $\Lambda_F$, that the granular internal coordinate structure is not noticed by the isospin vectors (nor by any human experiment). From the perspective of the isospin vectors it looks as if they are sitting on an internal coordinate structure which is rotationally invariant. They only feel the anti-ferromagnetic aversion towards their 3 fellows within one tetrahedron.\\
%and the ferromagnetic affinity towards their partners in the neighboring tetrahedrons. 
As a consequence of these considerations all 'star' configurations are energetically equivalent, and the symmetry breaking is spontaneous.\\
One can even go as far to say that there could be no coordinate alignment among the tetrahedrons at all. This would be in accord with the idea that the inter-tetrahedral coordinate (=gravitational) interactions are elastic and therefore can give rise to any relative coordinate orientation between neighboring tetrahedrons. The alignment of isospins could live with this option, because the only thing which matters for the SM SSB is that the 4 isospin vectors point in radial direction and build up the isomagnetically aligned multi-tetrahedral configuration, irrespective of what the coordinates of the tetrons are.\\
Further it may be noted that the question whether the SB is really spontaneous, is much easier to answer in what was called scenario C in section 1. In that case the electroweak phase transition III consists in a \textit{simultaneous} alignment of coordinate and isospin vectors as shown in fig. 2. In other words the tetrahedral star configuration consists in a coordinate and an isospin star where the coordinate and the associated isospin vector always point into the same radial directions. The transition to the ordered state is then necessarily spontaneous because all rotated (coordinate + isospins) tetrahedral star configurations are energetically equivalent.
\subsubsection{Is the electroweak phase transition first or second order?}\label{appii-ewpt}
Lattice calculations in the SM with one Higgs doublet give no definite answer to this question. The transition seems to be second order for $m_H \lesssim 120$GeV, while for $m_H \gtrsim 130$GeV one obtains a first order transition\cite{latticesm}. In the intermediate region it may be a cross-over\cite{crossover}. By contrast, in 2HDM models the situation is clearer, because the electroweak phase transition turns out to be first order\cite{lattice2hdm} and terms of order $\sim\Phi^3 T$ arise in the temperature dependent Higgs potential.\\
Since the 2HDM model (\ref{appii-qqq5}) arises as the low energy approximation of the tetron model, one may be content with this result, in particular because a first order transition is preferable for phenomenological reasons, cf. (\ref{appii535}) and \cite{crossover}. However, it should also be possible to directly determine the nature of the phase transition in the tetron model without recurring to an effective theory. To achieve this aim, a calculation in the framework specified in (\ref{appii529}) and by figs. 1 and 2 should be performed. If one looks at fig. 2, such an isomagnetic alignment is normally expected to be of second order. However, first order magnetic transitions are also known, in particular in connection with deformable structures\cite{beanvielzitiert}.
\subsubsection{What are the relevant scales in the model?}\label{appii530}
Naively, there are only 2 scales: the Fermi scale $\Lambda_F$ and the Planck scale $\Lambda_P$. The binding and crystallization process with coordinate alignment of tetrahedrons but erratic directions of isospin vectors corresponds to scales $\Lambda_r$ and $\Lambda_R$ both or order $\Lambda_P$, while $\Lambda_F$ is the scale where the isospin vectors align.\\
On a more sophisticated level some other scales might seem reasonable:\\
--the formation of tetrahedral 'molecules' from a tetron gas in 6 dimensions and of the hyper-crystal from the 'molecules' may happen at different scales $\Lambda_R$ and $\Lambda_r(t=0)$ where $t=0$ corresponds to big bang time when the hyper-crystal was formed. Furthermore, due to the elasticity of the DMESC, the crystal binding energy at that point was different than it is now: $\Lambda_r(0) \gg \Lambda_r(now)$. This corresponds to the fact that the average distance between neighboring tetrahedrons in the hyper-crystal has grown since the big bang: $\langle r(0)\rangle \ll \langle r(now)\rangle$ where according to (\ref{tmet6177}) we should identify $\langle r(now)\rangle$ with the present value of the Planck length. In other words one has $\Lambda_r(now)=\Lambda_P$ and thus a hierarchy of scales
\begin{eqnarray}  
\Lambda_R \geq \Lambda_r(0) \gg \Lambda_r(now)\approx \Lambda_P \gg \Lambda_F 
\label{posca}
\end{eqnarray}
--it is conceivable that the crystallization and the coordinate alignment of tetrahedrons do not happen at the same temperature, i.e. there is $\Lambda_P$ for the crystallization and another scale $\Lambda_A$ for the coordinate alignment fulfilling $\Lambda_P > \Lambda_A > \Lambda_F$. However, this would imply another phase transition in the early universe for which there is no indication. As explained in section 1 and (\ref{appii529}) it is best to assume that either $\Lambda_A \approx \Lambda_P$ or $\Lambda_A\approx \Lambda_F$ or that there is actually no coordinate alignment at all, only isospin alignment.\\
--there may be a separate scale $\Lambda_d$ for the pairing interaction of 2 tetrons on one tetrahedral site, as discussed in (\ref{appiiee5}), (\ref{appii4}) and (\ref{appii51}).
\subsubsection{What is the geometrical meaning of these scales?}\label{appii530uu}
According to fig. 2, r and R can be interpreted as lengths of certain tetron bonds within the discrete structure fig. 2. Namely, R is the fixed bond length of 2 tetrons within a tetrahedron, while r is the variable (elastic) bond length of 2 tetrahedrons in the hyper-crystal. By contrast, the Fermi scale measures the 'length' of isospin vectors of the ground state fig. 1.
\subsubsection{Why are the scales $\Lambda_R$ and $\Lambda_r$ so much larger than the quark and lepton masses? Why is gravity so weak as compared to the isomagnetic interactions? Why is the Planck scale so large in comparison to the Fermi scale?}\label{appii312}
These questions are variation of why gravity is so weak as compared to QED and the other particle physics interactions.\\
An important point to note is that the weakness of the gravitational forces and the strong coordinate forces among tetrons and tetrahedrons are related. The latter express themselves in the close packing and the associated extreme stiffness of the tetrahedrons in the hyper-crystal, which was deducted from the smallness of Newton's constant in the discussion after (\ref{tm33gg8}). In other words, it is the strong binding of tetrons together with the close-meshed packing of tetrahedrons which makes the hyper-crystal react to mignon excitations with only tiny distortions. In the Newtonian approximation, for example, writing $g_{00}$ in (\ref{newt588x}) as $g_{00}=-1-2\phi$, the gravitational potential is $\phi\approx 10^{-39}$ at the surface of a proton and not larger than $10^{-6}$ at the surface of the sun.\\
Not knowing the precise nature (\ref{appiift66}) of the fundamental interaction among tetrons, one can only say that gravity is some kind of remnant elastic interaction among the mignons, while the tetrons themselves are strongly bound to the hyper-crystal structure.\\
%According to this philosophy, the forces of gravity are so weak because elastic spacetime deformations {\it by mignons} do not cost the system too much energy.\\
On the other hand, the energies $\sim\Lambda_F$ involved in the isospin alignment are much smaller than the energies $\Lambda_r\approx \Lambda_P$ needed for the coordinate formation of the crystal. R and r are the length scales at which the iso-magnetic exchange integrals J have to be calculated, i.e. they are on the {\it abscissa} of the Bethe-Slater curve fig. 3, while the values of J are drawn on the {\it ordinate} of fig. 3 and always $\leq \Lambda_F$. The reason why one can have $J\ll \Lambda_P$ is explained in (\ref{appiie55}).
\subsubsection{Why are the exchange energies $J\sim O(GeV)$ in (\ref{mm3}) so much smaller than the tetron binding energies $\sim O(E_P)$?}\label{appiie55}
Exchange couplings/integrals within a tetrahedron generically are of the form
\begin{eqnarray} 
J=\int d^3 y_1 d^3 y_2 f_1(\vec y_1)f_2(\vec y_2)V(\vec y_1-\vec y_2) f_1(\vec y_2)f_2(\vec y_1)
\label{dd5581}
\end{eqnarray}
where $y_1$ and $y_2$ denote internal coordinates and $f_i$ the corresponding tetron wave functions. The value of J is the smaller, the smaller the overlap of the wave functions at different sites is. If the wave functions are very strongly concentrated, it is no problem to have $J\ll E_P$. Furthermore, as discussed in (\ref{appiift66}), there may be super-strong coordinate forces among tetrons which constitute $E_P$ but do not touch the isomagnetic interactions $\sim J$ among isospin vectors.
\subsubsection{Why is the inter-tetrahedral exchange coupling so much larger than the inner-tetrahedral one?}\label{appiimay}
In order to establish the SM SSB, the inter-tetrahedral exchange coupling $J_{inter}$ must be chosen to be as in (\ref{mm3dm333}) corresponding to a numerical value $J_{inter}\approx 7.8$ GeV. In order to understand why the inner-tetrahedral couplings (\ref{mm15ffff}) come out much smaller than this value, one has to note that $J_{inter}$ as defined in (\ref{mm3dm}) contains an implicit factor which counts the number of nearest neighbors of a given tetrahedron. Assuming that on the average within the elastic hyper-crystal there are always at least 6 neighboring tetrahedrons (and the next-to-nearest neighbors probably also count), this will naturally enlarge $J_{inter}$ by an order of magnitude. In other words, the inter-tetrahedral isomagnetic forces are not per se unnaturally large as compared to the 'inner' ones. Instead, originally, all interactions among 2 isospin vectors are the same generic order of magnitude of about 1 GeV.
\subsubsection{Do GUT theories play any role?}\label{appii531}
No. It is difficult to imagine why there should be other SSBs in the tetron model besides those described in section 1 and (\ref{appii206k})-(\ref{appii206m}). I see no reason for the proliferated Higgs sector characteristic for most GUT models, cf. (\ref{appii729}) and (\ref{appii729a}).
\subsubsection{Is there a unification of electroweak and strong couplings?}\label{appii532}
No. In the tetron model the QCD forces are on a less fundamental footing than the electroweak interactions.  For further details see (\ref{appii729}) and (\ref{appiiqcd})-(\ref{appii801}). The existence of a unification scale for electromagnetism and the weak interactions has been discussed in (\ref{appii206k}).
\subsubsection{What about domain walls?}\label{appii535}
Phase transitions in physics are usually associated with the formation of domains. However, domain structures have never been observed in cosmology.\\
In the tetron model, cosmological domains either appear as separate universes, which according to (\ref{appii816}) were created in $R^6$ via a condensation process similar to the one that has led to our own universe, or, if they are part of our own universe, they have long disappeared beyond our event horizon. To understand this in detail, one should first realize that one has to distinguish (i) domains arising at crystallization time (coordinate alignment of tetrahedrons) from (ii) those arising at the electroweak phase transition (tetrahedral alignment of isospins).\\
(i) In an ordinary crystal, one expects the appearance of domains with different values of the order parameter arising from concurrent nucleations of crystal germs in different points of space. 
%Note that one is talking here about the coordinate alignment of tetrahedrons in the process of crystal formation, not about the alignment of isospins.
In principle, this is also true for the $R^{6+1}$ space under consideration, and one would expect domains, where the ordered coordinate tetrahedrons are rotated by some angle as compared to those shown in fig. 2. 
Tetron model adapted cosmic inflation (\ref{appii702}) gives an argument, why the corresponding domain walls have moved so far away from us as not to be observable. Another, more absolute reason, why domains do not arise in the case at hand, has to do with the fact that the hyper-crystal grows into and occupies only a quasi 3-dimensional subspace of $R^{6}$. Therefore it intersects with other hyper-crystals from concurrent nucleation points, which grow into other 3-dimensional subspaces of $R^6$, in at most 1 point (because the intersection of 2 almost flat 3-dimensional submanifolds in $R^{6}$ in general is just 1 point). This means the result of the other nucleations will be different hyper-crystals, i.e. they correspond to different worlds whose intersection with our universe consists of at most one point. At this point there will be a defect within the hyper-crystal structure, cf. (\ref{appii816}).\\
Another possibility to avoid domain walls at the crystallization temperature is scenario C as described in section 1 where one can do without internal coordinate order.\\
%(nur wirklich wahrnehmbar wenn blackhole, sonst nur eine punktuelle Fehlstelle. warum sollte so eine fehlstelle ein black hole sein? transfer from one universe to another possible)
%There is a simpler (although less convincing) argument against observable domains namely that in the inflationary hyper-crystal formation process domains, if produced, have expanded together with the universe so that their walls are beyond our horizon. (argumentiert da nicht Kibble dagegen)\\
%zzz \\ 
(ii) In an ordinary ferromagnet, one expects the appearance of 'Weiss domains' with different ordering directions of spin vectors. In the case of isospin vectors such domains can in principle exist, too, and would differ by a global rotation of the isomagnetic tetrahedral 'star' configuration figs. 1 and 2. However, as discussed in (\ref{appii-ewpt}), the phase transition is first order, i.e. associated with a sudden release of latent heat, which blows up the micro-elastic continuum, i.e. triggers an inflationary process which in turn shifts domain walls outside the visible part of the cosmos.\\
Note that models with inflation near the electroweak scale have been discussed extensively in the literature\cite{ewinflat1,ewinflat2,ewinflat3,ewinflat4}.\\
Note added: in (\ref{appii529}) arguments have been given to show that the electroweak symmetry breaking (alignment of isospins) is truely spontaneous. It should be noted, however, that the non-existence of domain walls is much easier to understand in what was called scenario Q in (\ref{appii529}). The point is that in scenario Q the electroweak domains always coincide with the coordinate domains. There is then only one electroweak domain in the whole hyper-crystal because according to the discussion under item (i) different coordinate domains correspond to different universes (hyper-crystals).
%In other words, the universal coordinate alignment at big bang times would have pre-fixed the later alignment of isospins at the Fermi scale.
%(domain wall frage aendern und auch die Frage is the SB really spontaneous)
%für die Domain walls ist bessere Idee, dass sich zuerst die Koordinaten angeordnet haben gleich bei der Kristallbildung abgesehen von der Inflation, die sowieso die Domains sehr groß machen würde, wird dann jede andere Anordnung jeder andere Keim zu einem anderen weil nichtkollinearen Kristallteppich/Universum führen! Die dann zunächst erratischen Isospins müssen sich dann später nach diesen Koordinaten richten.
%Nachteil: wenn sich die Isospins dann nach den Koordinaten richten, ist die SSB nicht wirklich spontan. Aber ich halte das trotzdem für eine gute Erklärung!!!! Achtung: dies widerspricht der Frage in der Arbeit: is the SB really spontaneous, aber es klingt aber trotzdem vernünftig
%Laut Dove kann man auch Phasenübergang ohne SSB mit Landau beschreiben. Wobei hier trotz der Ausrichtung der Isospins an den Koordinaten ja eine Brechung der SU2 der erratischen Spins zur Tetraederkonfiguration vorliegt!!
\subsubsection{Why are quarks and leptons the same everywhere in the universe? Why does one electron look exactly like the other?}\label{appii6ey}
%yyy***(soll man es wirklich bringen? nicht in die journalversion)
A question which has bothered me already before I invented the tetron model. The answer: because the hyper-crystal is the same everywhere. It is built from tetrons in the same way and according to the same laws everywhere, and its mignon excitations are therefore the same everywhere in the universe.
\subsubsection{Are there excitations of the hyper-crystal besides the known quarks, leptons and scalar and vector bosons?}\label{appii536}
Yes. An incomplete listing includes:\\
--phinons. They are the analogs of phonons in a solid and have been described under more general circumstances in \cite{la4xz2}. In the case at hand there are 12 phinon states, that can be classified according to representations of the permutation group $S_4$. They travel as quasi-particles through the hyper-crystal in the same way as mignons do.
%zzz(als innere Dichteaenderungen sind es eigentlich sowas wie photon, Higgs und W??? aber da sind isospinoren involviert, bei phinons nicht, sondern Ortskoordinaten) 
%zzz(As internal density fluctuations they should be distinguished from gravitons which I consider as R31 'external' density. fluctuations.)
%zzz(aeussere Dichteaenderungen waeren wohl Gravitonen. Bei Gravitonen habe ich Zweifel, ob sie existieren, wegen der starken Plastikeigenschaft der Gravikraefte ... aber immerhin kommen sie aus den einsteingleichungen raus; auch die Tetronkoordinaten-WW im inneren koennten so sein. Dann gaebe es auch keine Phinonen.)
Phinon masses are expected to be much larger than mignon (quark/lepton) masses. While the mignon spectrum is lying at and below $\Lambda_F$, the phinon spectrum is concentrated towards the crystallization energy $\Lambda_P$. Note, this is not a very accurate characterization, in view of the fact that neutrino masses are so tiny with respect to the Fermi scale.  Note further, phinons are internal coordinate vibrations, and thus have to be distinguished from gravitational waves. An interesting question is whether mignon-phinon scattering is possible.\\
%zzz(welchen spin haben Phinonen? sind wohl auch Fermionen. aber schwingen da wirklich die Spinoren, oder nicht eher die inneren Ortskoordinaten? - unten sind es spinoren)\\
--isospin density waves: they are to be distinguished from phinons and from mignons. The vibrators in this case are similar to the isospin vectors (\ref{eq894}), however without the factor of $\vec \tau$, i.e. given by $\psi^\dagger (1\pm \gamma_5) \psi$. As far as I can judge these excitations correspond to a fourth family of fermions, i.e. a lepton-like and a quark-like isospin doublet, probably higher in mass, because they are not related to the other families by the $Z_3$ family quantum number inherent in the Shubnikov group $A_4+S(S_4-A_4)$. In particular, the fourth 'neutrino' is expected to be much heavier than the known neutrinos, because its mass is not suppressed by internal angular momentum conservation, cf. (\ref{appii605}).\\
%sondern durch Dichteerhaltung, aber die gilt wohl nicht
--one should mention scalar fields other than the standard Higgs boson. They are the components of the second Higgs doublet (\ref{phos1}), the lightest among them being the most promising candidate for dark matter, cf. (\ref{appii811}).
%In contrast to mignons, phinons and isospin density waves they are pairing involving 2 tetrahedrons (see the explanations in section 2.1). 
%-was sind die höchstenergetischen Elektronen/Protonen, die man je gemessen hat? nicht sehr hoch
\subsubsection{What about vibrations of $\bar \psi \gamma_\mu \vec\tau \psi$, $\bar \psi \sigma_{\mu\nu} \vec\tau \psi$ etc?}\label{appiiue}
These are other examples of higher mass excitations of the hyper-crystal.
%(Oder: letzlich nur die 4 dof eines psi=(U,D) schwingen koennen, entsprechend vecQ und der Dichte. Aber wie kann man dann mit 1+-gamma5 die zahl der dof verdoppeln? Außerdem warum sollte nicht jede Diracstrukur mit einer schwingung zu kombinieren sein und etwas neues ergeben? oder sind dies eben anregungen mit einem anderen externen spin quantnzahl?)
\subsubsection{Could quark and leptons be phinons?}\label{appii727}
or in other words: what is the advantage of using mignons with Shubnikov symmetry $A_4+S(S_4-A_4)$ over phinon excitations with symmetry group $A_4\times Z_2$ as advocated in \cite{la4xz2}?\\
The answer is that many of the attractive features of mignons are absent, like the explanation of the Higgs mechanism, of why $m_t \gg m_b$, of tiny neutrino masses etc.
%zzzauch Phinonen koennen unten Fermionen sein, da bei Eigenmoden wieder linkombs gebildet werden. Wirklich????
%aber SO3L statt SU2L, da keine Spinoren schwingen, sondernvektoren
%Probably not, because one would have an internal SO3L instead of SU2L, because no isospins are involved in phinon excitations.
\subsubsection{*Is it possible to understand the dynamics of the strong interactions from within the tetron approach?*}\label{appiiqcd}
I don't have a final answer to this question. From its very construction the tetron model is concerned mainly with the symmetries and interactions of electroweak physics. The colors of quarks arise merely as a byproduct, because they are interpreted as the 3 d.o.f. of a Shubnikov triplet representation. It is therefore clear that QCD with a color gauge group and SU(3) color triplets does not directly arise in the tetron model.\\
In section 1 it was argued that the phase transitions of a 6+1 dimensional spacetime filled with a condensing tetron gas supplies all relevant physics for the early universe and it may even account to understand the forces of gravity. As for the latter the suggestion is that it may arise from elastic forces between tetrahedrons which are the remnants of the fundamental 6-dimensional tetron-tetron coordinate interactions (\ref{appiift66}) and induce curvature and/or torsion effects on Minkowski space. One would like to interpret the strong interactions in a similar spirit, namely starting with the paradigm that there are no other interactions in the universe besides the ones among tetrons.\\
%On the other hand there is still a blind spot as concerns the understanding of the strong interactions within the tetron model. 
Gluons cannot be part of the isomagnetic geometry, because the SU(2)$\times$U(1) bundle connection allows only for the 4 electroweak gauge bosons. One may note, however, that a Shubnikov invariant mignon-mignon interaction is to be expected among the triplets, transforming as 
%(wg der Struktur des SU2xU1 Buendels kann es nur 4 Eichbosonen geben, dies spricht irgendwie gegen gluons)
\begin{eqnarray} 
T\times T = A+A'+A'' + 2T
\label{eqtkt}
\end{eqnarray}
As shown in \cite{lfound} this structure can be embedded into a SU$_c$(3) algebra where the color indices correspond to the 3 d.o.f. of the  triplet representations in (\ref{eq833hg}). %zzz(habe es nur fuer S4 gezeigt) 
%Unfotunately
Although as yet there is no proof that gluons and QCD gauge interactions can really arise from this line of reasoning, some hints will be given in (\ref{appii801}).
\subsubsection{Is color a part of isospin?}\label{appiiz00}
In the tetron model the color triplets of quarks transform according 3-dimensional representations of the Shubnikov group (\ref{eq8gs}), and this symmetry group is defined in terms of transformations within the same internal space as weak isospin. This could lead one to suspect that color in some sense is part of isospin. However, this is not the case. While the Shubnikov group is unbroken to the lowest energies, isospin symmetry corresponds to the free rotations of isovectors on each tetron site separately and is completely broken in the ordered state.
\subsubsection{Is there a connection between the QCD vacuum and the electroweak condensate?}\label{appii801}
Chiral symmetry breaking of the strong interactions is the appearance of a non-vanishing quark condensate 
\begin{eqnarray} 
\langle \bar u u \rangle  \approx \langle \bar d d \rangle  \approx \langle \bar s s \rangle  \approx -(0.25 GeV)^3
\label{qc66as}
\end{eqnarray}
which breaks $SU(2)_L\times SU(2)_R$ to the diagonal isospin $SU(2)_V$ group.\\
One difference as compared to the electroweak case is that these groups are global, not local symmetries.
%In the framework of QCD the values of the quark condensates roughly correspond to $\Lambda_{QCD}$. Furthermore, they are much smaller than that of the Higgs condensate. While the former extends over the strong interaction scale $\sim 1$ GeV, the latter is related to the Fermi scale.\\
Furthermore, according to (\ref{pho77}) the Higgs condensate is related to $U_\star$, i.e. the tetrahedral star configuration pointing outwards as in fig. 1. In contrast, the quark condensates are singlets w.r.t. the relevant quantum number (in this case color, while for the Higgs condensate it is isospin). In other words
\begin{eqnarray} 
\langle \bar q q\rangle=\langle \bar q_1 q_1 +\bar q_2 q_2 +\bar q_3 q_3 \rangle 
\label{qcdd}
\end{eqnarray}
does not correspond to a preferred direction or orientation in color space.\\
%, but has the property of an internal color density.\\
The quark condensates essentially are a measure of the nucleon masses, and in the QCD framework this role is taken over by the QCD Lambda parameter. Thus it appears that the nucleons get their mass not from the Higgs mechanism (mignon oscillations) but from the condensates (\ref{qc66as}). However, as discussed below, in the tetron model there may be a loose connection between the 2 mechanisms.\\
It is interesting to note that the strange condensate has about the same magnitude as up and down condensates, although its mignon mass (\ref{mm15ffg}) is much larger. This is an indication that chiral symmetry breaking is due to a {\it mignon triplet condensation} at 
%This is an indication that the strong interactions are due to an additional phase transition at 
temperatures of about 1 GeV.\\
%More precisely, one expects a {\it mignon triplet condensation} at that temperature, which leads to (\ref{qc66as}).\\
This phase transition is also responsible for the confinement of quarks\footnote{I have played with the idea that confinement arises from the stiffness of the DMESC but did not really find a good argument for that.}. The point is that each (Shubnikov) triplet state by definition defines a direction in internal space. This is nothing else than the oscillating isospin eigenvector which according to (\ref{mm3}) interacts via a Heisenberg Hamiltonian with other isospin vectors. Therefore, it is no accident that $\Lambda_{QCD}$ and the values of the inner-tetrahedral isospin exchange couplings (\ref{mm15f})-(\ref{mm15ffg}) [or the masses of the second family mignons] are of the same order of magnitude! Note that the corresponding attraction between 2 isotriplet eigenvectors should be equivalent to the effective description of the theory by low energy / lattice QCD.
%This point will become important for the following argument.
%In the tetron model the quark condensate is formed by triplets of mignons, i.e. of isospin excitations, which appear as triplets of the tetrahedral Shubnikov group. As for all tetrahedral groups, their triplet representations can be viewed to live in a space spanned by some (imaginary) tetrahedron.\footnote{This tetrahedron lives in color space and is not to be confused with the tetradedron in fig. 1. Furthermore, it only has a coordinate, not a 'magnetic' structure like the one in fig. 1.} The density (\ref{qcdd}) describes coordinate shifts in color space corresponding to the breaking of the color-internal translational symmetry. The Goldstone bosons corresponding to this SSB are the pions. Note that the breaking of the translational symmetry is induced by the discrete structure in color space due to the existence of the imaginary tetrahedron always accompanying Shubnikov triplet states.
%zzzThe mignon condensates correspond to an alignment of mignons q and qbar in their respective 3-dimensional representation spaces. This force is governed by the inner-tetrahedral exchange interactions with couplings O(1)GeV ... aber wieso ist es dann nicht lokal, wenn es inner ist?

%---------------------------------------------------------

\subsection{Questions about the Quark and Lepton Mass Spectrum and the CKM and PMNS mixing Matrices} 
Quark and lepton masses can be calculated as excitation frequencies of mignons via a straightforward procedure where the results are obtained quite naturally by considering isomagnetic interactions among the tetrons of one or two tetrahedrons only. Analytic expressions for masses in terms of internal Heisenberg and Dzyaloshinskii-Moriya exchange couplings have been derived in \cite{lmass} and reviewed in section 1.
\subsubsection{Using exchange couplings instead of Yukawas -- isn't one replacing one set of unknown parameters by another set and one effective theory (the Standard Model) by another one (the internal Heisenberg model)?}\label{appii601}
No, because the internal couplings in (\ref{mm3}) and (\ref{mm3dm}) can in principle be calculated from first principles as exchange integrals over internal space, just as in ordinary magnetism the exchange couplings of the Heisenberg model are in principle calculable from exchange integrals of electronic wave functions over physical space. What one needs to know is the underlying 6+1 dimensional dynamics of tetron interactions, cf. (\ref{appii5}), (\ref{appiift66}) and (\ref{appiiheuri}).
%\subsubsection{How can quarks and leptons be fermions if they excitation of 4 internal spin vectors?}
\subsubsection{Why are mignons spin-$\frac{1}{2}$ particles?}\label{appii602}
Since they are constructed from excitations of 'bosonic' isospin operators (\ref{eq89p}), one could be led to believe that they are bosons, just like magnons in ordinary ferromagnets are bosonic quasi-particles.\\
However, it is important to distinguish the behavior in internal space from that in Minkowski space. While mignons transform as Shubnikov singlets A and triplets T (i.e. {\it not} as projective representations) w.r.t. internal space, it is not hard to see that they are Dirac fermions w.r.t. Minkowski space.\\
%zzz(KEIN PROBLEM AUCH BEI NORMALEN BARYONEN baryon flavor SU3: 3x3x3=1+8+8+10. aber SU3 ist wie SU2???)
The point to note is that mignons are not bound states of tetrons but eigenmodes of their excitations. As such they are not tensor products but linear combinations of (fluctuations of) tetron fields $\psi_a^\alpha$ (where $a=1,2$ is the internal and $\alpha=0,1,2,3$ the Dirac index of the tetron).
%Antwort: kein psi1*psi2*psi3*psi4 bei Mignonbildung! sondern Eigenmoden sind immer Linkombs, auch bzgl der unteren Quantenzahlen, zb bei (2,2) von SU4=SU2xSU2 sind typische Eigenmoden psi(a,alpha)+-fi(a,alpha). 
%es geht um die einzelnen EigenModen, und da schwingt immer effektiv nur 1 Diracfermion!!!!
%If the mignons were bound states of 4 fermionic tetrons, then from taking tensor products they would show to be bosons.
Since each mignon is a vibration of {\it one} isospin eigenmode, one concludes that it must be a Dirac particle w.r.t. the Minkowski base space. Using the 'bosonic' isospin vectors $\vec Q=\psi^\dagger \vec \tau \psi$ is merely a tool to separate the isospin triplet vibrations from singlet density fluctuations of $\psi^\dagger \psi$, cf. (\ref{appii536}).
%For example, an isospin vector vibration 
%\begin{eqnarray} 
%\vec Q\sim (\cos (\omega t),\sin (\omega t),0)
%\label{ew97j}
%\end{eqnarray}
%originally corresponds to a vibration
%\begin{eqnarray} 
%\psi \sim (\cos (\omega t/2),i\sin (\omega t/2))
%\label{ew97k}
%\end{eqnarray}
%of the tetron $\psi$ field.\\
%Note that only the internal d.o.f. are given in (\ref{ew97k}). The Dirac spinor d.o.f. remain unchanged and make up for the fermion properties of the mignon.
Looked at from 'below', i.e. from the Minkowski base space, the tetron excitations are Dirac fermions. If such an excitation travels through the crystal as a quasi-particle, it can be either L or R, particle or anti-particle.
\subsubsection{Why are there exactly the quark and lepton states of the 3 generations?}\label{appii604}
%On each of the 4 sites of a tetrahedron there is a tetron-antitetron pair vibrating each with an independent isospin vector consisting of 3 vector components, so that there all in all 4x2x3=24 d.o.f., i.e. 24 eigenmodes to be expected.
The unbrocken (Shubnikov) symmetry group has only singlet and triplet representations, and the 8 independent isospin vectors on a tetrahedron lead exactly to the flavor spectrum of 24 quarks and leptons as given in (\ref{eq833hg}).
%Thus, in a wider sense the number of families corresponds to the number of dimensions of internal space, weak isospin arises from its covering group and the formation of tetron pairs the appearance of lepton singlets and quark triplets has to do with 
%isomagnetism in order to have a microscopic origin of isospin SU2 and the associated SSB.
\subsubsection{Are there other ground states than the tetrahedral one, which yield the appropriate quark and lepton spectrum (\ref{eq833hg})?}\label{appii524}
No. I scanned other geometries with 8 iso-magnetic vibrators and found that for most systems mignons appear in 2-dimensional representations\cite{shub,shub1}, not useful for the q/l spectrum of particle physics. This applies in particular to the configuration (m2) described in (\ref{appii701u}).  
%For example, an internal cube with 8 isospin vectors pointing outward leads to leptons doublets of leptons with identical masses. 8 ist sowieso nicht relevant
%(ALTE PHILO:So really needs pair of particle sitting on each tetrahedral site.)
\subsubsection{Should one really use covariant isospin vectors (\ref{m1u122}) containing both particle and anti-particle contributions as vibrators?}\label{appii618}
Yes, one should. It is important for the vanishing electric charge of the hyper-crystal, for the formation of gauge bosons out of tetron-antitetron pairs, and in general to maintain relativistic covariance as seen in (\ref{eq8rela}) and figs. 4 and 5. Technically it is important for the mass calculations corresponding to mignon mass terms $\langle 0| T(\bar q q)|0\rangle$.
\subsubsection{Why not use only $\vec Q=\vec Q_L+\vec Q_R$ as vibrators instead of $\vec Q_L$ and $\vec Q_R$ separately?}\label{appii619}
%Pluspunkt: Dann muesste doch Teilchen/Antiteilchen/Links/Rechts alles richtig rauskommen.
In order to obtain the 3 families of quarks and leptons one would have to consider systems with 8 instead of 4 tetrons. This possibility has been discussed in (\ref{appii51}). The main counterargument is that the $SU(2)_L$ vibrators $\vec Q_L$ are needed to account for the observed parity violation of the weak interaction.
\subsubsection{Is it okay to start the mass calculations using chiral $SU(2)_L\times SU(2)_R$ symmetry quantities $\vec S=\vec Q_L$ and $\vec T=\vec Q_R$?}\label{appii620}
Yes. This is just the way the SM works. Non-zero fermion masses are developed via SSB starting from a massless, i.e. $SU(2)_L\times SU(2)_R$ chirally symmetric theory.\\
%Aber ist SU2L nicht gebrochen, bevor die Massen auftreten? Nein, ohne Fermionmassen hat man chirale Symmetrie.
The role of custodial SU(2) in the tetron model is discussed in (\ref{appii-cu4}).
\subsubsection{Can the simple Heisenberg interaction (\ref{mm3}) really explain the full q/l mass spectrum with its extreme hierarchies?}\label{appii605xy}
No. As shown in \cite{lmass} the masses of some of the fermions get contributions from other physical sources, namely\\
--the top mass is dominated by a contribution of order $\Lambda_F$ which stems from the symmetry breaking {\it inter}-tetrahedral interactions (\ref{mm3dm}). Physically it arises because the top quark corresponds to the 3 eigenmodes which 'disturb' the global ground state in the strongest possible way. This disturbance is also responsible for the hierarchy observed in the CKM matrix elements.\\
--only strange-, charm- and muon-mass are dominated by anti-ferromagnetic exchange couplings within one tetrahedron, and thus can be obtained from the {\it inner}-tetrahedral exchange couplings (\ref{mm3}) alone.\\
--down-quark, up-quark and electron are left massless by the Heisenberg and DM interactions (\ref{mm3}) and (\ref{mm3dm}). They get their relatively small masses from energetically favored torsion contributions\cite{lmass}.\\
%, i.e. vibrations of the internal spin vectors of the generic form $d\vec Q /dt \sim \vec Q$.\\
--neutrino masses are protected by internal angular momentum conservation, i.e. by the internal rotational symmetry, cf. (\ref{appii605}). The way how they acquire their tiny mass values is described in \cite{lmass}.
\subsubsection{Is there a mismatch between the internal Heisenberg interaction (\ref{mm3}) for a single tetrahedron and the DM interaction (\ref{mm3dm}) involving neighboring tetrahedrons?}\label{appii-qq32}
This is a technical question concerning the calculation of quark and lepton masses presented in \cite{lmass}, and the answer is that technically one can treat the isospin vectors from the neighbors as if they were vectors of the original tetrahedron.\\
Denoting 2 neighboring tetrahedrons as primed and unprimed and starting with interactions of the form $\sim \vec S_i \vec S'_j$, one would expect the doubling of the number of eigenmodes. However, due to the symmetry between the 2 tetrahedons, i.e. between the $\vec S$ and $\vec S'$, the modes arrange in pairs of identical energy. In other words, the doubling of modes is a trivial one.\\
To understand this in more detail consider the first term (Heisenberg contribution) in the inter-tetrahedral interaction (\ref{mm3dm}) and assume that the inner-tetrahedral distances are much smaller than the inter-tetrahedral ones. This is in accord with arguments given in (\ref{appii530}), that after the long time of cosmic expansion one expects $r\gg R$ in fig. 2. This assumption implies that the couplings $J_{inter}$ of an $\vec S_i$ to all isospin vectors in the neighboring tetrahedron are identical (as anticipated in (\ref{mm3dm})) and one obtains
\begin{eqnarray}  
\frac{d\vec S_i}{dt}= J_{inter} \; [ \vec S_i^0 \times \sum_{j=1}^4 \vec S'_j ]
\label{eqq22ox}
\end{eqnarray}
where the superscript 0 denotes ground state values. A second set of equations is obtained for $d\vec S'_i/dt$ by exchanging the role of the 2 tetrahedrons, i.e. the primed and unprimed quantities. An obvious set of solutions to these equations fulfills $\vec S_i=\vec S'_i$, i.e. the vibrations are completely in step, and one obtains the trivial doubling of modes mentioned above.\\ 
As for the DM part (second term) in (\ref{mm3dm}) a similar argument can be given. Following the results in \cite{lmass} this leads to the conclusion that the top quark receives the overwhelming mass contribution from the inter-tetrahedral interaction (\ref{mm3dm}).\\
It is interesting to note that in a cosmological scenario where one would have $r\ll R$ the mignon mass spectrum would turn out quite different because in that case it is natural to assume that the coupling of $\vec S_i$ to $\vec S'_i$ is much larger than to the other $\vec S'_j$. %Therefore $d\vec S_i/dt\sim \vec S_i \vec S'_i$ corresponding to a much different mass spectrum of mignons.
\subsubsection{How are the CKM and PMNS matrix elements obtained in the tetron model?}\label{appii-ck4}
In the SM the possibility of inter-family mixing arises because interaction eigenstates are distinguished from mass eigenstates. In the tetron model there are the original modes which correspond to small vibrations of the isospin vectors, and on the other hand the eigenmodes of the vibrations whose eigenfrequencies correspond to the quark and lepton masses. The original modes $\delta\vec Q_{iL/R}$ are directly related to the SM interaction eigenstates because $\vec Q_L$ according to (\ref{eq894}) defines the isospin 'charge' of the left handed SU(2) current appearing in the Glashow theory.\\
%(warum dann nicht auch t und b mixing, weil ja auch t-b-mischterme im tetron modell hamiltonian vorhanden sind? weil wir ja im next subsection zeigen, dass die G4 quantenzahlen sehr restriktiv bei mischungen sind)\\
Therefore, in order to obtain the CKM and PMNS matrices in the tetron model, one can make use of the results of the diagonalization procedure from the $\delta\vec Q_{iL/R}$ to the mass/frequency eigenmodes. Since the mixing in the right-handed sector is not observable, only projections onto the $\delta\vec Q_{iL}$ need to be treated.
%Antwort: beide auf S denn technisch gehen wir bei dem 24x24 Problem so vor, dass wir 4 L und 4 R Schwinger weglassen, und S sind nun einmal die L-Schwinger, die allein das VCKM bestimmen.
%Erkenntnis: fuer Neutrinos braucht man T als Teilchenvektor-Teil von QR, denn man mischt ja S+T; fuer b-quarks braucht man T als Antiteilchenvektor-Teil von QR, denn sie sollen ja nicht durch QL Schwingungen große Masse kriegen.  
%Erkenntnis vecSi die WW-Eigenzustaende Warum sind die vecSi die WW-Eigenzustaende?
%WW-Eigenzustaende dadurch definiert, dass qabar*W*tau*gammue*(1-g5)*qb, wobei a und b der Generationenindex
%de facto nur relevant fuer taux und tauy also uabar...dbbar --> CKM=VLu*VLdkreuz
%nb: die Yuk-massenterme Y*qbar*phi*q haben kein tau, sind fuer dies Argument nur indirekt von Bedeutung
%dem entspricht erhaltene 'Ladung' unsere Si=psikreuz*tau*(1-gam5)*psi, wobei die qa ja die schwingenden psi sind, also in die Si direkt die Eigenzustaende entsprechen
%Wir leiten ja unser Model aus NJL ab, aus S*S=(psibar*tau*psi)*(psibar*tau*psi) wobei bei den erhaltenen Ladungen S ein psidagger statt psibar auftritt, und das gammue verschwindet auch beim Uebergang vom Strom zur Ladung S!!!! 
\subsubsection{Why are interfamily interactions suppressed in the tetron model?}\label{appii-sup4}
It is a prevalent problem in many composite models, to explain why - apart from CKM mixing effects - transitions between fermions of different families do not exist, e.g. why $\mu\rightarrow e\gamma$ is forbidden.\\
In the tetron model this fact can be understood from the symmetry of states. Since the gauge bosons W, Z and $\gamma$ (\ref{appii214n}) behave trivially under the Shubnikov group, the same must be true for the fermion-antifermion conglomerates, which get pair produced from them.\\
%\begin{eqnarray} 
%\delta \sum_i \delta \vec Q^i_{L,R} \equiv 0 
%\label{glgnur5}
%\end{eqnarray}
%on each site of the tetrahedrons separately? nein besser fuer die eigenmoden
%deltaQL=deltaQR gilt identisch für gamma,W,Z, da phasig deltaS und antiphasig deltaSc bzw deltaTc entgegengesetztes Vorzeichen haben. siehe Erkenntnisteil) 
%da deltaS und antiphasig deltaSc bzw deltaTc entgegengesetztes Vorzeichen haben dies setzt eigentlich voraus, dass im einen tetraeder aa schwingen und im anderen bb???? Nein es koennen auch schwingungen von aa ein kleines bb machen
%die mignon massen entstehen in einem tetraeder aus SS und TT WW.
%die mignon interactions hingegen erfolgen durch verbindung 2er tetraeder via den VB, und die VB sind tetronanregungen der form psibar*tau*(1-g5)*psi
Analyzing products of representations of (\ref{e33hgxx}), it turns out that all inter-family conglomerates behave non-trivially under Shubnikov transformations. For example, a combination of a $\mu^-$- and an $e^+$-mignon transforms as a non-trivial singlet and therefore cannot become a photon.\\
In order to actually carry out this analysis, the exposition in (\ref{appii610}) is useful. It may then be noted that the antiparticle of $A'_{1+is}$ transforms as $A''_{1-is}$.
%(AUCH FUER DIE 3DIM T-DARSTELLUNGEN IST DAS NICHT RICHTIG!!!!denn
%1   1+is   1-is
%MAL
%(1   1+is   1-is)stern
%gibt bei interfamily niemals ein Singlet\\
%von der s-struktur her passt dann nur 1. mit 1. familie, 2. mit 2. usw. denn wenn zb 1. mit 3. wuerde immer ein s-anteil bleiben, oder bei 2. mit 3. haette man (1+is)**2, waehrend bei 2. mit 2. (1+is)*(1-is)=A\\
%bei den 1dim Darstellungen passt es auch:\\
%A(1)+A'(1+is)+A''(1-is) mit cc : A(1)+A''(1-is)+A'(1+is)\\
%zb 2. mit 2. familie ist A'(1+is)x A''(1-is)=A,\\
%hingegen 2. mit 3. ist A'(1+is)xA'(1+is)=A''((1+is)**2)\\
%eine folgerung ist, dass W+ und W- nicht A' und A'' sein koennen. Das waere ohnedies unpassend, weil zb bei der 1. Familie AxA' zu nehmen waere, die familien waeren also durcheinander.)
\subsubsection{How can the smallness of neutrino masses be understood?}\label{appii605}
Neutrinos are interpreted as internal Goldstone particles of the breaking of the tetronic isospin by the formation of the discrete structure fig. 1. The associated conserved Noether charge is given by the total internal angular momentum $\vec \Sigma$ defined in (\ref{tm32}) which implies the existence of 3 zero-frequency modes. This is analogous to how magnons are interpreted as Goldstone modes in ordinary magnetism, except that here one is considering the physics of spin waves in the internal spaces.\\
While in ordinary ferromagnets after magnetization a U(1) symmetry about the z-axis survives, in the given frustrated configuration fig. 1 all three SO(3) generators give rise to Goldstone bosons, to be identified as the internal magnons corresponding to the 3 neutrino species.
\subsubsection{Neutrinos are fermions. How can they be Goldstone modes?}\label{appii606}
One has to distinguish the dynamics in internal from that in physical space. In physical space the neutrinos are fermions, but they are (Goldstone) bosons w.r.t. the dynamics in internal space, because in internal space they are described by (bosonic) excitations of the total internal angular momentum $\vec \Sigma$ defined in (\ref{tm32}) which is the conserved quantity associated with the internal rotational symmetry.\\
%(aber ich habe woanders argumentiert, dass in Wirklichkeit nicht vecS, sondern psi schwingt)
As discussed before, none of the representations in (\ref{eq833hg}) are projective representations of the Shubnikov group, so all quarks and leptons are 'bosonic' w.r.t. the internal dynamics, cf. (\ref{appii602}) and (\ref{appii610}).
%zzz(Fuer dies Argument mit interner Dynamik scheint man doch innere Zeit zu brauchen, also C statt Tinnen reicht evtl nicht aus) 
\subsubsection{How do neutrinos obtain their tiny non-zero masses?}\label{appii607}
As Goldstone modes neutrinos are strongly protected to getting masses. However, as proven in \cite{lmass} the observed non-zero neutrino masses can be generated on the phenomenological level by tiny torsional interactions which violate (\ref{tm31}). These can also be used to accommodate appropriate PMNS mixing values. Physically the existence of such interactions is a signal for the activity in isospin space of small anisotropic forces. This is discussed in more detail in (\ref{appii523}).
\subsubsection{Are neutrinos Dirac or Majorana particles?}\label{appii-dima}
They are Dirac fermions. Like all other quark and lepton flavors they inherit this property from the tetrons, cf. (\ref{appii602}).
\subsubsection{How is weak isospin realized on the mignon level?}\label{appii4}
The first fact to note is that the weak isospin of the SM cannot be directly identified with the tetronic isospin SU(2). As discussed below in connection with (\ref{eq150s}), the construction of mignon isospin relies on exchanging the roles of $\vec S$ and $\vec T$, i.e. of the $SU(2)_L$ and the $SU(2)_R$ sector.\\
%This can be made even more explicit by calculating the masses of quarks and leptons from vibrations $\delta \vec S=\delta \vec Q_L$ and $\delta \vec T=\delta \vec Q_R$ of the ground state fig. 1 
The most appropriate way to construct quark and lepton isospin is to assume the existence of a tetron $\psi_L$ and an antitetron $(\chi_R)^c=(\chi^c)_L$ on each tetrahedral site i=1,2,3,4, as depicted in figs. 4 and 5. Here the charge conjugation operator is defined as usual for a Dirac spinor F, i.e. $F^c=C\gamma_0F^*$.\\
%Note since there are isospin degrees of freedom, charge conjugation involves an additional internal time inversion (\ref{eq8it}).\\
One can either interpret fig. 4 in such a way that there are 2 particles (a tetron and an antitetron) on each site, or by saying that one is considering the tetron and the antitetron contribution of a single field to the isospin vectors $\vec Q_L$ and $\vec Q_R$, cf. the discussion in (\ref{appii51}).\\
%In the ground state they are denoted by $\langle U_{\star L}\rangle$ and $\langle D^c_{\star R}\rangle$ 
%Note that although $D^c_R$ is left-handed, it transforms under SU(2)$_R$ and thus can be used to build a $\vec Q_R$ vibrator.\\
Although $(\chi_R)^c$ is left-handed, it transforms under SU(2)$_R$ and thus can be used to build a $\vec Q_R$ vibrator. For the question why tetrons and antitetrons do not annihilate inside the hyper-crystal see (\ref{appii304}).\\
I also tried an approach without antiparticles, i.e. with $\chi_R$ instead of $(\chi_R)^c$, but have abandoned this for the following reasons:\\
--since tetrons are fermions, the wave function for each pair would have to respect the Pauli principle. While the spatial part must be symmetric (the argument is similar as for in the helium atom), the spin and the isospin part are both expected to be anti-symmetric singlets, and this would violate the Pauli principle. If one of the tetrons is an antiparticle, one does not run into this problem of two identical fermions.\\
--$A(\nu_e)$ and $A(e)$, $T(u)$ and $T(d)$ etc would not be true isospin partners but components of Kramers doublets. The exchange $\vec Q_L \leftrightarrow \vec Q_R$ induces an isospin exchange $U \leftrightarrow D$ only if one antiparticle is involved. The reason is that charge conjugation compensates for the internal time reversal active in a Kramers doublet. This will be explained in more detail in (\ref{appii-ssu8}).\\
%The appearance of a D-isospinor instead of U is standard when considering antiparticles. It means that for the antitetron the tetrahedral star points inward instead of outward. The geometrical situation on site i is depicted in fig. 4.
--finally, since the Higgs field and the gauge bosons have been interpreted as tetron-antitetron excitations, it seems a good idea to have antitetrons already appearing in the ground state.\\
The isospin vectors, whose vevs are depicted in fig. 4, are given by 
%This is because a mass term is of the form $\bar q_R q_L+c.c.$, and a charge conjugation transition (instead of parity) may be used to accomplish this exchange. 
\begin{eqnarray}
\vec Q_L=\frac{1}{2}\, \psi_L^\dagger \vec \tau \psi_L 
\qquad \qquad
\vec Q_R=\frac{1}{2}\, (\chi_R)^{c\dagger}\vec \tau (\chi_R)^c 
%\langle \vec Q_L\rangle=\frac{1}{2}\, U_{\star L}^\dagger \vec \tau U_{\star L} 
%\qquad \qquad
%\langle \vec Q_R\rangle=\frac{1}{2}\, D_{\star R}^{c\dagger}\vec \tau D^c_{\star R} 
\label{eq894gg}
\end{eqnarray}
A transition between them is then necessarily accompanied by an exchange of isospins U and D via\cite{goran}
\begin{eqnarray} 
(U_L,D_L)\rightarrow(-D_R^c,U_R^c)
\label{eq150s}
\end{eqnarray}
%where the radial stars have been left out for convenience.\\
If one strips off the Dirac structure, (\ref{eq150s}) is identical to an internal time reversal (\ref{eq8it}), which was considered in connection with the Shubnikov transformations in section 1. One concludes that the Shubnikov symmetry (\ref{eq8gs}) can be defined by using charge conjugation and without introducing the concept of an internal time, as anticipated in the discussion after (\ref{eq8rela}). For more details on the action of discrete symmetries like C, P and T on the tetron ground state see (\ref{appiicpt}). It is interesting to note that the internal reflection operators which exchange the elements of $A_4$ and $S_4-A_4$ comprise isospin transformations, charge conjugation as well as transitions between left and right.\\
%(As argued above, the full charge conjugation operator C acts in isospin space in the same way as internal time reversal. Furthermore, C has the interesting property that it exchanges $SU(2)_L$ and $SU(2)_R$\cite{goran}. In the present case this amounts to saying it interchanges the roles of $\vec Q_L$ and $\vec Q_R$.)
%These considerations are in agreement with how isospin partners like $A_1(e)$ and $A_1(\nu)$ in (\ref{eq833hg}) arise from each other. Furthermore, they imply that $\vec T$ in (\ref{eq894}) should better be defined with a minus sign, i.e. $\vec T=-\vec Q_R$. This can best be seen by remembering that charge conjugation exchanges the role of tetron and antitetron operators and thus reverses the sign of isospin vectors according to (\ref{m1u122}).\\
%It can also be obtained by explicitly applying charge conjugation according to on spinor level
%$\psi_L \rightarrow \psi_R^c\equiv C\gamma_0 \psi_R$
%and taking into account that 
%$\vec Q_R=\psi_R\vec\tau \psi_R=-\psi_R^c\vec\tau \psi_R^c$.
To summarize, it has been shown that the transition between weak isospin partners like $A_1(e)$ and $A_1(\nu_e)$ can be obtained by exchanging $\vec Q_{Li}$ and $\vec Q_{Ri}$ on the tetrahedral sites i. In the actual calculation of mignon eigenstates, it turns out that the top-quark is predominantly an $\vec S=\vec Q_L$ excitation while the b-quark is $\vec T=\vec Q_R$. On the other hand the neutrinos are given by vibrations of the conserved quantity $\sum_i (\vec S_i+\vec T_i)$, while charged leptons are approximately excitations of the $\sum_i (\vec S_i-\vec T_i)$ combination.\\
The connection between the attributes 'left-handed' and 'up-type' (and similarly 'right-handed' and 'down-type') is of fundamental importance in the tetron model. It relies on the chiral property of the internal ground state fig. 5 and the octonion induced form of the tetron interaction (\ref{ewrr82}) and (\ref{appii5}). Furthermore, it is at the heart of the tetron model explanation of weak parity violation (\ref{appii5}) and of the large value of the top quark mass from (\ref{mm3dm}) as compared to the other q/l masses, cf. (\ref{appii614}).
\begin{figure}
\begin{center}
\epsfig{file=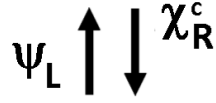,height=1.5cm}
\bigskip
\caption{Ground state configuration of the tetron-antitetron pair on a single tetrahedral site. The vertical axis corresponds to the radial direction in the sense of (\ref{radsp1}), i.e. isospin pointing  outwards=up means U, inwards=down means D. Drawn are the ground state values of isospin vectors $\vec Q_L$ and $\vec Q_R$ which in (\ref{eq894gg}) are defined in terms of $\psi_L$ and $(\chi_R)^c$, respectively. Considering the tetrahedron as a whole, the configuration is shown in fig. 5 and is chiral, both in internal and physical space. The internal chirality is flipped by $U\leftrightarrow D$, the external by $L\leftrightarrow R$. Note that $(\chi_R)^c$ is left-handed, but transforms under SU(2)$_R$. Therefore and since the configuration shown is preferred over that of opposite chiralities through the sign of the interaction (\ref{ewip81}), $(\chi_R)^c$ does not take part in the symmetry-breaking {\it inter}-tetrahedral interactions (\ref{mm3dm}).}
%(Ja!, da DRc ein linksh su2R feld ist) (Was hier abgebildet ist, koennte einfach Wplus sein. Wenn man L und R UND D und U vertauscht, kriegt man ein kongomerat was gerade das antiteilchen ist, also Wminus. wo ist das dann im kristall?????? Wenn man nur U und D ODER L und R vertauscht, kriegt man entgegengesetze innenchiralitaet?) (ausserdem hat D elektr ladung! Dann ist der ganze tetraeder geladen!!!!! DRc muss eher was up-artiges sein)
\nonumber
\end{center}
\end{figure}
\begin{figure}
\begin{center}
\epsfig{file=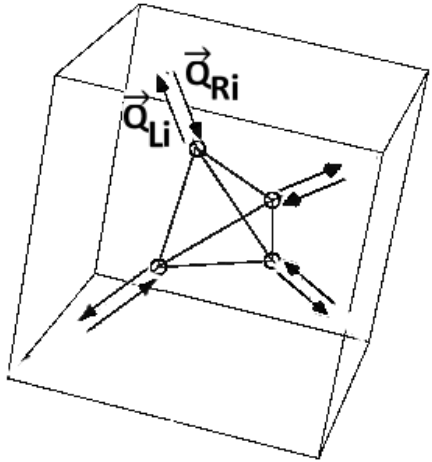,height=6.4cm}
%\bigskip
\caption{The local ground state of the tetron model with 8 internal spin vectors $\vec Q_{Li}$ (pointing outwards) and $\vec Q_{Ri}$ (pointing inwards), i=1,2,3,4, accounts for 3$\times$8 d.o.f. corresponding to 24 quarks and leptons according to (\ref{eq833hg}). Due to the anti-particle nature of the isospin vectors pointing inwards, the depicted configuration is chiral and internal parity maximally violated.}
%For details on the action of discrete symmetries C, P and T on this state see (\ref{appiicpt}).}
\nonumber
%\label{figac1} 
\end{center}
\end{figure}
\subsubsection{The difference between tetron isospin and 'custodial' SU(2)}\label{appii-cu4}
In the bosonic sector of the SM (i.e. without quarks and leptons) there is a global vectorlike SU(2) symmetry which remains intact after the SSB. This is usually called 'custodial SU(2)', and it should not be mixed up with the isospin SU(2) of the 3-dimensional internal space considered in the tetron model. While custodial SU(2) arises on the level of mignons, the tetron model SU(2) exists on the level of tetrons and comprises not only electroweak but also color and family symmetries. While the dominant corrections from custodial SU(2) breaking are $\sim m_t-m_b$, tetron SU(2) breaking terms generate the neutrino masses, cf. (\ref{appii605})ff. More details on the relation between tetron and electroweak(=mignon) SU(2) can be found in (\ref{appiiz00}) and (\ref{appii4}).
\subsubsection{Do pairs of ordinary and true Shubnikov representations form isospin doublets?}\label{appii610}
No. To explain this point, I start with the remark, that 'true' Shubnikov representations have nothing to do with 'projective' representations mentioned above. While the latter are representations of the covering group, a true Shubnikov representation is a representation of $G_4=A_4+S(S_4-A_4)$ which is not a representation of $A_4$. True Shubnikov representations are labeled by an index s in the following. The Shubnikov group $G_4$ under consideration has the property that for each ordinary representation $D$ of $G_4$ there is exactly one true Shubnikov representation $D_s$. If one puts an isospin along the z-direction, true Shubnikov representations are related to the excitations of the y-components of the isospin vectors, while $Q_x$ and $Q_z$ correspond to the ordinary representations.\\
Analyzing the representations appearing in (\ref{eq833hg}) one finds that $A$ and $A_s$, $T$ and $T_s$ arise in the combinations
\begin{eqnarray} 
A(\nu_{e})+A'_{1+is}(\nu_{\mu}) +A''_{1-is}(\nu_{\tau}) &+& 
T(u)+T_{1+is}(c)+T_{1-is}(t)+ \nonumber \\
A(e)+A'_{1+is}(\mu)+A''_{1-is}(\tau) &+& 
T(d)+T_{1+is}(s)+T_{1-is}(b) 
\label{e33hgxx}
\end{eqnarray}
%(hier fehlt die genaue herleitung dass A' A'' usw auftreten)\\
where the notation $1\pm is$ means that the second and third family are generated by excitations of $Q_x \pm iQ_y$, while the first family corresponds to excitations of $Q_z$. According to this result an isospin doublet is not given by a pair $(A,A_s)$ or $(T,T_s)$ of an ordinary representation and a Shubnikov representation. Instead, the rows $A,A_{1+is},A_{1-is}$ and $T,T_{1+is},T_{1-is}$ correspond to particles of the 1., 2. and 3. family.\\
Finally nota bene that the classification (\ref{e33hgxx}) can be used to show that interfamily transitions are suppressed in the tetron model, as explained in (\ref{appii-sup4}).
%How this comes about, is made explicit in (\ref{appii-ssu8}). 
\subsubsection{A simple mathematical understanding of all the 24 Shubnikov states in (\ref{e33hgxx})}\label{appii-ssu8}
%in other words: is there a simple mathematical description of mignons?\\
First of all note that, although mignons are actually precessions of internal angular momentum vectors, I sometimes call them 'vibrations' or simply 'excitations' in this paper.\\
In this section the 24 mignon precessions are derived successively starting with the 3 mignons arising from a single tetron, then going over to 2 tetrons and ending with the 8 tetron configuration fig. 5. The results of the discussion will amount to the following statements:\\
--the existence of 3 families reflects the 3 dimensions of the internal space, as already claimed in the first paper on the subject\cite{lfirst}.\\
--weak isospin of quarks and leptons corresponds to isospin transformations of tetron and anti-tetron within one tetrahedral site i=1,2,3 or 4.\\
--the color d.o.f. arise from the 4-fold structure of the tetrahedrons.\\  
By slightly abusing notation, mignon excitations of a single tetron can be described as isospinors of the form
\begin{eqnarray} 
\psi=(\langle U\rangle+\delta U ,\langle D\rangle+\delta D)=(1 +\delta U ,\delta D)
\label{ewan2}
\end{eqnarray}
$\langle U\rangle$ and $\langle D\rangle$ are the isospinor values in the ground state. For the convenience of the following discussion, $\langle U\rangle$ can be put to 1 and $\langle D\rangle$ assumed to vanish.
%, and $\delta U$ and $\delta D$ should be interpreted as radial quantities in the sense of (\ref{radsp1}). $\langle U\rangle$ 
$\delta U$ and $\delta D$ contain a time dependence $\sim \exp (i\omega t)$ where $\omega$ is the mass/energy/frequency of the precession. The corresponding fluctuation of the isospin vector $\vec Q=\psi^\dagger\vec\tau\psi$ is given by
\begin{eqnarray} 
\delta \vec Q=\vec Q -\langle \vec Q \rangle 
%=(\Re \delta D + \delta D^\dagger ,i(\delta D-\delta D^\dagger),\delta U + \delta U^\dagger) +O(\delta^2)
=\begin{pmatrix}
\operatorname{Re} \delta D  \\
\operatorname{Im} \delta D \\
\operatorname{Re} \delta U
\end{pmatrix}
+O(\delta^2)
\label{ewan55}
\end{eqnarray}
where $\langle \vec Q \rangle=(0,0,1)$.\\
It may be noted, that on the tetrahedron one has non-vanishing vevs $\langle \vec Q \rangle \neq 0$ only for the single isospin vectors, whereas for the tetrahedron as a whole the total internal angular momentum vanishes in the ground state, i.e.
\begin{eqnarray} 
\sum_{i=1}^4 \langle \vec Q_i \rangle =0
\label{ea6n2}
\end{eqnarray}
and the only relevant vev becomes the Higgs vev as described in (\ref{appii528})-(\ref{appii-vev}).\\
%The Heisenberg isomagnetic energy of this state is
%\begin{eqnarray} 
%H_H=\vec Q \vec Q = \frac{1}{2}\psi^\dagger\psi = 1+2\operatorname{Re} \delta U
%+|\delta U|^2 +|\delta D|^2 
%\label{ewa73}
%\end{eqnarray}
%NUR RELEVANT WENN DIE BEIDEN Q VERSCHIEDEN WAEREN
Coming back to the single tetron states (\ref{ewan55}), the 3 vibrations $\operatorname{Re} \delta D$, $\operatorname{Im} \delta D$ and $\operatorname{Re} \delta U$ in the x, y and z-direction essentially correspond to the third, second and first family of quarks and leptons. The y-vibration is a true Shubnikov excitation in the sense of (\ref{appii610}).\\
%(wo ist Im delta U? Ausserdem ist normalerweis x+iy also deltaD das magnon, also wie soll y dann die Shubnikovdarstellung sein?; sind die linearen term relevant, nicht besser die quadartischen? WEITER UNTEN IN DIESME KAPITEL)\\
The color quantum number arises as soon as all four tetrahedral sites are taken into account. In place of one vibrator (\ref{ewan2}) one then has four, and arrives at one singlet A (the lepton) and one triplet T (the quark) for each family and for each isospin value $\pm$1/2. More details about the origin of color in the tetron model can be found in (\ref{appiiqcd}) and (\ref{appiiz00}).\\
%This means for example that it points into the 'z'-direction (=the radial direction) for a pure $\delta U$ excitation and gives rise to the first family of quarks and leptons. On the other hand, $\delta D$ excitations lead to isospinors which vibrate in x- and y-directions and give rise to the second and third family in (\ref{e33hgxx}). Note, the y-vibration is the true Shubnikov excitation in.\\
So far, we have ignored the fact that there are 2 vectors $\vec Q_L$ and $\vec Q_R$ that vibrate on each tetrahedral site instead of one. These vectors are given in (\ref{eq894gg}), and the doubling of mignon states which they induce gives rise to weak isospin of quarks and leptons, i.e. to partners like $A(\nu_{e})$ and $A(e)$ in (\ref{e33hgxx}).\\
In explicit terms the mignons from the 2 tetrons fig. 4 on a tetrahedral site are given by isospinors of the form
\begin{eqnarray} 
\psi_L=(U_L,D_L)=(1+\delta U_L ,\delta D_L) 
\qquad
\chi_R^c=(-D_R^c,U_R^c) =(-\delta D_R^c,-1+\delta U_R^c) 
\label{ewan2rr}
\end{eqnarray}
and the corresponding isospin vectors are 
\begin{eqnarray} 
\vec Q_L - (0,0,\frac{1}{2}  ) 
=\begin{pmatrix}
\operatorname{Re} \delta D_L  \\
\operatorname{Im} \delta D_L \\
\operatorname{Re} \delta U_L
\end{pmatrix}
\qquad \qquad
\vec Q_R - (0,0,-\frac{1}{2}) 
=\begin{pmatrix}
\operatorname{Re} \delta U_R^c  \\
\operatorname{Im} \delta U_R^c \\
\operatorname{Re} \delta D_R^c
\end{pmatrix}
\label{ewan55rrr}
\end{eqnarray}
which makes the weak isospin transformation between the 2 kinds of mignons evident.\\
Actually, for a more precise analysis, terms of order $\delta^2$ or, equivalently, fluctuations of $\operatorname{Im} \delta U$ have to be included in the analysis. For example, they enter in 
\begin{eqnarray} 
\delta Q_x=\operatorname{Re} \delta D +\frac{1}{2} [\delta U^\dagger \delta D +c.c.]
\label{ewuuuu1}
\end{eqnarray}
or in $Q_z$, because $Q_z=\psi^\dagger \psi/2-|\delta D|^2$ with the tetron density given by
\begin{eqnarray} 
\psi^\dagger \psi=|1+\delta U|^2 + |\delta D|^2
\label{ewuuuu2}
\end{eqnarray}
This makes the analysis of the weak isospin transition somewhat more complicated than appears in (\ref{ewan55rrr}).\\
%Qz=psik*psi/2-|dD|**2 aus dichteschwingung und |dD|**2 zusammen\\
%Qx=RedD+(dUstern*dD+c.c)/2 ist nur in leading order RedD, enthaelt auch ImdU etc\\
%Qy=ImdD-i*(dUstern*dD-c.c)/2 ist nur in leading order ImdD\\
%mit ansatz exp(iwt) ist die y-komponente ein sin(wt), also ungerade unter zeit\\
%Die linkombs Qx+iy ~ exp(iwt) haben T[Qx+iy]~exp(-iwt), also das c.c.\\
%DAS MACHT DEN WEAK ISOSPIN UEBERGANG U--Uc und D--Dc in der letzten formel von 2419 schwieriger\\
%wenn man vecQL und vecQR kombiniert betrachtet, treten ImdDL und ImdURc gleichberechtigt auf.)
The complete set of the 24 quark and lepton eigenstates is worked out in \cite{lmass}.
%If one correctly distinguishes $\vec Q_L$ and $\vec Q_R$, weak isospin partners like $A(\nu_{e})$ and $A(e)$ are obtained. This is explained in detail in (\ref{appii4}).\\
%As explained in (\ref{appii4}), by including right and left handed vibrators weak isospin partners are obtained, 
%weil sie sein sinus statt cosinus verhalten wie die beiden anderen hat, und unter cc ihr vorzeichen wechselt, also im wesentlichen imaginaer ist
%deltaSz fuer deltaU (1.famlie) und deltaSx+iSy fuer deltaD (2+3.familie)
%deltaQRc=deltaT liefert den Isospinpartner zu deltaQL=deltaS, 
%weil deltaDc=deltaUstern und deltaUc=-deltaDstern
\subsubsection{The role of CPT symmetry in the tetron model}\label{appiicpt}
%(irgendwohin - aber habe ich auch schon woanders erwaehnt: free tetrons in 6+1 dimensions there is no question about tetron or antitetron, because transforms under the real 8 spinor rep of so61. Only when the 3+1 dim hypercrystal is formed, the 8 breaks according to into a particle and an antiparticle of so31.)\\
In this section I want to show that CPT plays an eminent role in the relativistic generalization (\ref{eq8rela}) of the Shubnikov group (\ref{eq8gs}). The point is that in the presence of relativistic particles, CPT supersedes the PT symmetry inherent in non-relativistic (iso)-magnetic phenomena.\\
%(erklaeren, warum Pin*T erhalten ist bei tetraeder, obwohl Punktspiegelung keine symm des tetraeders ist)\\
Actually for this argument to work, the parity transformation P is to be interpreted as the product $P_{in}P_{au}$ of internal and external parity, where $P_{in}$ reverses the sign of the internal and $P_{au}$ that of the physical coordinates. In other words, CPT is the CPT transformation not in Minkowski space (this would be C$P_{au}$T) but in the underlying 6+1 dimensional spacetime. Therefore this invariance will sometimes be called CPT6 to distinguish it from ordinary CPT in Minkowski space.\\
We start our line of reasoning by summarizing what we already know about C, P and T in the tetron model:\\
-in (\ref{appii5}) the (maximal external) parity violation observed in the weak interactions was related to the maximal violation of internal parity $P_{in}$ induced by the ground state fig. 1. This relation arises via the appearance of products $\gamma_5\vec\tau$ in the tetron dynamics.\\
-in (\ref{appii4}) charge conjugation was discussed as a necessary ingredient, if one is working in a relativistic environment and includes antitetrons in the ground state as done in figs. 4 and 5. Due to the antiparticle nature of the isospin vectors pointing inwards, internal parity in fig. 5 is maximally violated just as it is in fig. 1.\\
-it is well known that time reversal is an essential ingredient in any kind of magnetic phenomena. One may even go as far to say that magnetism is the physics of the breaking of T symmetry. For example, T appears in the nonrelativistic version (\ref{eq8gs}) of our Shubnikov symmetry. T itself is broken as in any magnetic alignment, but the product $TP_{in}$ is an implicit part of the symmetry (\ref{eq8gs}).\\
The action of CPT6 on an isospin vector $\vec Q_L$ pointing up(=outwards) as in figs. 4 and 5 is given by 
\begin{eqnarray}
\vec Q_L \xrightarrow{C} -\vec Q_L \xrightarrow{P_{au}}
-\vec Q_R \xrightarrow{P_{in}} \vec Q_R \xrightarrow{T} -\vec Q_R  
\label{m1arr12}
\end{eqnarray}
As discussed in section 1, T and $P_{in}$ reverse the orientation of isospin vectors, while $P_{au}$ exchanges L and R. C also reverses the orientation of an isospin vector, because antiparticle contributions enter in isospin vectors $\vec Q$ with an opposite sign, cf. (\ref{m1u122}).\\
Eq. (\ref{m1arr12}) shows that the ground state fig. 5 consisting of 4 tetrons and 4 anti-tetrons fulfilling $\langle\vec Q_{Li}\rangle =- \langle\vec Q_{Ri}\rangle$ on each tetrahedral site i=1,2,3,4 is CPT6 invariant.
%that CPT6 invariance is equivalent to demanding that the local ground state must be formed by $\vec Q_L$ and $\vec Q_R$ of opposite orientation like in figs. 4 and 5, more precisely a tetron and an anti-tetron are involved, and one has $\langle\vec Q_{Li}\rangle =- \langle\vec Q_{Ri}\rangle$ on each tetrahedral site i. 
In contrast, if the local ground state would be $\langle\vec Q_{Li}\rangle = \langle\vec Q_{Ri}\rangle$, i.e. if the isospin vectors would have the same orientation, one would have PT or CP symmetry, but CPT6 would be violated.\\
%gut: (Note this is due to the existence of a tetron and an antitetron on each tetraehedral site i, where $\langle\vec Q_{Li}\rangle \neq 0$ arises from a tetron and $\langle\vec Q_{Ri}\rangle\neq 0$ from an antitetron contribution. Since it was argued in , that only radial U isospinors contribute to the vevs ... aU=bU aus dem higgs vev führt zu gam5 beitrag=0. siehe zettel.)\\
%If no antiparticle was involved, the $\vec Q_L$ and $\vec Q_R$ in the ground state would have to be identical for internal pv.  
%welches sind die vevs? einmal a(1-g5)a mit teilchenopertor a, und -b(1+g5)b mit antiteilchen b daher <psik*tau*psi>=-<psik*tau*g5*psi> soll man nicht schließen. ICH DENKE JETZT DASS DAS FALSCH IST UND DAS SM GIBT JA EINEN VEV OHNE GAM5 VOR (wenn auch nur für die summe der tetraederpunkte). der sm-vev enthaelt a und b term.
%In summary one can state that all discrete symmetries C, T, $P_{in}$ and $P_{au}$ are separately broken, but that some of their products like $P_{in}P_{au}$ (=6-dimensional parity) and $CP_{au}$ (=CP in Minkowski space) are symmetries of the ground state figs. 4 and 5.\\
Note, this discussion of the ground state's discrete symmetries has to be distinguished from the CP violation in the CKM matrix. As discussed in \cite{lmass}, the latter arises from the complex nature of the exchange couplings and the resulting phases in the mixing of mignon states. 
%inzwischen glaube ich, dass fuer den vev C*Pa erhalten ist, da auch T*Pi erhalten ist, da T UND Pi jeden vektor umdrehen, egal ob er L oder R hat oder Teilchen oder AT. Da CPaPiT erhalten ist, ist dann auch CPa erhalten.\\
%insgesamt kann man die gruppe als A4+T*(S4-A4) oder A4+C*Pa*(S4-A4) oder A4+CPT*(S4-A4)/Pi schreiben, wobei die letzteren beiden auch den Teilchen und den Antiteilchenvektor an einer site vertauschen, waehrend die normale shubgruppe A4+T*(S4-A4) nur innerhalb der teilchen bzw antiteilchen vektoren wirkt.\\
%vecQLa--Pa-->vecQRa--Pi-->-vecQRa--C-->-vecQRb--T-->vecQRb\\
%wobei vecQLa=ak*tau*(1-g5)*a=vecQL mod b terme\\
%vecQRb=bk*tau*(1+g5)*b=-vecQR mod a terme\\
%damit ist die vacuum config parallel wenn man mit vecQRb arbeitet unad antiparallel, wenn man mit vecQR arbeitet
\subsubsection{Why is top so heavy, why not bottom?}\label{appii614}
Up to tiny CKM and V+A mixing effects, the top quark corresponds to vibrations of $\vec \Sigma_L=\sum_i \vec Q_{Li}$. This vector plays a special role in the tetron model, which has to do with the chiral nature of the SSB and with the relation between internal and external parity violation\cite{lhiggs} induced by (\ref{ewip81}), and because nature has chosen to break internal parity, i.e. prefers the state fig. 1 with isospin vectors pointing outwards over the one with those pointing inwards. The top-mignon is defined as that internal precession, where all isospin vibrations act against the SSB alignment in the strongest possible way.
%zzz(mostly weil jLR doch eine Beimischung macht)
%zzz(entweder weil mb aus Abweichungen der inter-WW von HH+DM kommt, zb HH+g*DM mit laufender Kopplung g oder weil b eine kleine L-Beimischung hat - oder eine V+A.)
%haengt letztlich mit innerer PV und ground state bevorzugung alle Pfeile nach aussen zusammen. groundstate hat nur eine (radiale) isospinkomponente nach außen nicht nach innen. Letztlich schlaegt sich das in Bevorzugung top-quark nieder 
%(aber oben habe ich top als die 3 rot-Schwingungen identifiziert, die gibt es fuer R als auch fuer L sind die rot-schwingungen des chiralen Tetr immer L?)
%\subsubsection{Is there a correlation between up-type of isospin and left-handedness?}\label{appii615}
%Yes. The correlation has been described in (\ref{appii4}).
%dieser Zusammenhang kommt in der Rechnung nur raus, weil man alle Spins auf eine einzige Quantisierungsachse dreht. Dem entspricht die Benutzung radialer Spinoren. Nur in einem solchen System ist der VEV von der Form Fermiskala=<UbarU> und gleich der 4fachen Laenge des Isospinvektors (siehe ein Zettel). Aber die Laenge von piz enthaelt ein gamma5. Diese Laenge darf nicht mit der geom Groesse R des Koordinatentetraeders verwechselt werden.
\subsubsection{Why is $m_b$ much smaller than $m_t$ and from where does it get its predominant contributions?}\label{appii723err}
In contrast to the top quark, the b-quark is mostly an excitation of the $\sum_i \vec Q_{Ri}$, and therefore does not get a contribution from the SU(2)$_L$ Dzyaloshinskii-Moriya interaction (\ref{mm3dm}).\\
%However, there are non-leading inter-tetrahedral SSB interactions in addition to (\ref{mm3dm}), which provide contributions to $m_b$. Those terms typically involve right-handed isospin vectors and are correlated to the existence of the second Higgs doublet (\ref{phos1}).
I have considered several sources for an $m_b$ value of order 5 GeV, the most straightforward being the existence of a tiny right handed (=V+A) current. Actually, in the calculation\cite{lmass} it is more difficult to accommodate the bottom mass than the values of the lighter quark masses. This is because contributions to $m_b$ are naturally associated to the right handed version of (\ref{mm3dm}) where $\vec T$ instead of $\vec S$ appears, and this V+A contribution must be radically suppressed according to the argument that weak parity violation is due to the internal chirality of the tetrahedrons, as discussed in (\ref{appii5}). Thus from a general point of view the introduction of a small V+A current does not serve as a fully satisfactory explanation for a noteworthy bottom mass.\\
%***(vielleicht muss man doch dem 2.higgs einen vev geben, also kein inert higgs aber auch hier waere das ziemlich ad hoc recht vev fuer 2.higgs wirklich fuer b-mass bottom mass ist eigentlich ungeloest vielleicht ein mischungseffekt a la ckm)\\
On the other hand one is talking here about a $m_b/m_t=$3\% effect, and a 3\% V+A correction to the leading V-A coupling is compatible with present experimental bounds\cite{nuemix}. Such a correction can then be used to account for the observed value of $m_b$\cite{lmass}.
\subsubsection{If up type quarks arise mainly from vibrations of $\vec Q_L$, how can their right-handed version be produced?}\label{appii616}
%(eine variation dieser frage habe ich woanders?)
As shown in \cite{lmass}, lepton states originate dominantly from vibrations of the form $\vec S \pm \vec T$, while up and down quark states are related to vibrations of $\vec S$ and $\vec T$, respectively, where $\vec S=\vec Q_L$ and $\vec T=\vec Q_R$. Therefore one might suspect that the helicities of quarks and leptons generated in this way are restricted, too. However, it must be noted that an excitation $\delta$ of a left-handed isospin vector can in principle vibrate into any chiral direction. The same is true for right handed vectors $\vec Q_R$.

\subsection{Connections to Gravitation and Cosmology}
This section relies on the interpretation of the world as an elastic system of internal tetrahedrons. According to section 1 the tetrahedrons form an extremely dense monolayer mesh which can buckle and bulge into the surrounding higher dimensional space thus inducing the forces of gravity. The following list of questions and answers shows how this picture may be incorporated within a larger, cosmological framework.\\
% As evident from fig. 2, the full universe can be generated by exact repetition of this discrete structure. The following list of questions and answers sheds light on the nature of this repetition and furthermore shows how the tetron particle model may be incorporated within a larger, cosmological framework.\\
Throughout it will be assumed that after crystallization spacetime is homogeneous and isotropic, i.e. it can be described by a Friedmann / Robertson-Walker (FLRW) metric
\begin{equation} 
ds^2-c^2 dt^2=-a^2(t) [\frac{(d\vec x)^2}{1-k\vec x^2}+\vec x^2d\Omega^2]
\label{flrw3a}
\end{equation}
%(nicht mit r oder R arbeiten, da dies bei mir der tetraederabstand) 
where $a(t)$ is the dimensionless 'scale factor' and k is the Gaussian curvature of the space (not of spacetime) at the time when a(t)=1. k has units of length$^{-2}$ and is positive, zero or negative for an elliptical, Euclidean or hyperbolic universe, respectively. The time where a(t)=1 is often chosen to be the presence; however, in (\ref{appiiharm}) it will be chosen to be the time when the universe has reached its equilibrium.\\
%Rewriting this as 
%\begin{equation} 
%ds^2=dt^2-a^2(t) \gamma_{ij}dy^i dy^j
%\label{flrw3b}
%\end{equation}
%one has 
%\begin{equation} 
%\gamma_{ij} = \delta_{ij} + k\frac{y_i y_j}{1-k y_n y^n}
%\label{flrw5b}
%\end{equation}
%with $i,j=1,2,3$. The $y^i$ are the 'comoving' coordinates, from which the physical coordinates are obtained as $x^i=a(t)y^i$. 
Note, the $x^i$ in (\ref{flrw3a}) are 'comoving' coordinates, from which the physical coordinates are obtained as $a(t)x^i$. 
Accordingly, the physical velocity of an object can be decomposed as 
\begin{equation} 
v^i=a(t)\frac{dx^i}{dt}+\frac{da}{dt} x^i=w^i + H a x^i
\label{flrw3c}
\end{equation}
where the second term introducing the Hubble parameter $H=\dot a /a$ is called the 'Hubble flow' and the first term $w^i$ defines the 'peculiar velocity' of an object, i.e. its velocity relative to the Hubble flow.\\
Note that the FLRW metric fulfills $g_{00}=1$ and $g_{0i}=0$, i.e. it corresponds to a metric in a 'synchronous gauge'\cite{bema}. Such a metric has a particular simple interpretation in the tetron model, because the only non-trivial elements are the spatial $g_{ij}$. These can be identified with variable 'longitudinal' distances among the tetrahedrons in the 3-dimensional elastic system which is our physical universe. This point will be further discussed in connection with (\ref{appiiharm}) and fig. \ref{figchain2}a.
%Note that the ideas presented here are more speculative than most of what was discussed before.
\subsubsection{Is the tetrahedral ground state stable or metastable?}\label{appii701u}
%(gehört nicht in graviteil?; trotzdem ein netter einstieg)
In other words, are there isomagnetic states X with a lower energy? If yes, this might threaten our world when such a state would be produced in high energy collisions. More precisely, what could happen is that at collider energies of order $\Lambda_F$ the ordered state fig. 2 is locally destroyed and in the process of re-ordering of isospin vectors a germ of the state X appears. Energy would then be released, which would destroy the ordering in the neighborhood of the original collision, so that a chain reaction would start at the end of which the metastable ground state would be completely replaced 
by the stable one.\\
In the microscopic model it is possible to compare the energies of all isomagnetic configurations with 4 isovectors to the energy of fig. 1. Essentially, one deals with a system of 4 isospin vectors interacting via an 'anti-ferromagnetic' coupling $J_A  > 0$. The local energy of any such state X is given by
\begin{equation} 
E(X) = J_A \,[\, \vec{Q}_1 \vec{Q}_2 + \vec{Q}_1\vec{Q}_3+
\vec{Q}_1 \vec{Q}_4 + \vec{Q}_2\vec{Q}_3+
\vec{Q}_2 \vec{Q}_4 + \vec{Q}_3\vec{Q}_4 \, ]
\label{muruz}
\end{equation}
With this input one may run over all possible ground state configurations of isospin vectors. As a result, one finds 2 minima with exactly the same energy. One minimum (m1) corresponds to fig. 1 while the other (called m2) is characterized by the 4 isospins arranged into 2 pairs of opposite orientation as depicted in fig. 6 of \cite{pyrochlor}.\\
This conclusion remains unchanged, if one considers separate left- and right-handed isospin vectors on each tetrahedral site as in (\ref{mm3}) and fig. 5. However, it may get changed, if the {\it inter}-tetrahedral energies are different. In general, one expects roughly identical inter-tetrahedral energies, because in both cases (m1) and (m2) the same number of isospin pairs are aligned in a ferromagnetic way, with identical exchange couplings $J_{inter}$. However, in reality there may be small differences between the $J_{inter}$ values for the (m1) and the (m2) configuration due to the different geometries of the two ground states.  
\subsubsection{Is physical space discrete, i.e. is there a granular structure of physical space in addition to the discrete tetrahedral structure in internal space?}\label{appii701}
Most probably yes. Although the discrete structure of physical space is not compelling and the distance r between two tetrahedrons could be identically zero, the discussions in section 1 suggest that r has a tiny non-vanishing value of the order of the Planck length. Note, due to the elasticity of the system only an average $\langle r \rangle$ can be given, in accord with (\ref{tmet6177}).
%possible that the pairs extend from the internal fiber into full 6-dimensional space and thus set the scale for a possible tiny granular structure of Minkowski space.(kann das Plancklaenge sein ... wohl nicht denn sonst wuerde phys Verstaendnis der Paarbildung zusammenbrechen))
%zzzSEHR GUT: auch in einem Galileiraum $R^{6+1}$ darf es kein ausgezeichnetes Bezugssystem geben laut Descartes. Gitter im Grenzfall verschwindender Gitterkonstante sollte Galileisymmetrie restaurieren in dem Sinne das alle gegeneinander bewegte Bezugssysteme gleichberechtigt sind. Sobald man aber die Gitterkonstante aufloesen kann, merkt man dass doch ein System bevorzugt ist!!?? Aber merkt man das nicht AUCH dem Kontinuum an, weil die bewegten Teilchen den Widerstand des Gitter-aethers ueberwinden muessten???? bei mir wohl nicht: wenn ein Mignon im Ruhsystem des Gitters erzeugt wird, braucht man Energie um es in Bewegung zu setzen, aber danach gibt es seine Energie von Tetraeder zu Tetraeder weiter ohne dass der Widerstand eines aethers da waere. also: NUR wenn man Gitterabstand aufloesen koennte wuerde man ausgezeichnetes System sehen.Aber da Gitterabstand am Anfang des Kosmos noch viel kleiner war als Lpl heute, ist das schwieirg.
\subsubsection{How can such a granular structure be compatible with Lorentz invariance?}\label{appii703}
In other words: how can the elastic continuum of tetrahedrons be Lorentz invariant?\\
The point to note is that in the tetron model all physical objects that we know are superpositions of excitations which travel as quasi-particles through the hyper-crystal. This holds true even for photons and also for ourselves as well as for all experiments and 'reference frames' we can preparate. In a quantum mechanical framework such excitations always have a wave nature and are to be described by (generalizations of) the d'Alembert wave equation. This fact alone fixes the system of waves, which constitutes our physical environment, to be Lorentzian, because the d'Alembert operator 
\begin{equation} 
\Box = \frac{\partial^2}{ \partial x_1^2}
+\frac{\partial^2}{ \partial x_2^2}
+\frac{\partial^2}{ \partial x_3^2}
  - \frac{1}{c^2}\frac{\partial^2}{ \partial t^2}
\label{weqmur}
\end{equation}
leaves the squared 4-momentum $p^2$ invariant. In particular, if one wave packet is emitted from another one, their velocities add up according to the rules of Lorentz transformations.\\
For massive particles which move at velocities $< c$, the operator is modified to $\Box -m^2 c^2/\hbar^2$ corresponding to a 'dispersion relation' $p^2=m^2 c^2/\hbar^2$.
It is important for this argument (and in general to retain the Lorentz structure) that in all those wave equations there is a universal maximum speed c that fixes the relation between space and time. This is defined to be the speed of the massless excitations (\ref{appii210}) and according to (\ref{tm33bxc}) is given by the ratio of 'lattice constant' $L_P$ and 'hopping time' $T_P$, cf. the discussion after (\ref{tm38ddem22}).\\
The value of c is universal for all SM particles because all of them arise from the same isomagnetic interactions introduced in section 1. This is analyzed in more detail in (\ref{appii7m8}), and the question of (metrical) velocities larger than c will be discussed in (\ref{appii80d5}). The question why gravitational waves propagate at c will be answered in (\ref{appii80d3}).\\
It should be stressed, that in particle physics interactions usually only the excitations move. The tetrahedrons stay fixed at their location in the hyper-crystal. They only move in connection with metrical changes induced by gravity, for example when the universe expands after the crystallization or on a much tinier level in any kind of gravitational interaction, cf. the discussion after (\ref{newt588x}) and in (\ref{appii80d5}).
\subsubsection{Is time discrete?}\label{appii720}
Time is an averaged construct induced by the superposition of many elementary thermodynamics processes. In the tetron model it can be interpreted to have a granularity of order $T_P$ -- at least from the standpoint of the quasi-particles which constitute our physical environment, because $T_P$ is the minimum time it takes a quasi-particle to jump from one tetrahedron to the other. In our world of quasi-particles the duration of any observable physical process can never fall below this value.
%NB: whether time is discrete or not does not affect the arguments in (\ref{appii703}). 
%Within (general) relativity, however, both time and space are continuous and treated on a purely descriptive level.
%(ich glaube nicht an die diskreten microelastischen Teile in der Zeit; man kann bei klein genugem gitter gar nicht unterscheiden.)\\
%(ABER IST NICHT PROBLEM, dass räumlich diskret automatisch in zeitlich diskret reinmischt? NEIN, weil Lorentz setzt nur voraus dass alle Abstände so klein sind, dass wir sie nicht sehen.)
%The minimum unit of time in ordinary quantum mechanics is $T_P$. With h(t) and G(t) its changes are analogous to those of $L_P$. 
\subsubsection{Does quantum mechanics arise from material properties of the hyper crystal?}\label{appii1co}
My claim is yes. The proof will be given in the following section (\ref{appii1}). It relies on the fact that ordinary matter (including the photon) consists of internal excitations traveling as quasi-particles on a discrete structure with Planck length lattice spacing.
\subsubsection{Why is the Planck length the natural lattice spacing for the hyper-crystal?}\label{appii1}
The short answer: the Planck length $L_P$ arises as a lower limit on $\Delta x$ in the 'generalized' Heisenberg uncertainty relation (\ref{mur}), which includes the effects of gravity. On the other hand in the tetron model {\it all} known particles including the photon are interpreted as excitations with an extension of at least one lattice spacing $\langle r \rangle$, i.e. they are quasi-particle waves with wavelength $\lambda > L_P$. Since every physical experiment necessarily makes use solely of these quasi-particles, its resolution cannot be better than r. This strongly suggests $r\approx L_P$, because if r would be smaller than $L_P$, r instead of $L_P$ would determine the uncertainties of quantum mechanics.\\
Due to the extremely small lattice spacings, spacetime as we perceive it effectively looks like a continuum. This guarantees local rotational symmetry. The item of invariance under Lorentz boosts is discussed in (\ref{appii703}).\\
%zzz(nb: could still be that lattice spacing is lets say r=10**8GeV and that one can probe LP with other means. es ware noch moeglich dass die tetraeder WW nix mit dem Photon zu tun haben, dann wuerden sich r und R als weitere Skalen zwischen Planck und Fermi schieben. ABER wuerde dann nicht r statt LPL die Quantenechanikgroessen bestimmen? IN NAECHSTER FRAGE BEANTWORTET\\
%An extended answer: according to the Kopenhagn interpretation of quantum mechanics Planck's constant is due to an uncertainty necessarily arising from the process of measuring. In the present model all known particles (quarks, leptons and vector bosons) are interpreted as excitations on the Planck lattice hyper-crystal and thus will always extend over at least one lattice spacing r. Therefore measurements involving physical particles cannot be more accurate than r. This uncertainty is assumed to imply the quantum mechanical uncertainty and to fix the value of Planck's constant in the following way:\\
Extended answer: in ordinary quantum mechanics there is a fixed dimensionful quantity h which relates the canonical Fourier variables of frequency to energy and inverse distance to momentum, i.e. it transforms the spacetime quantities x and t into physically 'active' quantities
\begin{equation} 
\vec p=\hbar \vec k \qquad \qquad E=\hbar\omega
\label{murkss}
\end{equation}
%zzz(aber ist die Unschaerfe das einzige feature der QM, das man erklaeren muss mit lattice?).
%zzz\footnote{The situation is in parallel to the Newton constant which according to the Einstein equations allows for the Umrechnung from energy ($T_{\mu\nu}$) to geometry ($R_{\mu\nu}$).}
These relations are the reason why h appears in the uncertainty principle $\Delta x \Delta p > \hbar$ which otherwise would just be the Cauchy-Schwarz inequality known from the Fourier analysis of waves, i.e. $\Delta x \Delta k > 1$ with no dimension on the r.h.s. I consider them as further evidence that h is a material property of the hyper-crystal not valid outside of it, cf. the discussion later in this section and at the end of section 1.\\
The 'generalized' uncertainty relation includes gravitational effects of the photon on a test particle\cite{adlerpl}.
%(gegenueber der normalen Unschrelation fehlt irgendwie ein Faktor einhalb. dies hab ich in der kleinert formel unten schon beruecksichtigt)
These occur because the general relativistic effect from the photon adds to the uncertainty about the test particle. The Heisenberg relation is then modified to
\begin{equation} 
\Delta x \geq \frac{\hbar}{\Delta p}+L_P^2 \frac{\Delta p}{\hbar} =  \lambda [1+ (\frac{L_P}{\lambda})^2]
\label{mur}
\end{equation}
where $\lambda$ is the photon wavelength to be identified with the limit of resolution.\\
Eq. (\ref{mur}) can be derived e.g. by extending the 'Heisenberg microscope' thought experiment (which imagines a photon to measure x and p of a probe particle) to include gravitational effects of the photon\cite{adlerpl}. Without gravity, the position of the particle can be determined to an accuracy of about the wavelength of the light used, and this determines the ordinary uncertainty relation. If one includes gravity, there is a tiny gravitational force from the photon acting on the particle, because the photon has an effective mass 
\begin{equation} 
m_\gamma=\frac{h\nu}{c^2}= \frac{h}{c\lambda}  
\label{murzuz89}
\end{equation}
and this force will accelerate the particle, making the already uncertain position of the particle a bit more uncertain. The gravitational uncertainty can be estimated as follows: using Newton's law the additional acceleration due to gravity is given by $a=Gm_\gamma/r^2$, so that the relevant length can be estimated to be $aT_P^2=L_P^2/\lambda$. This is precisely the second term in (\ref{mur}).\\
This result has quite a natural interpretation, because the gravitational effect of the photon modifies the average spacing $\langle r \rangle =L_P$ between the tetrahedrons involved in the interaction and thus the resolution $\Delta p=\lambda$ of the photon. The modification factor can be derived from the general relativistic length contraction according to (\ref{newt588x}) by inserting the photon's gravitational mass (\ref{murzuz89}) and amounts to  
\begin{equation} 
1-\frac{\phi}{c^2} = 1+\frac{Gh}{|\vec x|\lambda c^3 }
\label{newt3x}
\end{equation}
Since the relevant regime of discussion is distances $|\vec x|\approx\lambda$, the factor (\ref{newt3x}) agrees with the factor in (\ref{mur}).\\
In any case eq. (\ref{mur}) implies that $\Delta x$ considered as a function of $\lambda$ takes on a minimum value
\begin{equation} 
\Delta x \geq L_P
\label{mur22}
\end{equation}
corresponding to a photon with wavelength $L_P$. In the usual folklore, this is interpreted in such a way that at distances/wavelengths smaller than $L_P$ all matter dissolves into quantum fluctuations and the laws of physics do not have a meaning any more.\\
In the tetron model, however, the interpretation is a little different. First of all, one should be reluctant in just adding up uncertainties and 'deriving' from that a minimum position uncertainty. 
%as shown later after (\ref{mur5}), there is a change of sign in (\ref{mur}), so that the argument about the minimum of $\Delta x$ does not really apply. 
Rather, the condition $\Delta x > L_P$ arises in the tetron model from the limitation of wavelengths of quasi-particles which holds true inside the hyper-crystal. In particular, since the photon is such a quasi-particle (a correlation among 2 tetrahedrons), the minimal wave length of photons, that can be produced and used in experiments, roughly corresponds to one lattice spacing $L_P$, and it is this minimal wavelength which absolutely restricts the precision of any experiment.\\
It may be noted that this limitation is valid only in the realm of quasi-particles (in which we ourselves exist), not of tetrons. On the other hand, this remark does not preclude, that there are analogous restrictions for tetrons, with maybe a different Planck constant, cf. the discussion in (\ref{appiift66}). In this connection it is worth mentioning that on the hyper-crystal the value of h is given by (\ref{tm33bxh}) where the Planck energy $\Lambda_P$ corresponds to the binding energy of the tetrahedrons and $T_P$ is the hopping time needed to absorb and re-emit the photon. Obviously, such a value of h can only be valid within the hyper-crystal and does not at all apply to, let's say, tetrons in a tetron gas before the crystallization.\\ 
Actually, within the microscopic model one can do better than just augmenting arguments in favor of (\ref{mur22}). Starting from the dispersion (\ref{tm38dd}) of a quasi-particle photon one can extent the Fourier analysis on the Planck lattice to derive a 
%At first sight the linear momentum dependence of the gravity contribution in (\ref{mur}) does not fit exactly into this picture. However, it can be interpreted in such a way that 
%As a consequence, the gravitational contribution to the uncertainty $\Delta x$ will be increased by a relative factor $L_P/\lambda$ as compared to the original value $L_P$. This explains the form of the second term in (\ref{mur}).\\ 
%These considerations may be compared to the Fourier wave analysis which has led to the dispersion relation (\ref{tm38dd}) and which also provides a 
modified uncertainty relation
\begin{equation} 
\Delta x \Delta k \geq 
%|\langle [x,k]\rangle |= 
|\langle \cos (k L_P) \rangle |
\label{mur5}
\end{equation}
Eq. (\ref{mur5}) may be evaluated in the limit of long wavelengths to obtain the analogon of (\ref{mur})
\begin{equation} 
\Delta x  \geq \frac{\hbar}{\Delta p}-L_P^2 \frac{\Delta p}{2\hbar}
\label{mur5aa}
\end{equation}
No need to worry about the factor 1/2, because the second term in (\ref{mur}) is only an estimate anyhow. More important seems the change in sign of the second term. However, as explained above, in the tetron model one is not reliant on the minimum argument based on (\ref{mur}).\\
Note further, that for wave vectors near the border of the first Brillouin zone, i.e. for wavelengths $\lambda\sim L_P$ the r.h.s. of (\ref{mur5}) vanishes. Therefore it seems that lattice quantum mechanics at Planck scale energies can exhibit classical, non-quantum mechanical behavior. This, however, will happen only for extremely high photon energies, in a region where photons cease to exist because their quasi-particle nature forbids wavelengths $\leq L_P$.
\subsubsection{How is physical space defined within the 6+1 dimensional world? How is it distinguished from the internal dimensions?}\label{appii709}
The aligned tetrahedrons define a 3-dimensional subspace of $R^6$. Everything orthogonal gives physical space.
%\subsubsection{How did the separation of internal and external space arise in the first place?}\label{appii708}
%There is Because of the way the crystal is built from aligned tetrahedrons. This happens because the internal and external structures communicate with each other -- which should be no surprise, e.g. in view of the explanation of weak parity violation within the tetron model in (\ref{appii5}) and \cite{lhiggs}.\\
%Note that I am talking here about the internal coordinate alignment, which was relevant at high temperatures shortly after crystallization, not about the isospin alignment.\\
\subsubsection{What is the exact lattice structure of the hyper-crystal?}\label{appii721}
This question is obsolete because of the elastic nature of inter-tetrahedral coordinate bonds. It is an irregular lattice of tetrahedrons in the base 3+1 dimensional spacetime without a discrete symmetry. As discussed in various places - e.g. after (\ref{eqfe}) in section 1, in (\ref{appii7a7}) and (\ref{appii534sf}) - there are more similarities to a fluid than to a crystal.
\subsubsection{Then why is this structure called a 'crystal'?}\label{appii7a7}
Because there is a rigid tetrahedral structure in the {\it internal} directions. Concerning physical space it is a disordered system which resembles a plastics or a fluid. However, the quasi-particles in this space follow an approximate Lorentz symmetry at low energies, cf. (\ref{appii703}).
\subsubsection{Why is there no growth of the crystal into the internal directions?}\label{appii534}
\begin{figure}
\begin{center}
\epsfig{file=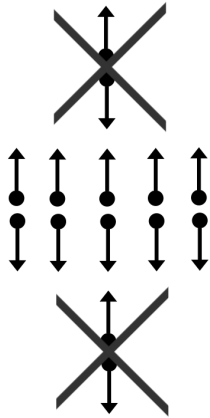,height=6.0cm}
%\bigskip
\caption{This picture illustrates the impossibility of the crystal to grow into the internal dimension for the 1+1 dimensional case. It is based on the assumption that the interactions among isospins of different internal molecules are ferromagnetic, so that adjacent molecules repel each other. The argument can be extended to the realistic 3+3 dimensional case. Note the definition of 'internal' as given in (\ref{appii709}).}
\nonumber
\label{figrow1} 
\end{center}
\end{figure}

%One observation is that there seems to be no crystallographic space group belonging to $A_4+T(S_4+A_4)$.
This has to do with the form of the fundamental tetron interaction, cf. (\ref{appiift66}) and (\ref{appii307uu}), and its preference to form the spiky tetrahedral 'star' system fig. 1, which do not allow tetrahedrons to be stacked on top of each other in fig. 2.
%In more scholarly terms this corresponds to the fact that there is no space symmetry group with point symmetry $A_4+S(S_4-A_4)$.\\
%***(ich weiss immer noch nicht, ob letzteres wirklich stimmt NEIN es stimmt nicht laut Bilbao)
It is the main reason why internal dimensions need not be compactified, cf. (\ref{appii725}).\\
In the following I want to make the argument more explicit and start by considering the 1+1 dimensional configuration in fig. \ref{figrow1}, i.e. one internal and one physical dimension with 'di-atomic' molecules consisting of 2 tetrons instead of 4. While the isospin interactions within one molecule are anti-ferromagnetic, the interactions among isospins of {\it different} tetrahedrons are assumed to be ferromagnetic. (Remember that the tetron model understanding of the SM symmetry breaking in section 1 was based on this assumption.) Therefore the crystal structure depicted in fig. \ref{figrow1} can only grow in the horizontal(=physical) but not in the vertical(=internal) direction.\\
%Note that for a one-dimensional system of molecules with 2 atoms (x) and spins arranged in an antiferro way, i.e. of the form <--x1--x2--> the statement can trivially be proven, because the only possible crystal structure is ... <--x1--x2--> <--x3--x4--> ... 
%and if there is to be ferromagnetic attraction between different molecules (i.e. between x2 and x3), this cannot be a stable configuration.
Actually, the essence of this argument can be transferred to the 3+3 dimensional case of tetrahedrons, because if there is to be a ferromagnetic arrangement, then this first of all builds up the physical base space. If one further tries to arrange in parallel e.g. isospins $\vec S_1$ and $\vec S'_1$ of 2 tetrahedrons $t$ and $t'$ in the vertical(=internal) direction, the ferromagnetic interaction $J\vec S_1\vec S'_1$ will not be effectual because between $\vec S_1$ and $\vec S'_1$ there will always be the vector $\vec S_2+\vec S_3+\vec S_4\approx -\vec S_1$.
\subsubsection{*Where do the 6 spatial dimensions come from?*}\label{appii711}
I have only a partial answer to this question. A tetron spinor $\psi =(U,D)$ transforming as 8 under SO(6,1) can be interpreted as an octonion field living in 6+1 dimensions. Octonions form a rather unique mathematical structure\cite{conway,kantor,schaf}. They are the next thing to consider when complex numbers (used for amplitudes in quantum mechanics) and quaternions (used for rotations and spinors in physical space) are not ample enough to describe physical phenomena. The octonion nature of tetrons has been used in (\ref{appii5}) to determine the form of the isomagnetic tetron interactions. The splitting of $R^6$ into an internal space and physical space corresponds to a splitting of an octonion into two quaternions.\\
Note that if octonion multiplication should really be relevant for the behavior of tetrons, it seems somewhat more natural to have 7 instead of 6 spatial dimensions. This, however, is pure speculation. It is the product $\vec\tau\gamma_5$ appearing in (\ref{ewip81}), which provides the link between internal and physical space, necessary to explain particle physics phenomena such as parity violation of the weak interaction.
%from the chiral configuration fig. 1.
%\subsubsection{Is the dimension of spacetime really 6+1?}\label{appii710}
%We only know that the tetrahedral molecules extend into 3 internal dimensions which are orthogonal to 3+1 dimensional Minkowski space. In addition there could be other dimensions into which the hyper-crystal did not grow.
\subsubsection{How does the value for c come about? Why is it finite?}\label{appii7mx}
In the tetron model light is interpreted as a (massless) excitation of tetron-antitetron pairs, cf. (\ref{appii204}) and (\ref{appii210}). While the pairs themselves are bound over 2 tetrahedrons with fixed positions within the hyper-crystal, the excitations propagate through physical space as quasi-particles. In contrast to massive excitations, whose propagation was considered in (\ref{tm38ddem22}) and (\ref{tm38ddem2222}), photons cannot cling to one fixed internal tetrahedron, but are constantly moving at the speed c of light. Nevertheless, c is not infinite, because even a massless excitation needs a certain 'hopping time' $T_P$ to jump from one tetrahedron to another. Since the tetrahedrons are distributed over the hyper-crystal with average distance $L_P$, this corresponds to $c=L_P/T_P$ in agreement with (\ref{tm33b}).
%In some sense c has more to do with spacetime properties than with the internal interactions by which the photon is produced.
%It is a finite quantity mostly because these excitations have to jump from one tetrahedron to the other when running through physical space. The idea is that the photon jumps from one tetrahedron to the other when running through physical space. It takes the 'hopping time' TP for the photon to get absorbed and afterwards released by a single tetrahedron, in order to jump to the next tetrahedron at distance LP.\\
%According to (\ref{tm33bxc}) it takes the 'hopping' time $T_P$ for the photon to get absorbed and afterwards released by internal tetrahedrons.\\
%(gehoert hier noch nicht rein: as explained in (\ref{appii80d3}) for a graviton the situation is similar. Its speed is not determined by movements of tetrahedrons or physical particles, but they are elastic schwingungen of tetrahedrons within the DMESC.) 
%(man kann die Arbeit The quantum vacuum as the origin of the speed of light ...benutzen, um c aus den jumping größen zu berechnen. aber das sollte in einer eigenen subsection passieren. Wo ist TP in jener Arbeit? ich habe jetzt den eindruck dass die einfach nur trivialitaeten umformen)
\subsubsection{Why is c the universal limiting speed for all particles?}\label{appii7m8}
In the tetron model all SM particles are internal excitations whose interactions have the same type of isomagnetism as universal origin. Therefore it is of no surprise that a common maximum speed exists, to be identified with the ratio $L_P/T_P$, as explained in (\ref{appii7mx}). Furthermore, it was demonstrated at the end of section 1, that while the mignon masses are determined by isomagnetic interactions, their propagation proceeds via the elastic/spacetime properties of the DMESC.\\
For the interpretation of (metrical) velocities larger than c see (\ref{appii80d5}).
%(Aber Magnonen im FK haben doch andere (limiting) Geschwindigkeiten als Schallwellen oder Cooperpaare. unten weg: wenn das photon nur eins unter vielen Anregungszuständen ist, warum bestimmt seine Geschwindigkeit allein die SR? Antwort: Weil alle Anregungszustände letzlich auf dieselbe fundamentaleDynamik zurückgehen.) 
\subsubsection{Is there a rest system of the hyper-crystal?}\label{appii7r1}
At first sight the existence of such a system seems to contradict Einstein's principle of equivalence and special relativity - well established concepts which I do not want to question. Still I think the answer to the question is affirmative.\\
The point is that in the tetron model all material objects and all normal matter and gauge bosons including the photons from their very nature are quasi-particle waves, i.e. entities fulfilling Lorentz covariant Klein-Gordon equations, as discussed in (\ref{appii703}). As such they cannot distinguish an absolute rest system, i.e. they naturally fulfill Einstein's principle of equivalence. By contrast, the original tetron matter, which forms the fixed hyper-crystal ground state, is merely the carrier of those quasi-particles, and mostly invisible for human experiments.\\ %(according to (\ref{appii80d3}) lives so to say in a different realm of reality than ordinary matter.) 
This point of view is close to the Fitzgerald interpretation of special relativity\cite{barce} and allows for a rest system of the hyper-crystal. More details are given in the next answers.
\subsubsection{Why has a rest system never been observed in Michelson Morley type of experiments?}\label{appii7xx2}
The essence of the answer to this question has already been given at the end of (\ref{appii7r1}). An extended version can be found in \cite{barce}. In that interpretation of special relativity a Galilean ground state like fig. 2 is not a contradiction to special relativity but rather supplements it. It supplies the 6-dimensional framework which leads to the observed spectrum of physical particles.
\subsubsection{Can the hyper-crystal's rest system ever be identified?}\label{appii7r2}
It follows from (\ref{appii7r1}) that it is difficult to experimentally perceive the ground state fig. 2 which forms the rest system of the hyper-crystal - the reason being that our reality (ourselves, our detectors as well as all test particles) consists of quasi particle waves and not of tetrons themselves.\\
%(which remain moving uniformly and in a straight line on the hyper-crystal unless acted upon by a force). 
The puzzle cannot be resolved on the level of excitations. Even gravitational waves propagate at the speed of light, cf. (\ref{appii80d3}). However, it is well known that the metrical expansion of the universe (and metrical changes in general) can proceed at superluminal velocities, cf. (\ref{appii7a90}) and (\ref{appii80d3}).\\
%It is true that according to general relativity gravity travels through the universe at the speed of light ... aber das ist nur weil wir in einem lorentz system leben, und wir gar nichts wahrnehmen koennen, was schneller als licht ist. However, in an expanding universe there is the well known effect of superluminal motion of the metric.
In section 1 it was suggested that metrical changes correspond to displacements of the internal tetrahedrons within the elastic hyper-crystal, cf. the discussion after (\ref{tmet}). Applied to the standard model of cosmic expansion, this leads to the idea that {\it the hyper-crystal's rest system is given by the spatial coordinates of the FLRW metric} (\ref{flrw3a}). Of course one must be aware that due to the expansion factor a(t) this is an elastic 'rest system' permanently changing with cosmic time, but at least momentarily it can be considered to be at rest. The time dependence of a(t) defines the expansion of the elastic hyper-crystal, and what comes nearest to the notion of 'matter at rest' are the galaxies on the Hubble flow. The peculiar velocity (\ref{flrw3c}) gives their 'true' nontrivial motion with respect to the elastic rest system.\\
%All those excitations are waves because their dynamics is dictated by variants of the d'Alembert wave equations. These waves are travelling on a completely dark/opaque background, and as waves they cannot be related to this unknown rest system.
The underlying idea is that to a good approximation mignon-matter forming the galaxies was originally produced 'at rest' (at least on the average) during and shortly after the big bang / crystallization and has since been moving together with the expanding elastic hyper-crystal. This is in accord with the finding that most large lumps of matter, such as galaxy clusters, are nearly comoving with the Hubble flow. 
\subsubsection{Is there an internal time different from the ordinary time variable?}\label{appii717}
In magnetism, time reversal is a far-spread concept because it allows to reverse the orientation of the magnetization / of a spin vector. This way it enters the definition of magnetic point groups, for example the Shubnikov group (\ref{eq8it}).\\
However, as shown in (\ref{appii4}) and (\ref{appiicpt}), for the case of iso-magnetism one may replace the role of internal time reversal by charge conjugation, so no separate time is needed for the internal dimensions.
\subsubsection{Is there an absolute time in the hyper crystal?}\label{appii7uu}
Yes, it is given by the comoving Hubble time coordinate, i.e. the elapsed time since the big bang according to a clock of a comoving observer.
%Note that time has a profound meaning not only in the tetron model but in  big bang model, as opposed to models with a constant expansion rate, such as the now discredited steady-state scenarios. 
%(ABER: hubble time zusammen mit comoving coordinates sind ein Lorentz, kein Galilei system.)
%(hier vielleicht die konstruktion aus den arbeiten zitieren, um aus dem hyperkristall ein relativisisches gebilde zu machen, aber ich habe ADM auch schon in sect1 ziziert)
\subsubsection{What is the status of the Copernican and of the Cosmological Principle in the tetron model?}\label{appii7vv} 
The Copernican Principle states that no place in the universe is 'special' or preferred, while the Cosmological Principle demands that the universe looks the same in all directions (is isotropic) and contains everywhere roughly the same amount and mixture of material (is homogeneous). These principles are not questioned by the microscopic model, assuming a suitable uniformity of the hyper-crystal built from tetrons. However they will {\it not} be fulfilled at the edges of the DMESC where there may be steadily new accretions of tetrahedrons to the crystal. Those edges are therefore expected to be both 'special' and anisotropic places.
%WEGL(\subsubsection{Is the universe open or closed?}\label{appii7pppp}
%It is almost flat and spatially open, with edges. The question whether it is timely open or closed cannot be answered in the tetron model.
%(daneben gibt es aber die frage der zeitlichen krümmung, also ob es sich irgendwann mal wieder zusammenzieht. Das hängt von der Massendichte incl dark matter und evtl existierender kosm konstante ab. wobei stark vibrierende tetronen das universum komischerweise zusammenziehen)
\subsubsection{Is the original $R^{6,1}$ Lorentzean oder Galilean?}\label{appii704}
It was argued in section 1, that the speed of light, which is at the heart of Lorentz symmetry, is an intrinsic property of the hyper-crystal, not valid outside of it. Furthermore, the fact that metric expansion of our universe can proceed with velocities larger than c gives support for Galilean (or, in case of a SO(6,1) Lorentz type symmetry, with a limiting speed much larger than the speed of light).
\subsubsection{Do we know anything about the position of our universe in full $R^{6+1}$?}\label{appii523}
%Is there a nontrivial relative position of the internal tetrahedron w.r.t. to the isospin axes? 
%logik: QGesamt conserved --> mnue=0
%also folgt aus mnue ungleich 0 dass Qgesamt nicht conserved ist.
%wir haben aber gezeigt dass aniso-ww wie zz oder xx+yy zu nue massen führt 
%also ist dort Qges nicht conserved
%die frage ist nur ob nue massen durch anisos oder durch torsions ww entstehen!!!!
%nb: Qges=0 im Grundzustand also zwar dQges/dt=0 aber schwingt wohl nur um 0 herum 
%So far the x,y,z-coordinate axes of internal space have been defined in terms of the weak vector bosons $W^\mu_{x,y,z}$. In other words, they do not refer to the absolute position of the ground state tetrahedron in internal space but to directions defined by excitations like mignons and the gauge bosons, i.e. small variations of the ground state. The tetrahedrons of the universe (fig. 2) may therefore lie globally rotated relative to those axes. (dieser letzte satz ist mir nicht einsichtig, da auch die excitations sich ans koordsystem halten) It is fair to ask whether anything can be said about the corresponding rotation angles or even whether there is some anisotropy in the surrounding $R^{6+1}$ environment to which this refers (and how this might be experimentally detected).\\
To show up in experiments such an information requires {\it anisotropic} interactions, which would get their anisotropy from 6-dimensional structures which go beyond the hyper-crystal. It cannot be obtained from the dominant isospin interactions (\ref{mm3}) and (\ref{mm3dm}), because these are rotationally invariant, i.e. the same mass spectrum is obtained after a global rotation of the isospin axes.\\
A simple anisotropic Hamiltonian of isospin vectors $\vec Q_i$ would look like
\begin{eqnarray} 
H_a=- J_z \sum_{i \neq j}^4 Q_{iz} Q_{jz} -J_{xy} \sum_{i \neq j}^4 ( Q_{ix} Q_{jx} +  Q_{iy} Q_{jy}  )
\label{eani}
\end{eqnarray}
with $J_z \neq J_{xy}$.\\
Since they explicitly break internal rotational symmetry, anisotropic interactions like (\ref{eani}) violate internal angular momentum conservation (\ref{tm31}). As discussed in (\ref{appii605}), neutrinos are the Goldstone modes corresponding to that symmetry. More precisely, the 3 neutrino species were identified as the 3 vibrating components of total internal angular momentum. Therefore, the anisotropic Hamiltonian (\ref{eani}) contributes to the neutrino masses, and measuring the neutrino mass matrix has the potential to answer the present question.
%Quantitative details on the neutrino mass spectrum can be found in section 7 of \cite{lmass}.
%zzz(get modified under global rotations of isospin space - aber wie haengt das mit der custodial su2v zusammen? die ist fuer mb=mt bereits gebrochen. Dann waere mt-mb ein mass fuer aniso, da bei mir b und t bereits am Anfang der rechnung keine Isospinpartner sind. ---- jedoch gilt su2V im isospin raum der mignonen, der ein bisschen anders definiert ist als der der tetraeder.
\subsubsection{Is there a contribution from the tetron ground state configuration to the mass/energy density and to the expansion rate of the universe?}\label{appii7ww}
%In (\ref{appii701u}) the energy of the ground state fig. 2 came into focus and was compared to another possible ground state configuration. 
According to general relativity the mass/energy density of the universe determines its curvature, its expansion rate and its future. So one might expect that the energy of the ground state fig. 2 should also contribute. However, as argued in (\ref{appiien6}), this energy does not appear in energy balance equations of ordinary matter.\\
%the forces of gravity have been interpreted in section 1 solely as stress arising from disclinations and/or dislocations within the hyper-crystal, which in turn originate from the presence of quasi-particle states which constitute ordinary matter. By contrast, the DMESC in its ground state (=the arrangement fig. 2 without any excitation) is a sort of perfect fluid (\ref{ewgrpi}), and by definition without any stresses.\\
The main effect of the enormous tetron ground state energy is to initialize big bang inflation by releasing a vast amount of crystallization energy.
\subsubsection{How does all of this fit into inflationary cosmology?}\label{appii702}
As discussed in (\ref{appii206k}) and after (\ref{eqfe}), inflation in the microscopic model is associated to the crystallization process of tetrahedrons with an accelerated expansion due to the elastic nature of the bindings and the sudden release of crystallization energy.\\
The major signature of inflation is the exponential increase of the scale parameter a(t) in (\ref{flrw3a}). According to the Einstein equations a(t) has to fulfill
\begin{eqnarray} 
\ddot{a} =-\frac{4\pi}{3} G (\rho+3p/c^2) a 
\label{eani456}
\end{eqnarray}
where $\rho$ is the mass density and p the pressure, and the combination $\rho+3p/c^2$ corresponds to the trace of the energy momentum tensor. Eq. (\ref{eani456}) makes it clear that the exponential increase needed for inflation can be obtained for constant and negative $\rho+3p/c^2$.\\
What kind of matter fulfills such a condition? The immediate answer: matter undergoing a phase transitions. Indeed, it is normally assumed in models of inflation that a false vacuum decays in the framework of some abstract phase transition being active in the very early universe. However, the physical background of this phase transition is never specified.\\ 
In the tetron model the understanding is clearer, because the phase transition relies on the Landau free energy $\Delta F$ in (\ref{eqfe}). The order parameter is defined in terms of density fluctuations D of tetrahedrons  and is thus a material quantity. 
%(der folgende satz ist falsch, da er sich auf mignonmatter bezieht nicht auf tetronmatter: The trace of the energy momentum tensor can be obtained from $\Delta F$ via $p=-\Delta F$ and $\rho=-p/c^2$, provided one assumes sufficiently weakly fluctuating D.)
%ich habe einheiten gecheckt! energiedichte hat selbe einheit wie druck; also F als energiedicht interpretieren
Inflation(=crystallization) starts almost immediately after the big bang, at the time when the germ of the hyper-crystal comes into being. This point corresponds to the maximum value of the free energy curve (\ref{eqfe}) from where the system rolls down to its non-trivial minimum -- the moment when the latent heat is released.\\
A word of warning: although the qualitative features of an exponential expansion are well described by (\ref{eani456}), one should be aware that the Einstein equations are not valid close to the crystallization point. According to the discussions after (\ref{eqfe}) and after (\ref{tm33chi9}), the least one must expect is a temperature dependence of c and G which will modify the details of the description. Since I have not quantitatively estimated the amount of energy released in the crystallization, it is even possible, that the dominant part of the growth of the hyper-crystal arises from simultaneous accretion of tetrahedrons and only a minor part is due to subsequent expansion, i.e. to inflation in the proper sense. In that case the accretion would almost instantaneously produce a huge hyper-crystal, and expansion literally would start only at tetrahedral distances of let's say $r\geq 0.1 \, L_P$. I will call this possibility scenario X.
%(evtl auf bigbang und inflation verzichten; stattdessen Kristallbildung bei gitterabstand 0.2xLP oder so und danach moderate Expansion mit Abbremsen durch Materie und Wiederbeschleunigung weil die kontrahierende normale Materie sich immer mehr verduennt und nun die Konvergenz gegen a-unten-s überwiegt.
\subsubsection{What are the tetron model answers to the flatness and the horizon problem?}\label{appii714}
If one rejects scenario X, they are similar to those in ordinary inflationary models:\\
--flatness problem (the question why the universe is almost flat everywhere): due to the exponential expansion, triggered by the initial release of crystallization energy, the universe is much larger than anticipated.\\
%visible part of the universe was causally connected when the hyper-crstal was formed. auch bei mir ist Univ viel größer als beobachtet, wegen starker Ausdehnung durch Kristallwaerme. geht nur wenn das Kristall=der Kosmos ist. Die Kristallisation muss den >c Effekt der sich ausdehnenden Metrik erklaeren: sie findet im aeusseren Galileiraum statt und betrifft direkt die Metrik. deutet auf ein R61 Galileisystem hin\\
--horizon problem (the question why the universe looks almost the same everywhere): all parts of the universe were causally connected at the time when the hyper-crstal was born. Due to the subsequent exponential expansion they have lost their causal contact.\\
On the other hand, within scenario X, the universe is extremely large right from the beginning, and also flat because the interactions of tetrahedrons are such that they form a flat hyper-crystal. Concerning the horizon problem: what appears to be a causal contact shortly after the big bang, would be due to the fact that the physics and initial conditions are everywhere the same at the point of accretion.
%man fängt sich in scenario X aber ein problem mit domain walls ein (section domain walls)
\subsubsection{Is there an inflaton field?}\label{appii713}
Inflation was explained in section 1 and (\ref{appii702}) as arising from the latent heat released in the crystallization of the hyper-plastics. Therefore in the tetron model, the inflaton can be interpreted as the energy/density wave that carries the initial crystallization energy.
%There is some similarity to the way dark energy is interpreted in (\ref{appii81de}).
\subsubsection{How can the release of crystallization energy lead to metric velocities much larger than the speed of light?}\label{appii7a90}
It is well known that metrical changes proceeding faster than c do not contradict Einstein's equations, because in GR there are rules about matter moving through space, but there is no rule about space expanding faster than light. They necessarily occur in the standard cosmological model at times shortly after the big bang.\\
In the microscopic model metric velocities have a special interpretation due to the fact that an expanding space corresponds to elastic motion of tetrahedrons in the hyper-plastics, cf. (\ref{appii80d5}) and section 1. Since tetron velocities are not bounded by c, the release of a large amount of crystallization energy at the big bang can push this motion to faster than c. It is only excitations and quasi-particles whose propagation on the hyper-crystal is bounded by c, cf. (\ref{appii703}), (\ref{appii7mx}), (\ref{appii7m8}), (\ref{tm33chi8}) and (\ref{tm38ddem2222}).
\subsubsection{Nature of the forces among tetrahedrons}\label{appiiharm}
\begin{figure}
\begin{center}
\epsfig{file=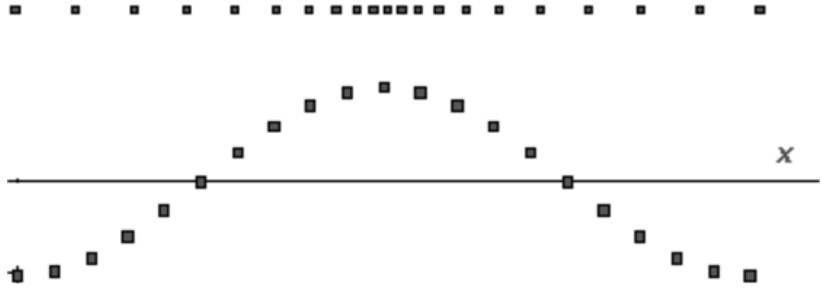,height=4.0cm}
%\bigskip
\caption{This picture illustrates 'longitudinal' vs 'transverse' curvature for a 1-dimensional chain of tetrahedrons. The tetrahedrons are drawn as tiny black squares. Physical space is represented by the x-axis, similar to fig. 2. Longitudinal displacements/accelerations (fig. \ref{figchain2}a) are felt as local or global contraction or expansion of physical space like in the FLRW metric. Transverse displacements (fig. \ref{figchain2}b) go into the internal dimension. Modifications of the hopping time (relevant for the Newtonian approximation) are not drawn in the figure.}
%(Note any curvature effect can only be felt by (mignon) matter present, and it must be intrinsic curvature.)}
\nonumber
\label{figchain2} 
\end{center}
\end{figure}

The elastic forces among the tetrahedrons are responsible for the curvature of spacetime, i.e. for the gravitational interactions. In principle, they should be calculable from the fundamental tetron interaction considered in (\ref{appiift66}). Because of the many particles involved, in practice this is not an easy task.\\
On a heuristic level the forces can be understood as depicted in fig. \ref{figchain2} for a linear chain of tetrahedrons. Longitudinal displacements/accelerations are felt as local or global contraction or expansion of flat physical space and can be used to understand the physics of the FLRW metric. Transverse displacements go into one of the 3 internal dimensions thus inducing genuine extrinsic spatial curvature.\\
Neither of the two can explain the appearance of a gravitational field in the Newtonian limit, i.e. the potential (\ref{newt2xu7}) appearing in the $g_{00}$ component of the metric (\ref{newt2x}). To account for that, a 'timely' curvature, i.e. variations of the hopping time $T_P$ are needed, i.e. the time a quasi-particle needs to travel(=be emitted, run, get absorbed) from one tetrahedron to its neighbor. These variations happen, because the presence of mass/energy %together with the elastic nature of the environment 
modifies the microscopic processes behind the hopping of any 'test excitation'. Note that the Newton effect has an extremely simple form - it is proportional to M and inversely proportional to r, and actually arises from the elastic nature of the tetronic environment in which the excitation propagates, cf. (\ref{tm38ddem2222}).\\
In more formal terms the FLRW cosmology relies on a longitudinal spatial expansion $r\rightarrow a(t)r$ in the sense of fig. \ref{figchain2}a, while the Newtonian limit (\ref{newt2x}) corresponds to a change of the local time and space variables according to
\begin{eqnarray} 
t&\rightarrow & t(1+\frac{GM}{c^2r}) \\
r&\rightarrow & r-r_p +\frac{GM}{c^2}\ln \frac{r}{r_p} 
\label{ealoc}
\end{eqnarray}
in the sense of (\ref{tmet}). $r_p$ arises from an integration constant and can be chosen to be the position of the local observer.\\
According to (\ref{ealoc}), switching on a Newtonian gravitational field, not only the hopping time is modified as discussed above, but also the tetrahedrons are concentrating near the source. For a comoving observer it appears that they have moved away from him, their distance increased by a factor $1+GM/c^2/r_p$.\\
Normally these effects are tiny, however for $r\rightarrow 0$ or $GM\sim c^2$ the Newton approximation breaks down, and the Schwarzschild metric
\begin{equation} 
ds^2=(1-\frac{2GM}{c^2r})(cdt)^2 - (1-\frac{2GM}{c^2r})^{-1}(dr)^2
\label{schsch}
\end{equation}
should better be used. Based on Kruskal's coordinates, a similar analysis as in the Newtonian case can be carried out in order to obtain the rearrangement of tetrahedrons near a black hole. This can also be used to analyze the behavior at $r=0$ where the hyper-crystal structure dissolves, cf. (\ref{appiibb}).\\
%bei räumlicher Krümmung biegt sich der Grundzustand in den inner space, bei zeitlicher krümmung nähern sich die tetraeder einander an.\\
%As in any medium composed of interacting microscopic entities, there must be a force among the tetrahedrons.
I will now end the discussion of the effects of point masses and return to the question to what extent the standard FLRW cosmology is modified by the tetron model. As described in section 1, the underlying idea is the existence of an elastic {\it and curved} 3-dimensional space embedded in R6. The possibility to have curvature is important here, otherwise one would have a flat elastic system which in the Einstein framework would correspond to 'pure gauge' configurations (\ref{dint45}), which do not affect ordinary mignon matter. Note that curvature in the time direction corresponds to accelerations of tetrahedrons within the elastic medium. It can be combined with spatial curvature in the way described in section 1 \cite{arnodesermisner,carter}.\\
According to the analysis in (\ref{appiift66}) the force among the tetrahedrons may be of such a form that these are eventually driven towards an equilibrium distance $r_0$ corresponding to a minimum of the inter-tetrahedral potential. In that case the force can be expanded in powers of $r-r_0$, and in the leading harmonic approximation it is of Hooke's form 
\begin{eqnarray} 
m_T \ddot{r}=-d_T(r-r_0)
\label{ddot81}
\end{eqnarray}
where $m_T$ is the tetrahedron mass and 
\begin{eqnarray} 
\omega_T^2=d_T/m_T
\label{ddot81zrw}
\end{eqnarray}
the frequency characteristic for the interaction.\\
In an elastic medium of many tetrahedrons this interaction corresponds to 
\begin{eqnarray} 
\rho_T \ddot{u}=\zeta u''
\label{ddot82}
\end{eqnarray}
where $\zeta$ is the Lame parameter introduced in (\ref{tm33gg8}), $u$ the deformation vector and $\rho_T$ the mass density (\ref{tm33chi8}) of tetrahedrons. Note that eq. (\ref{tm33chi8}) actually corresponds to a tetrahedron mass
\begin{eqnarray} 
m_T=\rho_T L_P^3=M_P
\label{fgot82}
\end{eqnarray}
In other words, the tetron masses and binding energies within one tetrahedron plus the binding energy of one tetrahedron within the hyper-crystal sum up to define the Planck mass $M_P$.\\
%(bei (\ref{dint45}) besser u genannt, aber in sect1 haben wir sowohl xi als auch xstrich genannt). 
%\rho_T =m_T/r^3 % \frac{m_T}{L_P^3}  
%\label{ddot83555}
%the mass density of tetrahedrons.\\
Since $\zeta$ can be traced back to Hooke's constant via 
\begin{eqnarray} 
d_T=\zeta \, r_0 
%d_T=\zeta L_P 
\label{ddot83444}
\end{eqnarray}
%where the Planck length $L_P$ corresponds to the average distance among the tetrahedrons.
it can be considered as Hooke's constant per unit of length.\\
%and c in (\ref{ddot83}) is actually the speed of light.\\
According to (\ref{ddot82}) one expects elastic waves at speed 
\begin{eqnarray} 
c^2=\zeta/\rho_T
\label{ddot83}
\end{eqnarray}
%(hier muss wohl ueberall c0 stehen, um mit der kleinertformel am ende des hauptteils zu passen - oder dort den 0-index weglassen, denn ich habe ja c2=zeta/rho in der formel vor kleinert verwendet)
in agreement with (\ref{tm33chi8}). It must be noted, however, that this kind of elastic wave can only be perceived by mignon matter, i.e. by human observers, if it is associated to a non-vanishing curvature, i.e. is not just a pure gauge transformation (\ref{dint45}). In that case it constitutes a gravitational wave.\\
There is another caveat in these considerations, and this concerns the use of the non-equilibrium quantities $\zeta$, $\rho_T$ etc in the above equations. These were introduced in section 1 to describe the properties of the universe today. However, since cosmic expansion has not reached the equilibrium $r_0$ but currently corresponds to an average distance $r=L_P < r_0$ between tetrahedrons, quantities $\zeta$ and $\rho_T$ in (\ref{ddot83444}) and (\ref{ddot83}) differ from the values (\ref{tm33chi8}) and (\ref{tm33chi9}) which they take at present. It is true that there will always be a minimum in the inter-tetrahedral potential, so that a harmonic approximation is reasonable; however, the values of the potential parameters certainly depend on cosmic time.\\
For reasons of simplicity I will ignore this point in the following; in other words, I will assume that
%we are reasonably near to the equilibrium so that
the time dependence of these parameters can be neglected. Under this assumption, one obtains 
\begin{eqnarray} 
\omega_T \equiv \sqrt{\frac{d_T}{m_T}} = \frac{1}{T_P}
\label{ddot84}
\end{eqnarray}
This is an extremely high frequency, which evidently does not play any role in the universe's expansion. 
%(zu hypothetisch: If stimulated at all, they die out at once because of viscosity effects. oder es sind reine eichungen, die von Mignonen nicht gesehen werden koennen)\\
%(es muss hier rein, dass es eine Rückstellkraft m*app=-D(a-as) geben muss, wobei m durch die Dichte der Tetraeder bestimmt ist, bzw rhoT*app=mu*ass Wellenglg, wobei wir die Wellen, da es reine eichungen sind, aber nicht sehen koennen. es gibt die wellen deshalb nicht, weil gar nicht um gleichgewichtslage schwingen kann und TP ist eh viel zu gross)\\
%(dies ist normale elastizität reine eichung, krümmungstensor=0, die wir nicht sehen koennen; ausserdem was ist mit der zeitl krümmung? die expansion des kuchens ist doch eigentlich elastisch, wird aber anscheinend nicht durch die reine eichung du+du beschrieben; ist wohl eher plastics? eher die 2.odnungsterme bzgl der zeit)\\
The point to note is that 
%$L_P$ - the present day average distance of tetrahedrons in the hyper-crystal - is not equal to the equilibrium distance $r_0$ of tetrahedrons. 
in the presence of the densely packed many other tetrahedrons, a given tetrahedron is not able to immediately attain the relaxed equilibrium distance $r_0$ to its neighbors. Instead, due to the large stiffness (\ref{tm33gg8}) of the DMESC, the characteristic time for this to happen is much larger. In other words, there exists another characteristic frequency $\omega_M \ll \omega_T$ contributing to the enhanced expansion of the hyper-crystal, so that (\ref{eani456}) should be replaced by 
\begin{eqnarray} 
\ddot{a} =-\frac{4\pi}{3} G (\rho+3p/c^2) a  -\omega_M^2 (a-a_0) \,\;\; [ +\Lambda_C c^2 a ]
\label{eanibuff}
\end{eqnarray}
where $\rho$ and p are the (mignon) matter density and pressure, respectively, and in a matter dominated universe one has $\rho\sim 1/a^3$ as usual.\\
For reasons of comparison I have also added a cosmological constant term. It may be noted that apart from the constant $\omega_M^2 a_0$ it is of the same form as the harmonic term. Actually, all of the terms in (\ref{eanibuff}) can be interpreted to arise from the total energy-momentum of the tetrahedral system, an issue which is further discussed in (\ref{appiien6}).\\
%(and thus can be interpreted as being part of it).
%The question to what extent the harmonic interactions in (\ref{ddot81}) and (\ref{eanibuff}) are compatible with the idea of general relativity is further discussed in (\ref{appiiharm1}). 
%One point to make is that usually a harmonic term is valid only for small displacements $a-a_0$ whereas GR is a 'finite strain theory' which claims validity beyond small displacements. Aber das habe ich oben bereits diskutiert, dass das Potenzial sich mit der cosmic time aendert, und also auch die Potenzialparameter.
It is important to note that for a universe expanding towards equilibrium $a_0$ one has $a<a_0$ so that the sign of the harmonic term is the same as that of a $\Lambda_C$ term with a positive cosmological constant. If one analyzes this more closely, it turns out that it is possible to accommodate the observed dark energy effect to a harmonic contribution $\omega_M\neq 0$ instead of a cosmological constant provided one chooses $\omega_M\approx 10^{-18}$Hz (very small as compared to $\omega_T\approx 10^{43}$Hz).\\
%the behavior of a(t) from this equation is similar as in the standard cosmological model with a (positive) $\Lambda_C$. However, this is true only up to times where 
The 2 terms on the rhs of (\ref{eanibuff}) are then of the same order. The first one leads to the well-known $t^{2/3}$ behavior of a(t) for a dust dominated universe. For larger values of a and t the second term becomes relevant, with $a\sim a_0 (1-\cos (\omega_M t))$. We live at $0\ll a \ll a_0$. This condition corresponds to times where the universe has started to re-accelerate but does not yet feel the equilibrium.\\
%At much later times the Hooke term will dominate and drive it towards $a_0$.\\
Note for a proper treatment of (\ref{eanibuff}) the temperature dependence of $a_0$ must be taken into account
\begin{eqnarray} 
a_0 \sim (T_c -T)^{-\frac{1}{6}}
\label{eanibuff1}
\end{eqnarray}
where $T<T_c$ and $T_c$ was introduced in (\ref{eqfe}). 
\subsubsection{How is the curved hyper-crystal embedded in $R^6$?}\label{appii5emp}
In this section I will show that it is not very difficult to embed the standard solutions of the Einstein equations like the FLRW cosmology and the Schwarzschild black hole into the flat 6(+1) dimensional space, which according to section 1 represents the basis of the tetron model. As will be seen, it is actually enough to have 1 additional spatial dimension instead of 3, i.e. an $R^4$ instead of an $R^6$; and as a matter of fact it remains unclear at this moment, whether this one additional dimension should be chosen to be one of the isospin directions or as the seventh spatial dimension on which I speculated in (\ref{appii711}).\\
Consider, for example, the case $k=1$ of a closed FLRW universe (\ref{flrw3a}), where the spatial part of the FLRW metric describes a 3-sphere $S^3$. In terms of $R^4$ coordinates this is given by 
\begin{eqnarray} 
u_1&=&a \cos \phi  \\
u_2&=&a \sin \phi \cos \theta_1 \\ 
u_3&=&a \sin \phi \sin \theta_1 \cos \theta_2 \\
u_4&=&a \sin \phi \sin \theta_1 \sin \theta_2 
\label{eqs3ss}
\end{eqnarray}
and the well known $S^3$ line element
\begin{eqnarray} 
ds^2=a^2 [d \phi^2 +\sin^2 \phi (d \theta_1^2+\sin^2 \theta_1 d\theta_2^2) ]
\label{eqs3see}
\end{eqnarray}
can be identified as the spatial part of the FLRW metric via the identification 
$\vec x^2=\sin^2 \phi$ because of $d \phi^2=d\vec x^2/(1-\vec x^2)$.\\
One is thus led to the conclusion, that such a structure - as well as small disturbances of it - can always be embedded into $R^4$ and therefore also in $R^6$ or $R^7$.\\
The situation is similar for the Schwarzschild geometry (\ref{schsch}). To understand that point, one may consider the Schwarzschild singularity at r=0 in the rest frame of the hyper-crystal. While the curvature in the time direction is due to a modification of the hopping time as explained in section 1, the spatial component of (\ref{schsch}) is obtained from local $R^4$ coordinates $u_1$, $u_2$, $u_3$ and 
\begin{eqnarray} 
u_4=2 \sqrt{r_b(r-r_b)}  
\label{eqs3sszu}
\end{eqnarray}
where $r^2=u_1^2+u_2^2+u_3^3$ and $r_b=\frac{2GM}{c^2}$ is the Schwarzschild radius.\\
Note that the Schwarzschild description of a black hole loses its validity at Planck scale energies $r<r_b$. What happens physically inside the black hole in that regime, is discussed in (\ref{appiibb}).
\subsubsection{Is there a similarity to the behavior of superfluids?}\label{appii534sf}
There is a certain similarity, but there are also appreciable differences.\\
The growth and expansion of the elastic hyper-crystal along a lower (in this case 3) dimensional structure is reminiscent of the behavior of superfluids, which can creep along arbitrary surfaces.
%and leads one to suspect that it shares some properties with them. 
There are 2 prototypes of superfluids whose representatives are He-3 and He-4. While in He-3 fermion condensates populate the macroscopic quantum state, in He-4 superfluidity is a consequence of a normal Bose-Einstein condensation of the He-4 bosons. There are actually speculations that associate gravity to a He-3 type of superfluid\cite{volovik}. In the tetron model, however, it is the tetrahedrons and not the tetrons themselves which are involved in the gravitational interactions. The tetrahedrons are bosons, so the ordinary Bose-Einstein approach to superfluidity could be useful to describe the growth and expansion of the hyper-crystal. More precisely, 
%Accepting the idea of a superfluid, 
the Bose-Einstein condensate of tetrahedrons within the hyper-crystal would
%by a 6-dimensional Schr\"odinger equation (\ref{ewrr86}), but 
obey a Gross-Pitaevskii equation in 3 dimensions\cite{pita}
\begin{eqnarray} 
i\hbar \frac{\partial}{\partial t} D(\vec r,t) = [-\frac{\hbar^2}{2m} \Delta + \alpha |D(\vec r,t)|^2 ] D(\vec r,t)
\label{ewgrpi}
\end{eqnarray}
%(was genau sind al und gamma? antwort: aus glg 16!!!!)
Although it fulfills a (nonlinear) Schr\"odinger equation and is sometimes called the 'wave function' of the condensate, D is a classical field and can be identified with the order parameter of the system, i.e. with the density fluctuations of tetrahedrons introduced in (\ref{eqfe}).\\
%(aber bei landau free energy tritt kein h auf?! ist bei dieser gravigeschichte wirklich h relevant? super ist immer ein makroskopiescher quantenzustand muss also h eine rolle spielen vielleicht das tetronische h)\\
%\\(frage zur superfluid: wie beschreibt man das creeping verhalten entlang der ebene)
One should, however, be hesitant to accept the idea of a superfluid, for the following reasons:\\
--quite in general superfluidity is a low temperature effect. There is no indication of an additional cosmic phase transition in the low energy regime.\\
--in the last 10 billion years the universe has expanded rather slowly. This is not what one would expect from a {\it super}fluid.\\
--in a superfluid there is a macroscopic quantum state governed by a Planck constant h, as appears in (\ref{ewgrpi}). This picture is different from the interpretation (\ref{appii1}) of quantum theory on the hyper-crystal and the model with elastic Hooke's inter-tetrahedral forces advocated in sections 1 and (\ref{appiiharm}).\\
As a result I prefer to consider the DMESC in its ground state to be a perfect fluid (not a superfluid), and by definition without any stresses. Only if mignons and other excitations are put into the system, this will produce stress and strain, thus modifying metric and curvature in the way discussed in section 1 and giving rise to the effects of gravity.
%(ist dies nicht genau was wir brauchen, weil laut einstein nur mignonmaterie die kruemmung veraendern kann. mein eigener ansatz ist, dass elastische effekte diuj+djui reine eichungen sind, also in der kruemmung eh nicht sichtbar. es gibt doch auch selbst-ww der metrik in den nichtlinearen Einsteingleichungen.)
%(und wie man normal hier zu curvature kommt, without too much buckling into the internal direction. More or less equivalent to the micro-elastic continuum interpretation of GR presented in section 1. ABER IST SUPERFLUID NICHT MEHR ALS ELAST CONTINUMM???? wenn es keine stresses gibt, erzeugen auch kruemmunengen keinen stress wie ist dann die einsteindynamik erklaerbar? auch die idee mit der D*(x-xs)**2 Elastizitaet wird dann durch pitaevsi ersetzt. ich sollte superfluid verwerfen.)
\subsubsection{Is energy conserved in the tetron dynamics?}\label{appiien6}
Yes. This is in contrast to general relativity, where energy is only 'covariantly' conserved, i.e. the energy momentum tensor fulfils $D^\mu T_{\mu\nu}=0$ where $D^\mu$ is the covariant derivative involving the Christoffel symbols. This point can be understood by considering the Einstein equation $G_{\mu\nu}=\kappa T_{\mu\nu}$, which relates the energy-momentum $T_{\mu\nu}$ of matter(=mignons) to the intrinsic curvature equivalent to displacements of tetrahedrons, in the following way: the Einstein equation can be re-interpreted by saying that the total energy momentum of the universe with contributions $G_{\mu\nu}$ (from the tetrahedrons) and $T_{\mu\nu}$ (from the mignons) is not only constant but actually zero, and that energy can be shifted from one side of the equation to the other, i.e. from mignons to tetrahedrons and vice versa. A famous example is the red shift of photons in the expanding universe where the photons lose energy to the FLRM metric.\\
In the tetron model, this is not the whole story. As discussed after (\ref{tm33chi9}), the vacuum energy density of the universe does not only consist in the (small) contribution from the cosmological constant responsible for dark energy, cf. (\ref{appii81de}), and the (somewhat larger) contribution from the QCD and electroweak symmetry breaking vacua, but there is also a contribution 
\begin{eqnarray} 
V_T \; \sim \; \Lambda_P^4
\label{tnt6}
\end{eqnarray}
to the vacuum energy from the tetronic ground state energy fig. 2. This contribution arises as the product of the energy of a bound tetrahedron ($\sim \Lambda_P$) and its inverse volume ($\sim \Lambda_P^3$) and is very large, because the tetrahedrons are closely packed and strongly bound within the hyper-crystal.\\
In the framework of quantum field theories the vacuum energy density arises as the sum of all zero point oscillations. For example, the vacuum energy for a free field is the sum of the ground state energies $\omega_k \sim\sqrt{k^2+m^2}$ of all oscillator modes k. For a cubic box of side-length $L_P$, one has to sum over integers $n=kL_P/2\pi$, and in the continuum limit this turns into an integral
\begin{eqnarray} 
\int \sqrt{k^2+m^2} d^3k \; \sim \; \Lambda_P^4
\label{dint81}
\end{eqnarray}
Using the Planck scale as a natural cutoff, this is again much larger than the cosmological constant and the energies of the QCD and electroweak symmetry breaking vacua.
%of the same order as the hyper-crystal's vacuum energy density. 
In field theory\cite{vacuumenergy} it is usually renormalized away, so that it does not appear in any particle or gravitational energy balance consideration. By contrast, in the tetron model it must be included and added to the ground state energy of the hyper-crystal, which according to (\ref{tnt6}) is of the same order of magnitude.\\
%This is related to the vacuum energy of quantum field theory\cite{vacuumenergy} and usually is renormalized away, so that it does not appear in any particle or gravitational energy balance consideration.\\
%On the level of tetrons it should certainly be included in the consideration, as well as 
There is also a third contribution to be included in tetron energy considerations, and that is elastic energy $\sim (\partial_\mu u_\nu + \partial_\nu u_\mu)^2$ from displacements of tetrahedrons with vanishing curvature. It is true that these contributions do not matter in the energy balance of ordinary matter, because mignons do not respond to these 'flat' elastic deformations (pure gauges) 
\begin{eqnarray}
g_{\mu\nu}=\eta_{\mu\nu}+ \partial_\mu u_\nu + \partial_\nu u_\mu 
\label{dint45}
\end{eqnarray}
For the tetron dynamics, however, they are relevant. 
%***(aber die hooketerme D*u tauchen doch in der einstein glg auf, sind sie nicht pure gauge? evtl bei der 0-komponente nicht, da hooke eine beschleunigung ist und daher eine kruemmung. Auch nicht, wenn sich das Dispalcement u irgendwie transversal in den 6d Raum verhaelt)
\subsubsection{We are living in a rather cold universe. Why don't we see a transition from the elastic hyper-crystal to a rigid crystalline structure at zero temperature?}\label{appii763kj}
One could speculate that the present temperatures of the universe are not small enough for this phase transition to occur, or that pressure is required like for the solidification of He-3.\\
In such a rigid crystal no shifts of the tetrahedrons would be allowed, and therefore one expects gravitation to completely disappear.\\
On the other hand, this remark is only true in the absence of mignons. When a mignon excitation is present, its mass/energy induces a stress in the hyper-crystal which at least locally re-liquidizes the system. In other words, the mignon's mass/energy is automatically accompanied by curvature, i.e. a by non-vanishing gravitational potential.
%(dies muss sich mit c ausdehnen und damit liquidiert es auch alle andere stellen im kristall wie eine stoerstelle eben)
\subsubsection{Are the internal spaces compact or infinite?}\label{appii725}
They are infinite, no assumption about compactification of internal spaces needs to be made. The reason why we cannot step into the internal dimensions is because we are built from quasi-particles and thus cannot leave the hyper-crystal(=our universe).\\
According to fig. 2, the hyper-crystal is restricted to a 3+1 dimensional 'surface' in $R^{6+1}$. Going away from this 'surface', internal space is empty, because according to (\ref{appii726}) and (\ref{appii534}) ordinary matter cannot dissipate into the internal dimensions. Exception: other hyper-crystals may have condensed at big bang times. In general these will lie skewed with respect to the one we live in and form separate 'universes', cf. (\ref{appii535}) and (\ref{appii816}).
\subsubsection{How can one avoid dissipation of energy into the internal dimensions?}\label{appii726}
%(aus einem normalen nichtrel Oberflaechen-Kristall dissipiert hauptsaechlich Waerme IR Strahlung)
Since internal space is infinite, matter and energy could in principle disappear into it. This could happen in the form of particles which move in the direction orthogonal to the hyper-crystal. However, in (\ref{appii204}) and (\ref{appii210}) I have taken the viewpoint that all observed particles including the photon are excitations of the crystal and as such cannot exist away from it. Furthermore, the tetrons, from which the crystal is made, are assumed to be so strongly bound, that they can be split off the hyper-crystal only by supply of Planck scale energies. 
\subsubsection{Was there a GUT era in the early universe where electroweak and strong couplings were unified?}\label{appii729}
No, because there is no GUT -- cf. (\ref{appii531}). In the tetron model the strong force has a different origin than the electroweak one (cf. \ref{appiiqcd}), and thus GUT unification seems unlikely.\\
%In general, all known forces, including gravity, stem from fundamental interactions among tetrons in 6+1 dimensions.
The proper history of the universe starts with the end of the crystallization process (=the inflation era), at which point electromagnetism and weak forces are unified, cf. (\ref{appii206k}). In the standard terminology this is the starting point of the radiation dominated era, with photons, effectively massless W/Z and dark matter as the dominant excitations. At the end of the electroweak era at temperatures of order $\Lambda_F$ there is the alignment of isospin vectors corresponding to the electroweak phase transition which gives masses to q/l and W/Z.
\subsubsection{Is there a unification of the SM and gravitational forces at the Planck scale?}\label{appii729a}
Not in the sense of supergravity and related models. Even at big bang temperatures gravity and SM forces have a very different nature. Although it is true that everything observed can eventually be derived from the fundamental forces among tetrons, the SM interactions trace back to interactions between isospin vectors of tetrons, whereas gravity is an elastic force between tetrahedrons which stems from remnant tetron coordinate interactions.\\
The question, why the forces of gravity appear to be much weaker than particle physics effects, has been treated in (\ref{appii312}).  
\subsubsection{Are there dark matter candidates in the model?}\label{appii811}
Yes, there are several possibilities:\\
%(ich tendiere jetzt zu ersterem. wg fehlendem WR gibt es vielleicht auch kein eta, und die spin3/2 usw hat man nur, wenn Minkraum ein S4 Gitter hat)
--further internal excitations of the crystal, like phinons, cf. (\ref{appii536}) and \cite{la4xz2}. 
%Could be relevant mass/energy store at high temperatures where the isospin alignment is destroyed.
Their interactions with mignons(=ordinary matter) is tiny, because they are not involved in the isomagnetic correlations giving rise to the SM.\\
%--excitations of spin 3/2 and 5/2 simlar to those discussed in \cite{lcos}.\\
%Ausserdem masses probably to high, typically of order of crystalliz energy.\\
--the pseudoscalar $\eta$ arising in the 2HDM ansatz (\ref{appii207}). Such a possibility is widely discussed in the literature \cite{inert1,inert2} provided the $\eta$ is inert, i.e. does not interact with quarks and leptons. This condition can be fulfilled in the present model essentially because of (\ref{phiss}). Note that right after inflation there is the radiation dominated era where the inert scalar is copiously produced together with a soup of many other tetron-antitetron bound state excitations (photon, W/Z, $\Phi$ and $\Phi'$).
%(nicht bringen: in \cite{lampeastro} it was suggested that the tetrahedral symmetry is relevant not only for internal but also for a discrete structure in physical space. (S8 aus 6 Teilchen. da das raumzeitgitter elastisch fluktuiert, hat es keine symm außerdem bich sowieso von 8 abgekommen, weil es 4*(Ql+QR) sind) This somehow contradicts the idea of an elastic structure, although it is possible that such a symmetry perceives in some average sense. In any case there would be a separate world of spin-$\frac{3}{2}$ and spin-$\frac{5}{2}$ excitation in addition to the spin-$\frac{1}{2}$ quarks and leptons. From the multiplicity of states one can infer that the density of ordinrary matter to spin-$\frac{3}{2}$ and spin-$\frac{5}{2}$ matter is roughly 1:5. Again there is the question, why not couple to the electroweak bosons.)
\subsubsection{Is there an explanation for dark energy?}\label{appii81de}
Observations indicate that the universe's expansion rate was decelerating until about 5 billion years ago, after which time the expansion began re-accelerating. Phenomenologically, this can be explained by including a cosmological constant $\Lambda_C$ in the theory (this amounts to saying that a volume in space has some intrinsic fundamental vacuum energy creating a pressure which makes the universe expand) or by a weakly fluctuating scalar 'quintessence' field (this does a similar job).\\
Furthermore, such a type of energy that is not matter or dark matter is also needed to explain the apparent flatness of the universe (absence of any detectable global curvature). According to that argument the contribution of dark energy should be more than twice as large as that of matter and dark matter together.\\
A third explanation of the dark energy effect is offered within the f(R,T) models (\ref{tmet5}) discussed in section 1 by suitable accommodation of the 11 phenomenological coupling constants\cite{wlu}.\\
%This is sometimes considered superior to the cosmological constant, because the value of $\Lambda_C$ needed for dark energy is of unnatural size.(aber ich dachte die size waere wie die krit density unddaher ok) It should be noted, though, that f(R,T) models with their many effective interactions offer only a phenomenological description of gravity.\\
To understand dark energy within the tetron model one should remember, that the micro-elastic forces that have initially induced cosmic expansion during the big bang crystallization process are still at work today. For example, there may still be accretions to the hyper-crystal at its edges which are setting free large amounts of crystallization energy. The energy is then transferred to the other tetrahedrons of the crystal in the form of weakly fluctuating energy/density waves traveling through the universe. This picture is well along the line of the quintessence idea mentioned above.\\
Another possibility is that the increased acceleration arises, because the average distance $L_P$ between tetrahedrons has not yet reached its equilibrium value $r_0$. This line of argument was followed in (\ref{appiiharm}) and gives rise to a self-contained explanation of the dark energy effect. 
\subsubsection{Are there other universes?}\label{appii816}
Probably yes. One may consider the original $R^{(6,1)}$ spacetime as a container of universes. When the tetron gas cooled down and temperatures reached the crystallization temperature (Planck energy), germs of 3+1-dimensional hyper-crystals came into being in various places of $R^{(6,1)}$, cf. (\ref{appii535}). These crystals then grew in their respective 3+1 dimensionsal subspaces, each of them making up for a separate $R^{(3,1)}$ spacetime. Since they are of low dimension as compared to the whole $R^{(6,1)}$, they hardly interfere with one another. If at all, they intersect in isolated (1-dimensional) points. At those points a defect in the isomagnetic and/or coordinate crystal structure will show up, because the isospin and/or coordinate vectors of the tetrons do not know how to orient themselves.
%\\It remains pure speculation whether matter/energy can be transferred from one hyper-crystal universe to another at those points.
%BLACK HOLE IST WAS ANDERES: grosse MATERIAL AGGREGATION This then yields an addendum to what happens in a black hole. It is sitting at the intersection point of 2 hyper-crystals. Not only is matter attracted to the black hole but it is blown into another universe, or maybe even used to build up a new hyper-crystal. 
%Black hole saugt Materie an und blaest in ein anderes Universum. Es ist Schnittpunkt 2er 3dim Kristalle. Materie stroemt an einer Seite ein, um auf der anderen ein neues Kristall aufzubauen. Denn i.a. ist der Schnittpunkt 2er R3 in R6 genau der 0-Punkt des Koordinatensystems!!!! siehe x1=x2=x3=0 und x4=x5=x6=0. (bei schiefliegenden zb x1=x2=x3=0 und 3 Gleichungen sumaixi=0, sumbixi=0,sumcixi=0 ergibt sich auch genau 1 Punkt als Schnittpunkt)
\subsubsection{On the interpretation of black holes in the tetron model}\label{appiibb}
From its very nature the discrete elastic hyper-crystal does not allow for mathematical singularities. It is true that black holes correspond to solutions of the Einstein equations, and the Einstein equations according to section 1 arise from the effective action (\ref{tmet5}) for the DMESC. However, these equations are not applicable at arbitrary small distances / high energies, where the discrete structure becomes perceptible.\\
It is generally believed, that if enough mass M is squeezed into a roughly spherical volume of size $r=GM/c^2$, it collapses into a black hole. In the Einstein theory, the geometry of a black hole can be understood via the Schwarzschild metric (\ref{schsch}). 
%regardless of internal pressure or other opposing forces.
What happens from the standpoint of the tetrahedrons is that inside the black hole's event horizon the crystal becomes extremely compressed, i.e. the distances between the tetrahedrons become smaller and smaller. At the same time the temperature strongly increases as more and more matter (mignons, gauge bosons and other quasi-particle excitations) is accreted. When the temperature exceeds the Fermi scale, the isospin alignment of the hyper-crystal gets lost and the accreted mignons decay to photons and weak gauge bosons. In some sense this scenario is reverse to the appearance of the 'radiation dominated' epoque of the big bang. Finally, if temperatures reach the order of the crystallization energy $\Lambda_P$, the DMESC structure completely dissolves and the tetrahedrons vaporize 
to form a gas which is set free into the full $R^{(6,1)}$, i.e. into the internal directions.\\
%There is then so much energy concentrated near one point, that the hyper-plastics dissolves and a tetron gas emerges 
%It is true that outside a black hole the temperature is small, but inside it becomes extremely hot, so that finally the tetrahedral structure is destroyed and free tetrons are released into 6-dimensional space where 
If enough energy would be available, this hot gas would become the germ of another hyper-crystallization process making up for another universe in the sense of (\ref{appii816}). One could then distinguish 'parent' and 'child' universes in an obvious sense. If our own universe would have been created as a child under such circumstances, the appearance of a big bang with a radially symmetric expansion like desribed by the FLRW metric (\ref{flrw3a}) would be somewhat easier to understand than in the picture developed in section 1 where accretions of tetrons to the edges of the hyper-crystal could in principle provoke deviations from an FLRW behavior, cf. (\ref{appii81de}).\\
For ordinary black holes, however, energies and temperatures are by far not high enough for such a process to start. Inside the black hole there may be a certain dissolution of the hyper-crystal; however due to the elasticity of the tetronic material this is reversible and can be repaired rather easily. Black holes are thus able to wander on the hyper-crystal with the same peculiar velocities as the galaxies in which they originally appeared.
%(aber wuerde nicht a(t) modifiziert, weil staendig neues gas aus dem black hole nachfliesst?)
%\subsubsection{Are black holes and the big bang spacetime singularities?}\label{appiibb7}
\subsubsection{Should gravity be quantized?}\label{appii806}
%Gravitonen sind Phononen des elastischen Kristalls. In contrast to phinons they modify the metric.
%Should gravity be quantized? nein, da wir so eine eff theory nicht quantisieren muessen, aber phononen muss man auch quantisieren, obwohl sie nur schallwellen sind.
%Aber gehören Phonon-Fluktuationen D eines Gitters nicht zur Translationsgruppe (also Poincaregruppe)? Das entspricht der Teleparallel Interpretation der Gravitation.
%\subsection{Questions about gravity}
As emphasized in (\ref{appii1}), the quantum behavior of nature is closely related to the granularity of physical space. Therefore it seems natural to believe that in circumstances where this discrete structure becomes relevant, gravitational effects should be treated in a quantum theoretical manner.\\ 
However, gravity in the tetron model is an effective interaction of (internal) tetrahedrons in an elastic/plastics system. Its description by the Einstein-Hilbert action or its generalization (\ref{tmet5}) is valid only at distances $\gg L_P$, i.e. looses its validity when probed at distances where the discrete structure becomes apparent, cf. the discussion in (\ref{appiibb}). Instead of 'quantizing' gravity one should quantize the fundamental interaction among tetrons.
\subsubsection{Why is the speed of gravitational waves equal to the speed of light? Why is there a universal maximum speed for all the objects in the universe?}\label{appii80d3}
%As discussed in the last question and in section 1, in the tetron model there are elastic waves which arise from vibrations of the rigid internal tetrahedrons inside the hyper-crystal. In the Einstein theory gravitational waves are transverse metrical waves, whose velocity is forced to be equal to the speed of light by the condition of local Lorentz invariance.\\
This question has already been answered in (\ref{appii7mx}), (\ref{appii7m8}) and at the end of section 1. In this section I want to add some more comprehensive remarks about the topic. In section 1 general relativity has been interpreted as an effective theory for an elastic system of internal tetrahedrons. Gravitational waves exist in this theory on the classical level, as solutions to the Einstein equations. They are metrical waves, whose velocity is forced to be equal to the speed of light by the condition of local Lorentz invariance. In the tetron model they can be associated to some of the density fluctuations of tetrahedrons\footnote{In distinction the {\it internal} translational excitations were coined phinons in \cite{la4xz2}, while the internal rotational excitations are mignons, i.e. quarks and leptons.} discussed in connection with (\ref{eqfe}).\\
%Since one would like to identify some of the elastic waves with the gravitational waves, 
In the tetron theory elastic waves propagate at the speed (\ref{ddot83}) of a typical excitation in an elastic medium, and thus at the same speed as photons and massless mignons, cf. (\ref{tm38ddem2222}). At first sight this looks like a rather amazing feature, because it was argued in (\ref{appii729a}) that the isomagnetic particle physics interactions of the photon and the elastic gravitational interactions do not have much in common.\\
In order to get a clearer understanding one should realize that the world according to the microscopic model falls apart into 2 rather disparate pieces:\\
-the realm of what philosophers would call emergent or appearance phenomena, i.e. isomagnetic quasi-particles like quarks, leptons, Higgs and gauge fields. Since all these excitations fulfill Lorentz invariant wave equations, any phenomenon and signal propagation in this sphere is necessarily limited by the speed of light.\\
-the realm of what could be called 'true' or tetron matter, consisting of tetrons, of aligned tetrahedrons and of the DMESC with its elastic/metric structure. This may rightfully be called $ ' \upsilon\lambda\eta \,\,\, \pi\rho\omega\tau\eta$. However, while for Aristotle this was more an idea than a concrete material, here it can be understood in a real, {\it materialistic} sense. Just for joke one could call it 'tatter' to distinguish it from the ordinary mignon material which remains m-atter.\\
Since the relevant scales $\Lambda_P \gg \Lambda_F$ are so vastly different, these two spheres do not have much in common. We ourselves live in the sphere of appearances and can perceive anything coming from the tatter sector only if suitable devices of mignon matter are patched in between. Gravity, for example, which originally corresponds to a shift of tetrahedron locations on the DMESC, becomes visible in our physical world only due to the small and stiff reaction to suitable conglomerations of m-atter. In particular, the physical effects of a gravitational wave can only be seen by plugging appearances in between, i.e. m-atter which 'rides' the gravitational waves. The principle of relativity states, that m-atter must respect local Lorentz invariance. Therefore, although the fundamental interactions among tetrons may proceed at larger velocities than c, the gravitational interactions between m-atter particles always appear to proceed at c.\\
Let me repeat that these arguments are supported by considering the dispersion relation for mignons (\ref{tm38ddem22}) and (\ref{tm38ddem2222}). As shown at the end of section 1, mignon propagation can be completely described by c, which itself according to (\ref{tm33chi8}) is given in terms of tetron variables as $c^2=\zeta/\rho_T$.
\subsubsection{How can metric velocities larger than c be interpreted in the tetron model?}\label{appii80d5}
This question is related to the discussion in (\ref{appii7r1}),(\ref{appii7xx2}), (\ref{appii7r2}), (\ref{appii7a90}) and (\ref{appii80d3}). According to (\ref{tm38ddem})-(\ref{tm38ddem2222}), c is the maximum speed for all the isomagnetic quasi-particles that build our known universe. However, this limit does not apply to tetrons nor to the bound tetrahedrons which make up the hyper-crystal and are the carriers of the quasi-particles. As evident from (\ref{tmet}), in the tetron model metrical changes are associated to displacements of tetrahedrons. The corresponding velocities of the tetrahedrons have been particularly large ($> c$) in the inflationary period shortly after the big bang (crystallization) where a lot of crystallization energy had been released.\\
According to (\ref{flrw3c}) one can roughly identify the metric velocity in an FLRW universe with the Hubble flow $Hd$. In the tetron model this can be interpreted as the relative velocity of 2 tetrahedrons at distance d.
%\subsubsection{Do velocities larger than c violate causality?}\label{appii8sd5}
%They would violate causality, only if signal transmission could proceed in our world of excitations at larger than c. This however by the argument given at the end of (\ref{appii80d3}) is not possible.

%-------------------------------------------------------------

\section{Conclusions}

%It is widely believed that the Standard Model of elementary particles is only an effective low energy theory valid below a certain energy scale, which is supposed to be of the order of 1-10 TeV. This view is based on the fact that the SM has many unknown parameters, most notably the quark and lepton masses, and one rather mysterious component, the so-called Higgs field, which is needed for the spontaneous symmetry breaking (SSB) to take place in the model.

%In the mainstream approaches to particle physics, a physical understanding of the underlying dynamics responsible for these effects is still lacking. For example, in (supersymmetric) grand unified theories fermion masses essentially remain free parameters. Furthermore, those models usually introduce many more additional degrees of freedom without much ambition to determine them from first principles. The point is that theories of that kind only extrapolate and extend the symmetries observed at low energies to small distances and, as can be concluded from the variety of theories floating around, that there is a strong amount of arbitrariness in this procedure.

%In my opinion it is obvious that a physical understanding of the masses and mixings is only possible in a microscopic theory. Superstring theories seem to offer such an understanding. However, although 'in principle' able to determine the masses as energies of string excitations, to my knowledge they have not come up with definite and verifiable predictions.

The present review is devoted to a model which tries to give a microscopic meaning to physical phenomena usually described by the Standard Model of elementary particles. By introducing an additional level of matter one is able to understand and to calculate known particle properties (like the quark and lepton masses and mixings) from first principles and furthermore to make predictions for future experiments. Most prominent among the latter are:\\
--the existence of a second Higgs doublet similar as in inert 2HDM models\cite{inert1,inert2}.\\
%, as discussed in (\ref{appii207})\\
--the existence of a fourth family of quarks and leptons.
%, as discussed in (\ref{appii536}). 
This family, however, is distinct from the other three, not only because it has a very massive neutrino but also because its couplings are not given by the SM. The point is that the 8 Dirac particles of this family do not arise from vibrations of iso-'magnetizations' $\psi^\dagger \vec\tau\psi$ but of internal 'densities' $\psi^\dagger \psi$, and therefore they do not obtain their masses via the Higgs mechanism and the SSB fig.2.      

After discussing particle physics properties, implications of the tetron model on the big bang and on phase transitions in the early universe have been elucidated. This has led to the idea that besides the\\
--isomagnetic interactions among aligned isospin vectors which are relevant for particle physics\\
one should consider 2 other forces:\\
--strong rigid coordinate forces among tetrons fixing the form and extension of the (internal) tetrahedrons.\\
--weak elastic forces between these tetrahedrons, which are the basis of the gravitational interactions.\\
All 3 types of forces are assumed to derive from one universal interaction among the tetrons, and it was suggested that one should use octonion multiplication and perhaps supersysmmetry as a guideline which eventually will lead to the correct renormalizable theory in 6+1 dimensions.

The various viewpoints on the model presented in the preceding sections have supplied a set of important requirements as to 
%have set interesting restrictions on 
the nature of tetron interaction. We already know that\\ 
--tetrons have a tendency to form excited pairs with antitetrons of neighboring tetrahedrons in the crystal with similarity to Cooper pairs or Frenkel excitons of solid state physics.\\
%There is on the one hand the (elastic) inter-tetrahedral coordinate interactions that keeps the crystal together (has to do with gravity), and on the other hand the inter-tetrahedral isospin interactions which have to do with the vector bosons and the Higgs particle.\\
--tetron bonds are extremely short, of the order of the Planck length.\\ 
--they get saturated in tetrahedral configurations.\\
--these configurations form 3-dimensional monolayer crystal structures, i.e. there is no stacking of tetrahedrons on top of each another, no growth of the hyper-crystal into internal directions.\\
--isospins within the quartets of tetrons are maximally frustrated (fig. 1).\\
%--that the spins of the 4 tetrons are identical and the mignon eigenfunctions carry this spin to become the spin of the quark and lepton under consideration.\\
--once tetrons are in such a saturated hyper-crystal configuration, there is a left-over elastic force among the internal tetrahedrons, which gives rise to the gravitational interactions.

%die wichtigsten der ungeloesten Fragen diskutieren

%highlights, dass gamma composite ist, nur wg U1fermion masselos und dass vor der Bildung von Photonen bzw des Kristalls gar kein Minkraum ist, sondern Galilei im 6dim Raum
%daraus folgt:No SR, no AR outside the crystal

%Starting from a new, fundamental fermion field in 6+1 dimensions it was shown that\\
In summary a new picture of the physical world was presented. It was shown that\\
--quarks and leptons can be interpreted as internal magnons of a discrete iso-magnetic structure and that their spectrum is due to a tetrahedral symmetry group which remains unbroken down to the lowest energies\\
--the SU(2)$\times$U(1) gauge group of the SM is related to an iso-magnetic 'Heisenberg' SU(2) and the SM SSB can be obtained from a global ordering of internal magnets\\
--the big bang is due to a crystallization process which has released a large amount of latent energy and has led to the formation of a rapidly expanding 3+1 dimensional universe\\  
--the physical world including the gauge bosons consists of excitations which travel as quasi-particles through the hyper-crystal. Since they are described by relativistic wave equations, this leads to the appearance of Minkowski space (and the associated Lorentz structure).\\
--an 'elastic' rest system of the hyper-crystal exists which is identical to the cosmological comoving coordinates. However, due to the quasi-particle nature of our physical world this cannot be observed in Micholsen-Morley type of experiments.

%-------------------------------------------------------------

\end{document}